\documentclass{article}

\usepackage{arxiv}

\usepackage[utf8]{inputenc} 
\usepackage[T1]{fontenc}    
\usepackage{hyperref}       
\usepackage{url}            
\usepackage{booktabs}       
\usepackage{amsfonts}       
\usepackage{nicefrac}       
\usepackage{microtype}      
\usepackage{lipsum}
\usepackage{graphicx}
\graphicspath{ {./images/} }
\usepackage[fleqn]{amsmath}
\usepackage{floatrow}
\usepackage{floatrow}
\usepackage[label font=bf,labelformat=simple]{subfig}
\usepackage{caption}

\title{Symmetry Effect on the Dynamic Behaviors of Sandwich Beams with
Periodic Face Sheets}

\author{
 Eshagh Farzaneh Joubaneh \\
  Department of Mechanical Engineering\\
   University of Vermont\\
  Burlington, VT 05405, USA \\
  \texttt{efarzane@uvm.edu} \\
   \And
 Jihong Ma \\
  Department of Mechanical Engineering\\
  University of Vermont\\
  Burlington, VT 05405, USA \\
  \texttt{Jihong.Ma@uvm.edu} \\
 }

\begin{document}
\maketitle
\begin{abstract}
We investigate theoretically and demonstrate experimentally, the effect of material and geometric periodicity in face sheets on the overall dynamics of sandwich beams. By treating the two face sheets and core as separate entities using the classic sandwich plate theory (CPT), in conjunction with Bloch’s theorem, and solving the governing equations with the transfer matrix method (TMM) and the generalized differential quadrature method (GDQM), we obtain a low-frequency \textit{total} bandgap in which both transverse and longitudinal waves are blocked in sandwich beams with glide-reflected (GR) face sheets. On the other hand, only \textit{partial} bandgaps blocking either transverse or longitudinal waves are present in the ones with mirror-reflected (MR) face sheets. Furthermore, a combination of full-scale finite element method (FEM) simulations and experimental characterization provide unequivocal evidence of these phenomenal wave features, as well as the accuracy of our theoretical results via defining separate displacement fields in comparison with effective periodic Timoshenko beam models. Our study indicates new mechanisms to suppress low-frequency elastic waves by tuning face sheet symmetry. Our explanation of wave behaviors from a structural symmetry point of view also provides a more comprehensive understanding of the symmetry effect on wave propagation in quasi-one-dimensional structures at large.
\end{abstract}


\section{Introduction}
Vibration is prevalent in engineering applications, such as civil, transportation, aerospace, and marine industries. Isolating or filtering vibrational waves has been a great challenge and has attracted significant attention. Among the various vibration reduction and filtering approaches, using sandwich structures to mitigate vibration is a popular choice due to their high specific stiffness and strength, exceptional energy absorption capability, and low structural weight \cite{birman2018review}, and has since been considerably developed for over half a century \cite{hoff1945buckling}. The light and soft but thick core, and strong yet thin face sheets constitute an overall sandwich structure with low weight and high rigidity  \cite{vinson2018behavior}. \\\indent
Various methods have been proposed to suppress vibrational wave propagation using sandwich structures. Many researchers have been focusing on adjusting core properties, such as using core materials with low density and high thickness \cite{tilbrook2006impulsive,karagiozova2009response} and with honeycomb patterns \cite{karagiozova2009response}. With a growing number of studies on acoustoelastic metamaterials and phononic crystals, researchers have started to embed these artificially architected structures with wave-manipulating capabilities in sandwich cores to further filter out waves \cite{chen2011dynamic,chen2017wave,zhu2014chiral,sharma2016local,Liu2018Broadband,MENG2017Small,chen2016wave,russillo2021free}. For example, Chen \textit{et al.} \cite{chen2011dynamic} and Russillo \textit{et al.} \cite{russillo2021free} implanted resonators in the core to create a forbidden range of frequencies for wave propagation, \textit{i.e.}, a bandgap, between the acoustic and optic phonon modes. Instead of treating face sheets, foam core, and resonators separately, the authors considered face sheets and core as one effective beam, with resonators discretely attached to the beam. Another popular mechanism of wave attenuation is via using periodic core materials in sandwich beams \cite{chen2013wave,chen2016sandwich,sheng2018vibration,guo2017flexural,ruzzene2003control,ruzzene2002wave,liu2014wave}. For example, Liu \textit{et al.} also sandwiched a two-dimensional chiral metacomposite into a beam frame to form low-frequency bandgaps \cite{liu2011wave}. Moreover, Gupta and Ruzzene used alternating quasiperiodic auxetic honeycomb core to create non-trivial bandgaps in the form of Hofstadter butterfly to modulate wave propagation \cite{Gupta2020Dynamics}. As we can see, most efforts to date have been devoted to the analysis of core properties on wave propagation with the notable exception of only a few studies of bandgap induction using face sheets attached with periodically stepped resonators via the finite element method (FEM) \cite{song2015reduction,song2019suppression}, despite the nontrivial role face sheets are playing in dynamics of sandwich beams.

In this paper, we shift our attention to the effect of face sheets with material and geometric periodicity on the overall dynamic behaviors of sandwich beams.
Inspired by the study of periodic flexural beams \cite{liu2012wave}, we start with the investigation of periodic face sheets that present a mirror-reflection (MR) symmetry about the axial direction of core (\textit{i.e.}, $x$-direction), as shown in Fig. 1 (a), (c), and (e), which can also be analyzed with equivalent Timoshenko beams as presented in Ref. \cite{liu2012wave}. Then we glide the bottom face sheet along the horizontal direction (\textit{i.e.}, $x$-direction) to obtain a glide-reflection (GR) symmetry about the core, $i.e.$, ${\textbf{\textit{T}}}_{G,0}$${\circ}$${\textbf{\textit{r}}}_{y=0}$, where $G$ is the glide magnitude along the $x$-direction, $y=0$ is the neutral axis of the sandwich beam. Such a glide introduces a complication, preventing us from treating our sandwich beams with an equivalent beam model. Hence, we will probe our systems with separate displacement fields using the classic sandwich plate theory (CPT). We obtain equations of motion using Hamilton's principle. Both the transfer matrix method (TMM) and generalized differential quadrature method (GDQM) are adopted, in conjunction with Bloch's theorem \cite{floquet1883equations,bloch1929quantenmechanik}, to obtain band diagrams for our sandwich beams with periodic face sheets. We then validate our theoretical results against ANSYS simulations and experimental characterization. \indent

This paper is arranged as follows. Section 2 presents governing equations and their solution derivation using TMM and GDQM. Comparison of results from theoretical analysis, numerical simulation, and experimental validation is discussed in Section 3. Section 4 concludes our findings of the relationship between face sheet symmetry and wave propagation in sandwich beams.

\section{Theory}
\subsection{Governing equations}

Consider an infinitely long three-layer sandwich beam in the Cartesian coordinate system, as shown in Fig.~\ref{F2}. The axial and transverse displacements in $x-$ and $y-$ directions at time instance, $\tau$, of each layer of the sandwich beam are denoted as $U_i(x,y,\tau)$ and $W_i(x,y,\tau)$, respectively with subscript $i \in \{c,t,b\}$, where $c$, $t$, and $b$ refer to the core, and top and bottom face sheet displacements, respectively. According to the CPT, face sheets and the core are treated as Euler-Bernoulli and Timoshenko beams, respectively. The longitudinal and transverse displacements for the core $(U_c$ and $W_c)$, top $(U_t$ and $W_t)$, and bottom $(U_b$ and $W_b)$ layers are expressed as

\begin{equation}
Core\:(\text{$-h_c/2<y<h_c/2$})\:\:\:\:\:\:\:\:\:\:\:\:\:\
\begin{array}{ll}
\end{array} \quad\left\{\! \begin{array}{ll}
U_c(x,y,\tau)=  u_c(x,\tau)+\varphi(x,\tau)y\\
W_c(x,y,\tau)= \displaystyle w_c(x,\tau)\end{array}\right.
\end{equation}
\begin{equation}
Top\:face\:sheet\:(\text{$-h_t/2<y<h_t/2$})\:\:\:\:\:\
\begin{array}{ll}
\end{array} \quad\left\{\! \begin{array}{ll}
U_t(x,y,\tau)=u_t(x,\tau)-yw_t'(x,\tau)\\
W_t(x,y,\tau)=w_t(x,\tau)\end{array}\right.
\end{equation}
\begin{equation}
Bottom\:face\:sheet\:(\text{$-h_b/2<y<h_b/2$})
\begin{array}{ll}
\end{array} \quad\left\{\! \begin{array}{ll}
U_b(x,y,\tau)=u_b(x,\tau)-yw_b'(x,\tau)\\
W_b(x,y,\tau)=w_b(x,\tau)\end{array}\right.
\end{equation}

The updated form of displacement fields by enforcing the displacement compatibility conditions at the layer interfaces, $i.e.,$ $U_c(x,\frac{h_c}{2},\tau)=U_t(x,-\frac{h_t}{2},\tau)$, $U_c(x,-\frac{h_c}{2},\tau)=U_b(x,\frac{h_b}{2},\tau)$, $W_c(x,\frac{h_c}{2},\tau)=W_t(x,-\frac{h_t}{2},\tau)$, and $W_c(x,-\frac{h_c}{2},\tau)=W_b(x,\frac{h_b}{2},\tau)$, are presented in Appendix A.  
\noindent

    \begin{figure}[H]\centering
\subfloat[]{\includegraphics[width=0.4\textwidth]{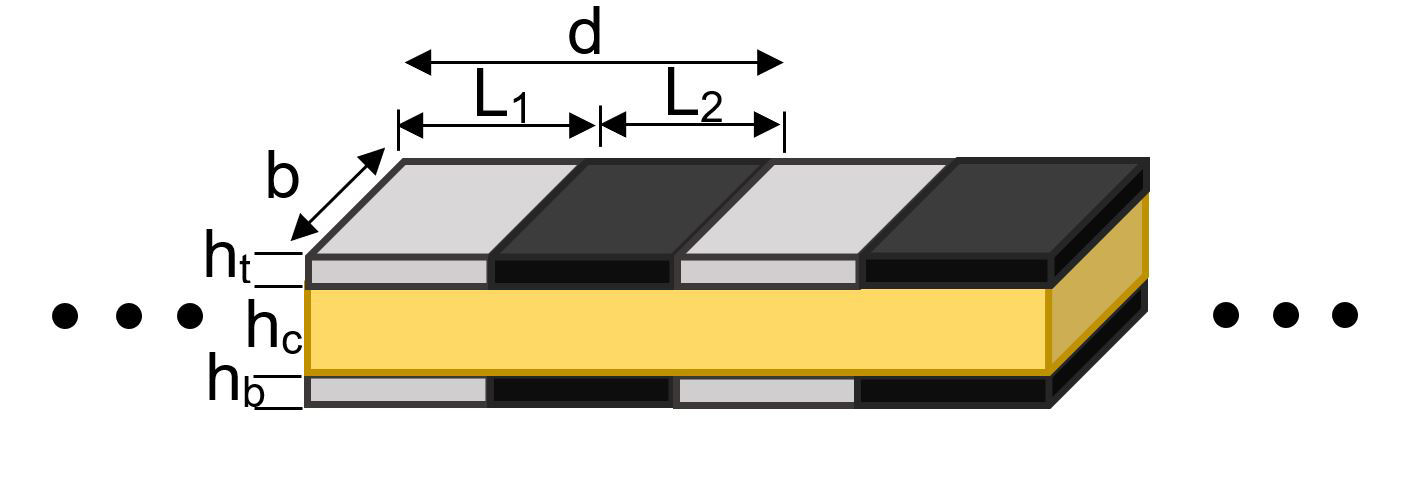}\label{fig:a1}}
\subfloat[]{\includegraphics[width=0.4\textwidth]{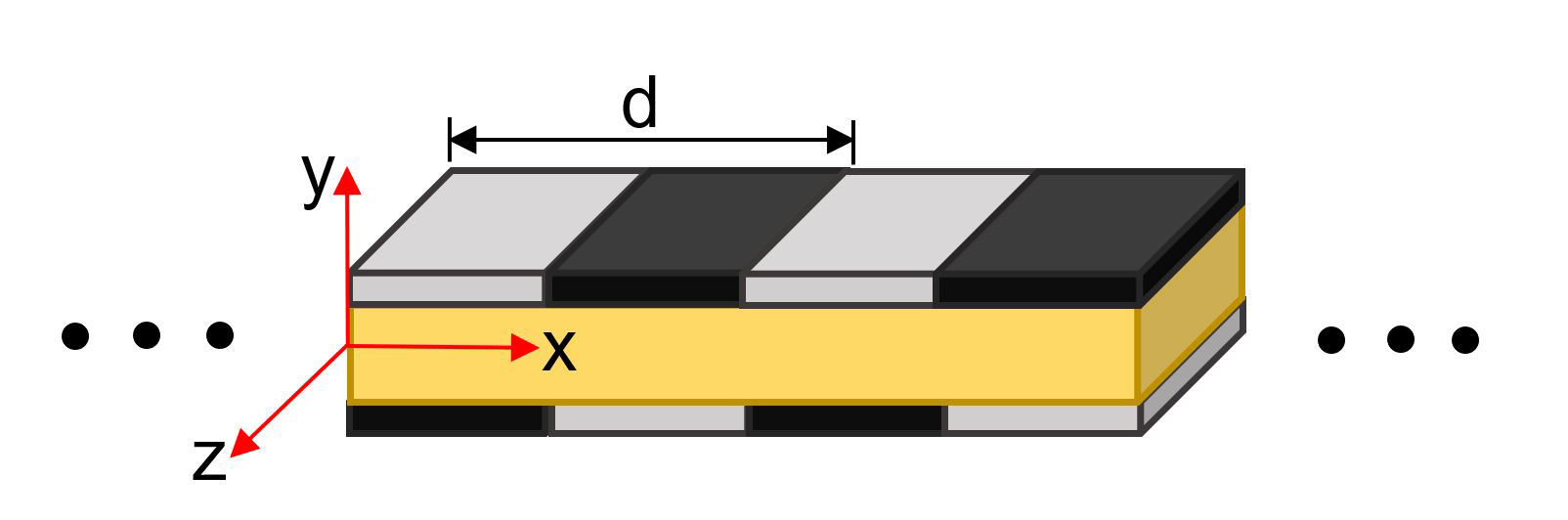}\label{fig:b1}}\hfil
\subfloat[]{\includegraphics[width=0.4\textwidth]{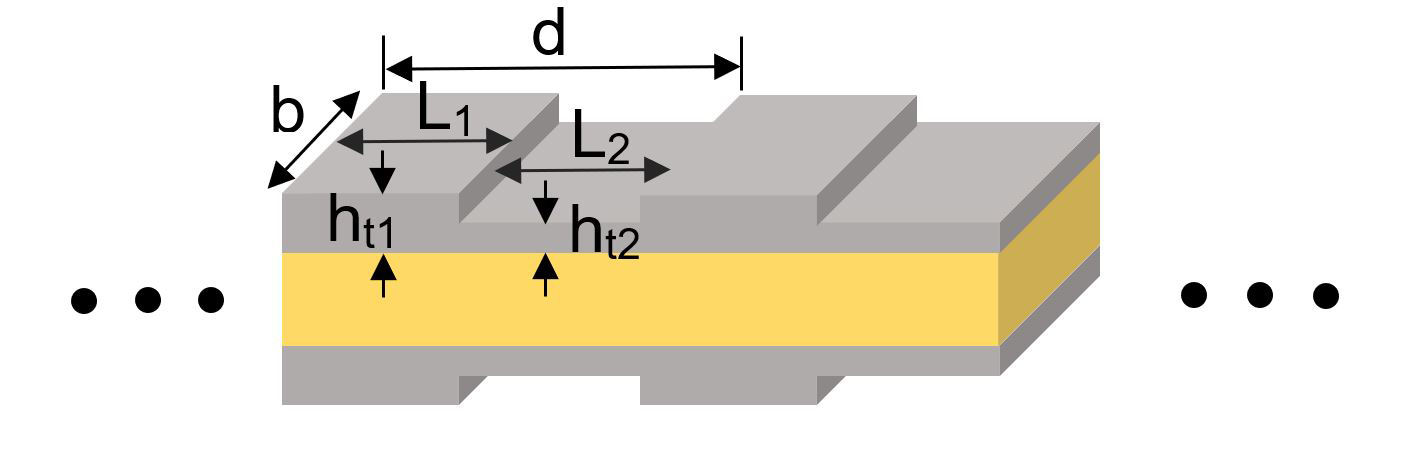}\label{fig:c1}}
\subfloat[]{\includegraphics[width=0.4\textwidth]{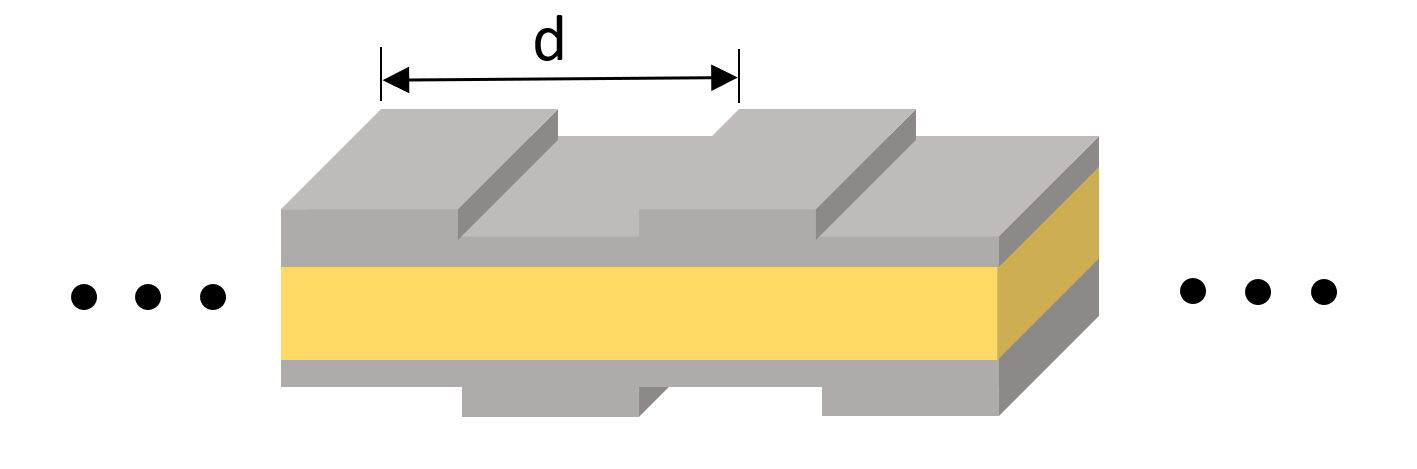}\label{fig:d1}}\hfil
\subfloat[]{\includegraphics[width=0.4\textwidth]{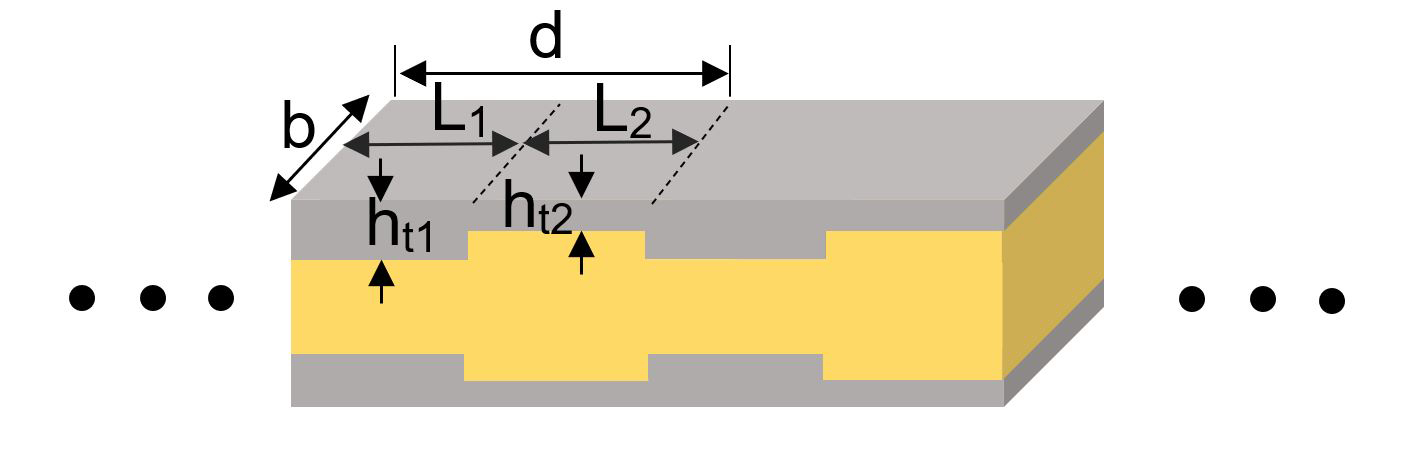}\label{fig:e1}}
\subfloat[]{\includegraphics[width=0.4\textwidth]{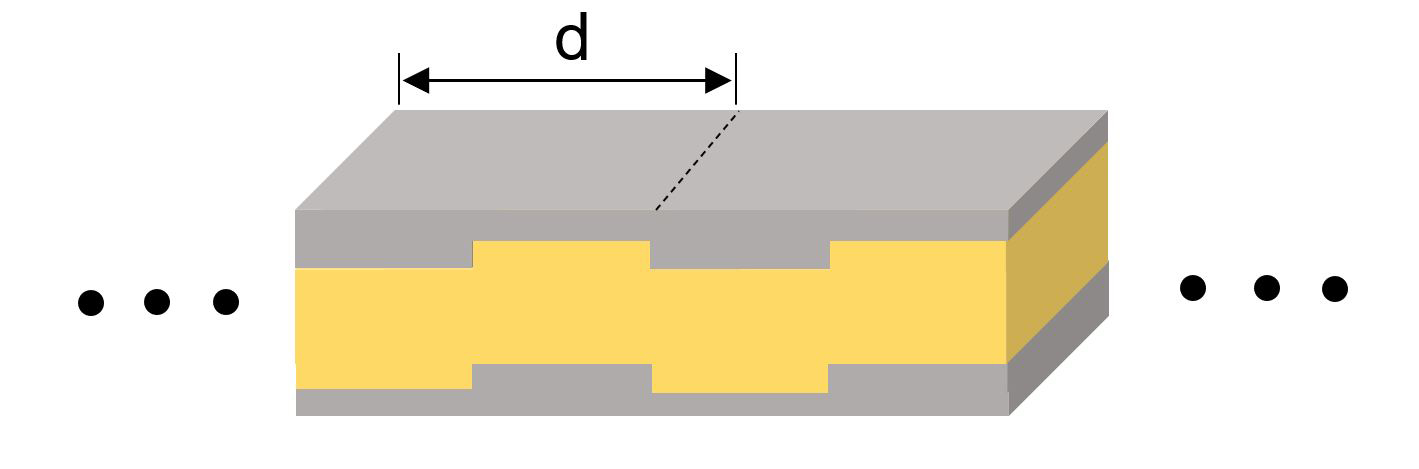}\label{fig:f1}}
\caption{Schematic representations of sandwich beams with different types of periodicity. Our sandwich beams are composed of unit cells with material (a,b) and geometric (c-f) periodicities in face sheets along the $x$ direction in the $x-y$ plane in which $z$ is in the beam width direction. Unit cells of the sandwich beam are composed of either mirror-reflected (a,c,e) or glide-reflected (b,d,f) face sheets with $G=d/2$. Here, $d$ is the lattice constant, $L_1$ and $L_2$ are the subunit cell's length, $h_c$, $h_t$, and $h_b$ are the thicknesses of the core and top and and bottom face sheets respectively, and $b$ is the width of the sandwich beam.}
    \label{F1}
    \end{figure}

 \begin{figure}[H]\centering
\subfloat[]{\includegraphics[width=3.5in]{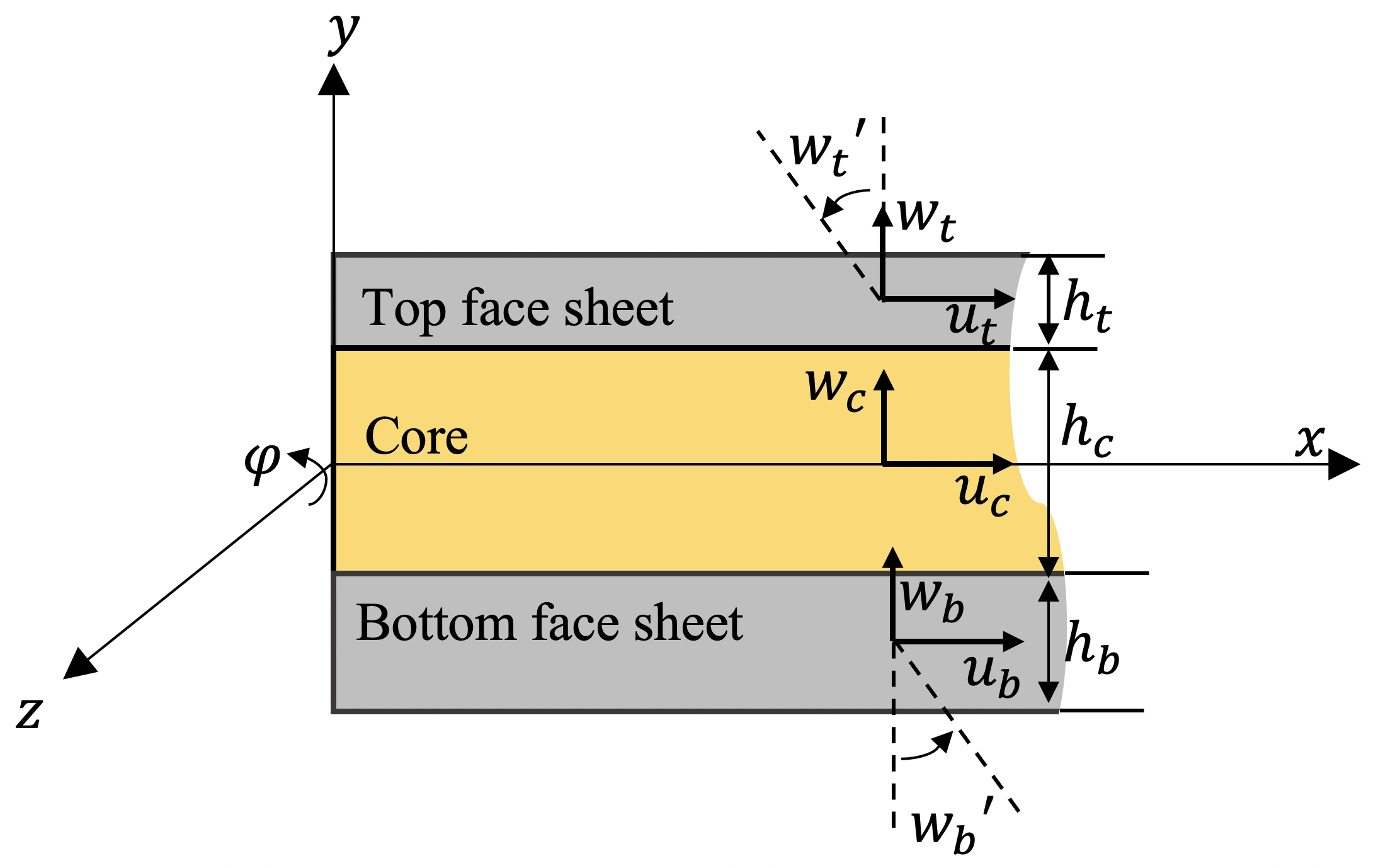}\label{fig:a2}}
\subfloat[]{\includegraphics[width=0.4\textwidth]{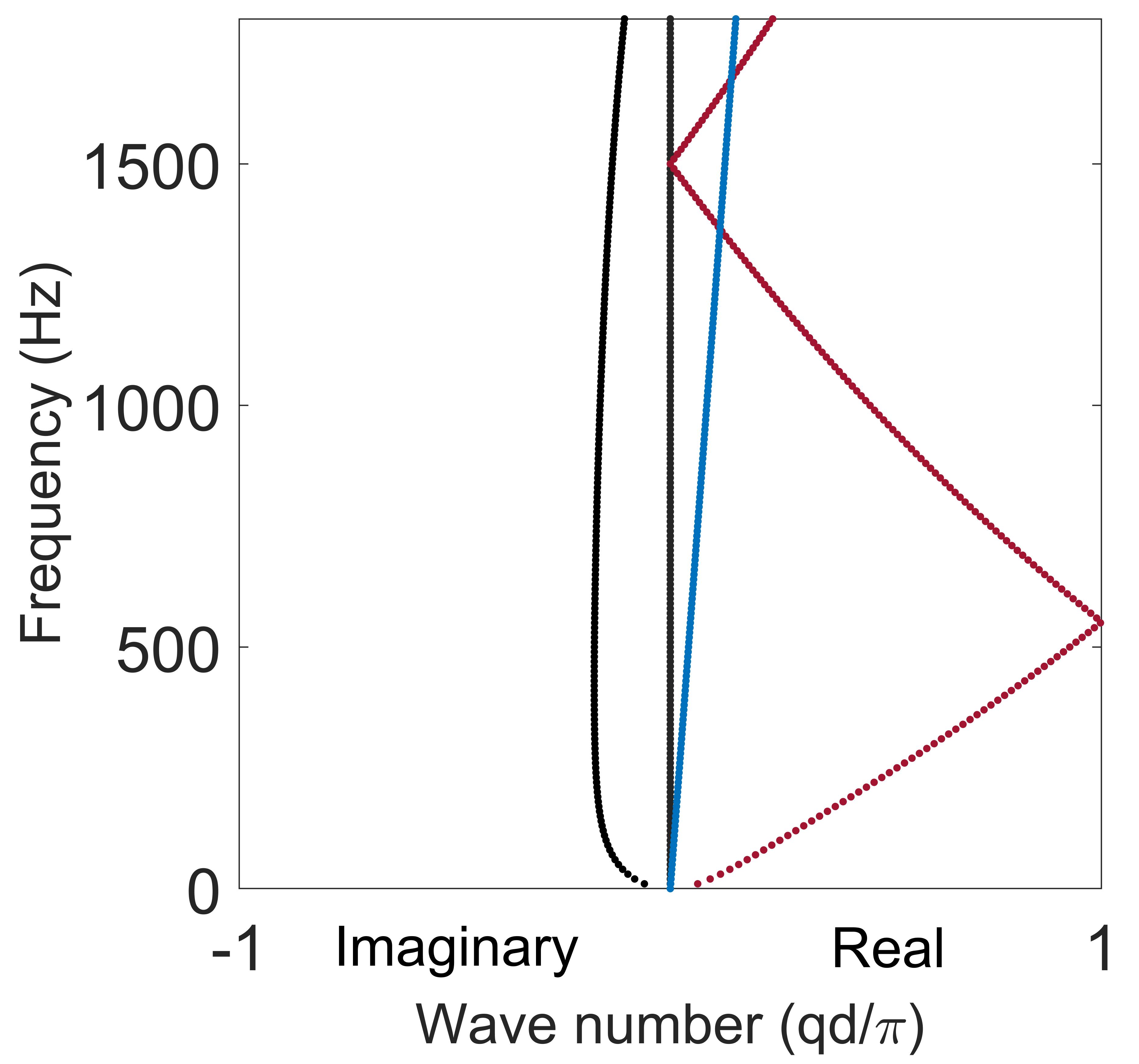}\label{fig:b2}}\hfil
\caption{(a) Coordinate system and notations for a generalized sandwich beam. Here, $u_c$, $u_t$, and  $u_b$ are longitudinal (along the $x$-axis) displacements of a point on the mid-plane of the core, top and bottom face sheets at location $x$, respectively, $\varphi$ denotes the angle of rotation of the mid-surface of the core, $w_t'$ and $w_b'$ are the slope of the mid-surface of the top and bottom face sheets, respectively. (b) A band diagram of a non-periodic sandwich beam.}    \label{F2}
    \end{figure}

Total potential $(V)$ and kinetic $(T)$ energies  of the sandwich beam are expressed as  
\begin{equation} 
V=V_{cN}+V_{tN}+V_{bN}+V_{cS} 
\end{equation} 
\begin{equation}
T=T_c+T_t+T_b
\end{equation} 
\noindent
where $V_{cN}$,$V_{tN}$, and $V_{bN}$ are the normal strain energy at the core and top and bottom face sheets, respectively, and $V_{cS}$ is the shear strain energy at the core. Kinetic energy at core and top and bottom layers are defined as $T_{c}$, $T_{t}$, and $T_{b}$, respectively. Detailed strain and kinetic energy expressions including the strain-displacement relationship are presented in Appendix A.


We can then plug the above expressions of $V$ and $T$ into the governing equations of motion derived using Hamilton's principle   \cite{Banerjee2005}

\begin{equation}
\delta \int^{\tau_2}_{\tau_1}{\int^L_0{\left(T-V\right)dxd\tau}}=0
\end{equation} 

\noindent where $\delta $ is the first variation operator, and ${\tau_1}$ and $\tau_2$ are time instances. The governing equations can then be expanded as follows \cite{Banerjee2005} 

\begin{eqnarray}
&&\left({A_1}^{\alpha}\frac{{\partial }^2}{\partial \tau^2}+{B_1}^{\alpha}\frac{{\partial }^2}{\partial x^2}+{B_2}^{\alpha}\right)u_t+\left({A_2}^{\alpha}\frac{{\partial }^2}{\partial \tau^2}+{B_3}^{\alpha}\frac{{\partial }^2}{\partial x^2}-{B_2}^{\alpha}\right)u_b+\left({A_3}^{\alpha}\frac{{\partial }^2}{\partial \tau^2}+{B_5}^{\alpha}\frac{{\partial }^2}{\partial x^2}-{B_4}^{\alpha}\right)\frac{\partial w}{\partial x}=0\nonumber\\
\end{eqnarray}
\begin{eqnarray}
&&\left({A_2}^{\alpha}\frac{{\partial }^2}{\partial \tau^2}+{B_3}^{\alpha}\frac{{\partial}^2}{\partial x^2}-{B_2}^{\alpha}\right)u_t+\left({A_4}^{\alpha}\frac{{\partial }^2}{\partial \tau^2}+{B_6}^{\alpha}\frac{{\partial }^2}{\partial x^2}+{B_2}^{\alpha}\right)u_b+\left({A_5}^{\alpha}\frac{{\partial }^2}{\partial \tau^2}+{B_7}^{\alpha}\frac{{\partial }^2}{\partial x^2}+{B_4}^{\alpha}\right)\frac{\partial w}{\partial x}=0\nonumber\\
\end{eqnarray}
\begin{eqnarray}
&&\left(-{A_3}^{\alpha}\frac{{\partial }^2}{\partial \tau^2}-{B_5}^{\alpha}\frac{{\partial }^2}{\partial x^2}+{B_4}^{\alpha}\right)\frac{\partial u_t}{\partial x}+\left(-{A_5}^{\alpha}\frac{{\partial }^2}{\partial \tau^2}-{B_7}^{\alpha}\frac{{\partial }^2}{\partial x^2}-{B_4}^{\alpha}\right)\frac{\partial u_b}{\partial x}\ \nonumber\\
&&+\left({A_6}^{\alpha}\frac{{\partial }^4}{\partial x^2\partial \tau^2}-{B_8}^{\alpha}\frac{{\partial }^4}{\partial x^4}+{B_9}^{\alpha}\frac{{\partial }^2}{\partial x^2}-{M_T}^{\alpha}\frac{{\partial }^2}{\partial \tau^2}\right)w=0
\end{eqnarray}

\noindent where coefficients ${A_i}^{\alpha}$, ${B_j}^{\alpha}$, and ${M_T}^{\alpha}$, in which $i=1, 2,...,6$ and $j=1, 2,...,8$, can be found in Appendix B. Here, $\alpha$ is the nominator for subunit cells and $\alpha=1,2,..,\beta$, in which $\beta$ is the total number of the subunit cells in each unit cell. From Hamilton's principle we can also obtain the axial forces ${P_t}^{\alpha}$ and ${P_b}^{\alpha}$, shear force ${S}^{\alpha}$, and bending moment $M^{\alpha}$ as follows

\begin{eqnarray}
&& {P_t}^{\alpha}={B_1}^{\alpha}\frac{\partial u_t}{\partial x}+{B_3}^{\alpha}\frac{\partial u_b}{\partial x}+{B_5}^{\alpha}\frac{{\partial }^2w}{\partial x^2} \nonumber\\ 
\end{eqnarray}
\begin{eqnarray}
&& {P_b}^{\alpha}={B_6}^{\alpha}\frac{\partial u_b}{\partial x}+{B_3}^{\alpha}\frac{\partial u_t}{\partial x}+{B_7}^{\alpha}\frac{{\partial }^2w}{\partial x^2} \nonumber\\ 
\end{eqnarray}
\begin{eqnarray}
&& S^{\alpha}=\left({A_3}^{\alpha}\frac{{\partial }^2}{\partial \tau^2}+{B_5}^{\alpha}\frac{{\partial }^2}{\partial x^2}-{B_4}^{\alpha}\right)u_t+\left({A_5}^{\alpha}\frac{{\partial }^2}{\partial \tau^2}+{B_7}^{\alpha}\frac{{\partial }^2}{\partial x^2}+{B_4}^{\alpha}\right)u_b+\left({-A_6}^{\alpha}\frac{{\partial }^2}{\partial \tau^2}\ +{B_8}^{\alpha}\frac{{\partial }^2}{\partial x^2}-{B_9}^{\alpha}\right)\frac{\partial w}{\partial x} \nonumber\\
\end{eqnarray}
\begin{eqnarray}
&& M^{\alpha}={B_5}^{\alpha}\frac{\partial u_t}{\partial x}+{B_7}^{\alpha}\frac{\partial u_b}{\partial x}+{B_8}^{\alpha}\frac{{\partial }^2w}{\partial x^2}
\end{eqnarray}

\subsection{Transfer Matrix Method}

In this section, we present our solutions of wave propagation in periodic sandwich beams using the transfer matrix method (TMM). Consider harmonic wave motions in a sandwich beam containing two subunit cells, the longitudinal displacements of mid-planes of top and bottom face sheets ($u_t$ and $u_b$) and transverse displacement ($w$) can be expressed as

\begin{eqnarray}
&u_t(x,\tau)=C^{(\alpha)}e^{i(q^{\left(\alpha\right)}x-\omega \tau)} \nonumber\\ 
& u_b(x,\tau)=D^{(\alpha)}e^{i(q^{\left(\alpha\right)}x-\omega \tau)} \nonumber\\ 
& w(x,\tau)=E^{(\alpha)}e^{i(q^{\left(\alpha\right)}x-\omega \tau)}
\end{eqnarray}

\noindent where $\alpha=I, II$, denoting the first and second subunit cells, respectively, $q$ is the wave number, and $\omega $ is frequency.

Substituting the above displacement variables into the governing equation, \textit{i.e.}, Eqs. (7-9), yields an 8th-degree polynomial equation (details are presented in Appendix C) indicating there are eight roots of wave numbers, $q_l$, where $l$=1,2,...,8, for each $\omega$. However, since all the degrees of monomials are even, we can obtain four pairs of roots with opposite signs. The displacements of subunit cell $\alpha$ at location $x$ can then be expressed with linear combinations of eight distinct functions corresponding to the eight solutions of $q_l$ as

\begin{eqnarray}
&&u_t(x,\tau)=\overline{u}_t(x)e^{-i \omega \tau} =({\sum_{l=1}^{l=8}}C^{\left(\alpha\right)}_le^{iq^{\left(\alpha\right)}_lx})e^{-i \omega \tau}\nonumber\\
&&u_b(x,\tau)=\overline{u}_b(x)e^{-i \omega \tau} =({\sum_{l=1}^{l=8}}D^{\left(\alpha\right)}_le^{iq^{\left(\alpha\right)}_lx})e^{-i \omega \tau}\nonumber\\
&&w(x,\tau)=\overline{w}(x)e^{-i \omega \tau} =({\sum_{l=1}^{l=8}}E^{\left(\alpha\right)}_le^{iq^{\left(\alpha\right)}_lx})e^{-i \omega \tau}
\end{eqnarray}

\noindent where $\overline{u}_t(x)$, $\overline{u}_b(x)$, and $\overline{w}(x)$ are the  spatial terms (amplitude), and $e^{-i \omega t}$ is the temporal term  of the wave vectors. Now, let's focus on the spatial terms only. By substituting Eq. (15) into Eqs. (7-13), we can express the longitudinal displacements [$\overline{u}_t(x)$ and $\overline{u}_b(x)$], axial ($\overline{P}_t$ and $\overline{P}_b$)  and shear ($\overline{S}$) forces, and bending moments ($\overline{M}$) in terms of the transverse displacement, $\overline{w}$:

\begin{equation} 
\mathit{\Psi }\left(x\right)=\mathit{\Phi_\alpha}\overline{w}(x)
\end{equation} 
where

\begin{eqnarray} 
\mathit{\Psi }=\left[ \begin{array}{c}
\overline{w}(x) \\ 
\overline{w}'(x) \\ 
{\overline{u}}_t(x) \\ 
{\overline{u}}_b(x) \\ 
{\overline{P}}_t(x) \\ 
{\overline{P}}_b(x) \\ 
\overline{S}(x) \\ 
\overline{M}(x) \end{array}
\right],\:\:\   \mathit{\Phi_\alpha}=\left[ \begin{array}{cccccccc}
1 & 1 & 1 & 1 & 1 & 1 & 1 & 1 \\ 
iq^{(\alpha)}_1 & iq^{(\alpha)}_2 & iq^{(\alpha)}_3 & iq^{(\alpha)}_4 & iq^{(\alpha)}_5 & iq^{(\alpha)}_6 & iq^{(\alpha)}_7 & iq^{(\alpha)}_8 \\ 
{\mathit{\Lambda }}^{(\alpha)}_1 & {\mathit{\Lambda }}^{(\alpha)}_2 & {\mathit{\Lambda }}^{(\alpha)}_3 & {\mathit{\Lambda }}^{(\alpha)}_4 & {\mathit{\Lambda }}^{(\alpha)}_5 & {\mathit{\Lambda }}^{(\alpha)}_6 & {\mathit{\Lambda }}^{(\alpha)}_7 & {\mathit{\Lambda }}^{(\alpha)}_8 \\ 
{\mathit{Z}}^{(\alpha)}_1 & {\mathit{Z}}^{(\alpha)}_2 & {\mathit{Z}}^{(\alpha)}_3 & {\mathit{Z}}^{(\alpha)}_4 & {\mathit{Z}}^{(\alpha)}_5 & {\mathit{Z}}^{(\alpha)}_6 & {\mathit{Z}}^{(\alpha)}_7 & {\mathit{Z}}^{(\alpha)}_8 \\ 
{\chi }^{(\alpha)}_1 & {\chi }^{(\alpha)}_2 & {\chi }^{(\alpha)}_3 & {\chi }^{(\alpha)}_4 & {\chi }^{(\alpha)}_5 & {\chi }^{(\alpha)}_6 & {\chi }^{(\alpha)}_7 & {\chi }^{(\alpha)}_8 \\ 
{\mathit{\Phi }}^{(\alpha)}_1 & {\mathit{\Phi }}^{(\alpha)}_2 & {\mathit{\Phi }}^{(\alpha)}_3 & {\mathit{\Phi }}^{(\alpha)}_4 & {\mathit{\Phi }}^{(\alpha)}_5 & {\mathit{\Phi }}^{(\alpha)}_6 & {\mathit{\Phi }}^{(\alpha)}_7 & {\mathit{\Phi }}^{(\alpha)}_8 \\ 
{\mathit{H}}^{(\alpha)}_1 & {\mathit{H}}^{(\alpha)}_2 & {\mathit{H}}^{(\alpha)}_3 & {\mathit{H}}^{(\alpha)}_4 & {\mathit{H}}^{(\alpha)}_5 & {\mathit{H}}^{(\alpha)}_6 & {\mathit{H}}^{(\alpha)}_7 & {\mathit{H}}^{(\alpha)}_8 \\ 
{\mathit{\Gamma }}^{(\alpha)}_1 & {\mathit{\Gamma }}^{(\alpha)}_2 & {\mathit{\Gamma }}^{(\alpha)}_3 & {\mathit{\Gamma }}^{(\alpha)}_4 & {\mathit{\Gamma }}^{(\alpha)}_5 & {\mathit{\Gamma }}^{(\alpha)}_6 & {\mathit{\Gamma }}^{(\alpha)}_7 & {\mathit{\Gamma }}^{(\alpha)}_8 \end{array}
\right],\:\:\ \overline{w}(x)= \left[ \begin{array}{c}
E^{\left(\alpha\right)}_1e^{iq^{\left(\alpha\right)}_1x} \\ 
E^{\left(\alpha\right)}_2e^{iq^{\left(\alpha\right)}_2x} \\ 
E^{\left(\alpha\right)}_3e^{iq^{\left(\alpha\right)}_3x} \\ 
E^{\left(\alpha\right)}_4e^{iq^{\left(\alpha\right)}_4x} \\ 
E^{\left(\alpha\right)}_5e^{iq^{\left(\alpha\right)}_5x} \\ 
E^{\left(\alpha\right)}_6e^{iq^{\left(\alpha\right)}_6x} \\ 
E^{\left(\alpha\right)}_7e^{iq^{\left(\alpha\right)}_7x} \\ 
E^{\left(j\right)}_8e^{iq^{\left(j\right)}_8x} \end{array}
\right] 
\end{eqnarray} 

\begin{figure}[ht!]
\centering
\begin{minipage}[c]{\textwidth}
\centering
    \includegraphics[width=6in]{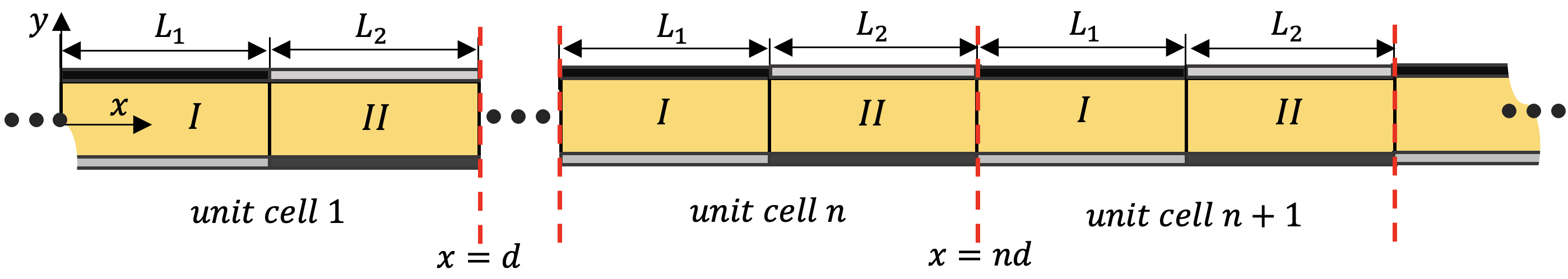}
\caption {Schematic of a sandwich beam with periodic face sheets with an infinite length, composed of unit cells with lattice constant $d$, subunit cell length $L_1$=$L_2$=$L$=$d/2$, in which $\textit{I}$ and $\textit{II}$ denote the first and second subunit cells, respectively. \label{F3}}
\end{minipage}
\end{figure}

Derivation of the matrix $\mathit{\Phi_\alpha}$ components is presented in Appendix D. Let's now apply the continuity boundary conditions at the interfaces between unit cells and subunit cells to Eq. (16), as shown in Fig. \ref{F3}.

The first continuity boundary condition on state vector $\mathit{\Psi}$ is applied at the interface between unit cells $n$ and $n+1$. Since this boundary is the interface between subunit $II$ in unit cell $n$ and subunit $I$ in unit cell $n+1$, we can obtain the following relationship between the two-state vectors in these two adjacent subunit cells \cite{zhang2018band}:

\begin{equation} 
\left[\mathit{\Upsilon }\right]{\mathit{\Psi }}^{(I)}_{n+1}=\left[\mathit{\Pi }\right]{\mathit{\Psi }}^{(II)}_n
\end{equation} 

The second continuity condition is applied at the interface between subunit cells $I$ and $II$ within the same unit cell $n+1$ as follows

\begin{equation} 
\left[\mathit{\Theta }\right]{\mathit{\Psi }}^{(I)}_{n+1}=\left[\mathit{\Delta }\right]{\mathit{\Psi }}^{(II)}_{n+1}
\end{equation} 

Combining Eqs. (18) and (19) yields the following equation  

\begin{equation} 
{\mathit{\Psi }}^{(II)}_{n+1}=\left[\mathit{T}\right]{\mathit{\Psi }}^{(II)}_{n}
\end{equation}

Applying Bloch's theorem for an infinite periodic sandwich beam, ${\mathit{\Psi }}^{(II)}_n$ satisfies \cite{bloch1929quantenmechanik}
\begin{equation}
{\mathit{\Psi }}^{(II)}_{n+1\ }=e^{iqd}{\mathit{\Psi }}^{(II)}_n 
\end{equation} 
where $q$ is the Bloch wave number in the $x$-direction and $d$ is the lattice constant. Substituting Eq. (20) into Eq. (21), we get
\begin{equation}
\left(T(\omega )-e^{iqd}I\right){\mathit{\Psi }}^{\left(II\right)}_n=0
\end{equation}

We can then obtain the solutions for $q$'s at a certain frequency, $\omega$, by solving the eigenvalues of the above equation  
\begin{equation}
\left|T-\lambda I\right|=0
\end{equation}
where $\lambda=e^{iqd}$. Band diagram can be obtained by calculating the values of $\lambda$ for a frequency range. Detailed expressions of the matrices $\mathit{\Gamma }$, $\mathit{\Pi }$, $\mathit{\Theta }$, and $\mathit{\Delta }$ are presented in Appendix E. 

\subsection{Generalized Differential Quadrature Method }

In addition to the TMM, in this paper, we are also presenting a numerical method for a broader application of problems. Among the various numerical methods, GDQM is among the most widely used for both free and forced vibration analysis of sandwich structures due to its effectiveness in solving partial differential equations via transformation into algebraic equations. Here, we will demonstrate the accuracy of GDQM in band diagram calculation of sandwich beams with periodic face sheets by comparing them with those obtained by TMM. We can achieve the transformation by approximating each variable at a given grid point as the linear weighted summation of its values at all of the discrete points in the domain of that variable. Let us consider function $f=f(\overline x)$ in the normalized domain $\overline x=\frac{x-x_{\alpha}}{L_{\alpha}}$ where $x_{\alpha}$ is the location of left edge of subunit cell $\alpha$ with length $L_{\alpha}$.

\begin{eqnarray}
& \displaystyle\frac{d^nf(\overline x)}{d{\overline {x}}^n}(at \; \overline x = \overline x_p)\cong\sum_{q=1}^{q=N}C_{pq}^{(n)}f(\overline {x}_q) \:\:\; \mbox{ for }
\:\:\;  p=1,2...,N
\end{eqnarray}

\noindent where $C_{pq}^{(n)}$ are the weighting coefficients for the $n$th order derivative $(n=0,1,...,4)$ and $N$ is the total number of domain points. The weighting coefficients, their derivatives, and grid point arrangement are presented in Appendix F.
The discretized form of Eq. (7) is shown below 

\begin{eqnarray}
&&\Big(-{A_1}^{\alpha}\omega^2+{B_2}^{\alpha}\Big){({{\overline{u}_t}})}_{p}^{\alpha}+\sum\limits_{q=1}^{N}\frac{{B_1}^{\alpha}}{{L_\alpha}^2}C_{pq}^{(2)}{({{\overline{u}_t}})}_{q}^{\alpha}+\Big(-A_2^{\alpha}\omega^2-{B_2}^{\alpha}\Big){({{\overline{u}_b}})}_{p}^{\alpha}+\sum\limits_{q=1}^{N}\frac{{B_3}^{\alpha}}{{L_{\alpha}}^2}C_{pq}^{(2)}{({{\overline{u}_b}})}_{q}^{\alpha}\nonumber\\
&&+\sum\limits_{q=1}^{N}\Big(\frac{{-A_3}^{\alpha}\omega^2-{B_4}^{\alpha}}{{L_\alpha}}C_{pq}^{(1)}+\frac{{B_5}^{\alpha}}{{L_\alpha}^3}C_{pq}^{(3)}\Big){({{\overline{w}}})}_{q}^{\alpha}=0  \nonumber\\
\end{eqnarray}

The other two discretized equations of motion [Eqs. (8) and (9)] are presented in Appendix F. The total discretized equations of motion can be written in a matrix form as follows 

\begin{equation}
K_DU+\omega^2M_DU_D=0
\end{equation}
\noindent where $K_D$ is a $(3N-8)\beta\times 3N\beta $ matrix, $M_D$ is a $(3N-8)\beta\times(3N-8)\beta$ matrix, $U$ is a 3$N\beta\times1$ vector, and $U_D$ is a $(3N-8)\beta\times1$ vector. In order to separate the equations that only include the nodes inside the domain of the subunit cells (the nodes except those that are in the boundaries of subunit cells) from the nodes that satisfy continuity conditions and the nodes that satisfy periodic boundary conditions, Eq. $(26)$ can be rewritten as \cite{cheng2018complex}

\begin{equation}
K_{DD}U_D+K_{DC}U_C+K_{DP}U_P+\omega^2M_DU_D=0
\end{equation}

\noindent where $K_{DD}$ is a $(3N-8)\beta\times(3N-8)\beta$ matrix, $K_{DC}$ is a $(3N-8)\beta\times8(\beta-1)$ matrix, $U_C$ is a $8(\beta-1)\times1$ vector, $K_{DP}$ is $(3N-8)\beta\times8$ matrix, and $U_P$ is a $8\times1$ vector. \noindent

Continuity conditions on variables $\overline{u}_t$, $\overline{u}_b$, $\overline{w}$, and $\overline{w}'$ between the subsequent subunit cells $\alpha$ and $\alpha+1$ ($\alpha=1,..,\beta$), in which $\beta$ is the total number of subunit cells, are as follows

\begin{equation}
{({{\overline{u}_t}})}_{N-2}^{\alpha}={({{\overline{u}_t}})}_{1}^{\alpha+1}
\end{equation}
\begin{equation}
{({{\overline{u}_b}})}_{N-1}^{\alpha}={({{\overline{u}_b}})}_{2}^{\alpha+1}
\end{equation}
\begin{equation}
{({{\overline{w}}})}_{N}^{\alpha}={({{\overline{w}}})}_{3}^{\alpha+1}
\end{equation}
\begin{equation}
\sum\limits_{q=1}^{N}\frac{C_{Nq}^{(1)}}{L_{\alpha}}{({{\overline{w}}})}_{q}^{\alpha}=\sum\limits_{q=1}^{N}\frac{C_{1q}^{(1)}}{L_{\alpha+1}}{({{\overline{w}}})}_{q}^{\alpha+1}
\end{equation}

The continuity conditions on $P_t$, $P_b$, $M$, $S$ are presented in Appendix F. The total discretized continuity conditions can be written in a matrix form as follows

\begin{equation}
K_CU=0
\end{equation}

\noindent where $K_C$ is a $8(\beta-1)\times3\beta N$ matrix. Eq. (37) can be decomposed as the following form
\begin{equation}
K_{CD}U_{D}+K_{CC}U_{C}+K_{CP}U_{P}=0
\end{equation}

\noindent where $K_{CD}$ is a $8(\beta-1)\times(3N-8)\beta$ matrix, $K_{CC}$ is a $8(\beta-1)\times8(\beta-1)$ matrix, and $K_{CP}$ is a $8(\beta-1)\times8$ matrix. \noindent

Periodic boundary conditions on variables $u_t$, $u_b$, $w$, and $w'$ between the left side of first subunit cell $I$ and right side of the last subunit cell $\beta$ are as

\begin{equation}
\lambda{({{\overline{u}_t}})}_{1}^{I}={({{\overline{u}_b}})}_{N}^{\beta}
\end{equation}
\begin{equation}
\lambda{({{\overline{u}_b}})}_{1}^{I}={({{\overline{u}_b}})}_{N}^{\beta}
\end{equation}
\begin{equation}
\lambda{({{\overline{w}}})}_{1}^{I}={({{\overline{w}}})}_{N}^{\beta}
\end{equation}

\begin{equation}
\lambda\sum\limits_{q=1}^{N}\frac{C_{1q}^{(1)}}{L_{I}}{({{\overline{w}}})}_{q}^{I}=\sum\limits_{q=1}^{N}\frac{C_{Nq}^{(1)}}{L_{\beta}}{({{\overline{w}}})}_{q}^{\beta}
\end{equation}

Periodic boundary conditions on $\overline{P}_t$, $\overline{P}_b$, $\overline{M}$, and $\overline{V}$, between the left side of first subunit cell $I$ and right side of the last subunit cell $\beta$ are presented in Appendix F.

The total discretized periodic equations are written in a matrix form as

\begin{equation}
K_P(\lambda)U=0
\end{equation}
where $K_P$ is a $8\times 3 \beta N$ matrix. Eq. (38) can be decomposed as the following form
\begin{equation}
K_{PD}U_{D}+K_{PC}U_{C}+K_{PP}U_{P}=0
\end{equation}
\noindent where $K_{PD}$ is a $8\times(3N-8)\beta$ matrix, $K_{PC}$ is a $8\times8(\beta-1)$ matrix, and $K_{PP}$ is a $8\times8$ matrix. \noindent
Combining Eqs.(27), (33), and (39) yields the following matrix

\begin{equation}
\begin{aligned}
\left[ \begin{array}{ccc}
K_{DD}-\omega^2M_D & K_{DC}& K_{DP}\\ 
K_{CD} & K_{CC} & K_{CP} \\ 
K_{PD}(\lambda) & K_{PC}(\lambda) & K_{PP}(\lambda) \\
 \end{array}
\right]
\left[ \begin{array}{ccc}
U_D\\ 
U_C\\ 
U_P \end{array}
\right]=\left[ \begin{array}{ccc}
0\\ 
0\\ 
0\end{array}
\right]
\end{aligned}
\end{equation}

We can obtain the solutions for $\lambda$ at a certain frequency, $\omega$, by solving the eigenvalues of the above equation. 

\section{Results and Discussion}

\begin{figure}[H]\centering
\subfloat[]{\includegraphics[width=7cm,height=6cm]{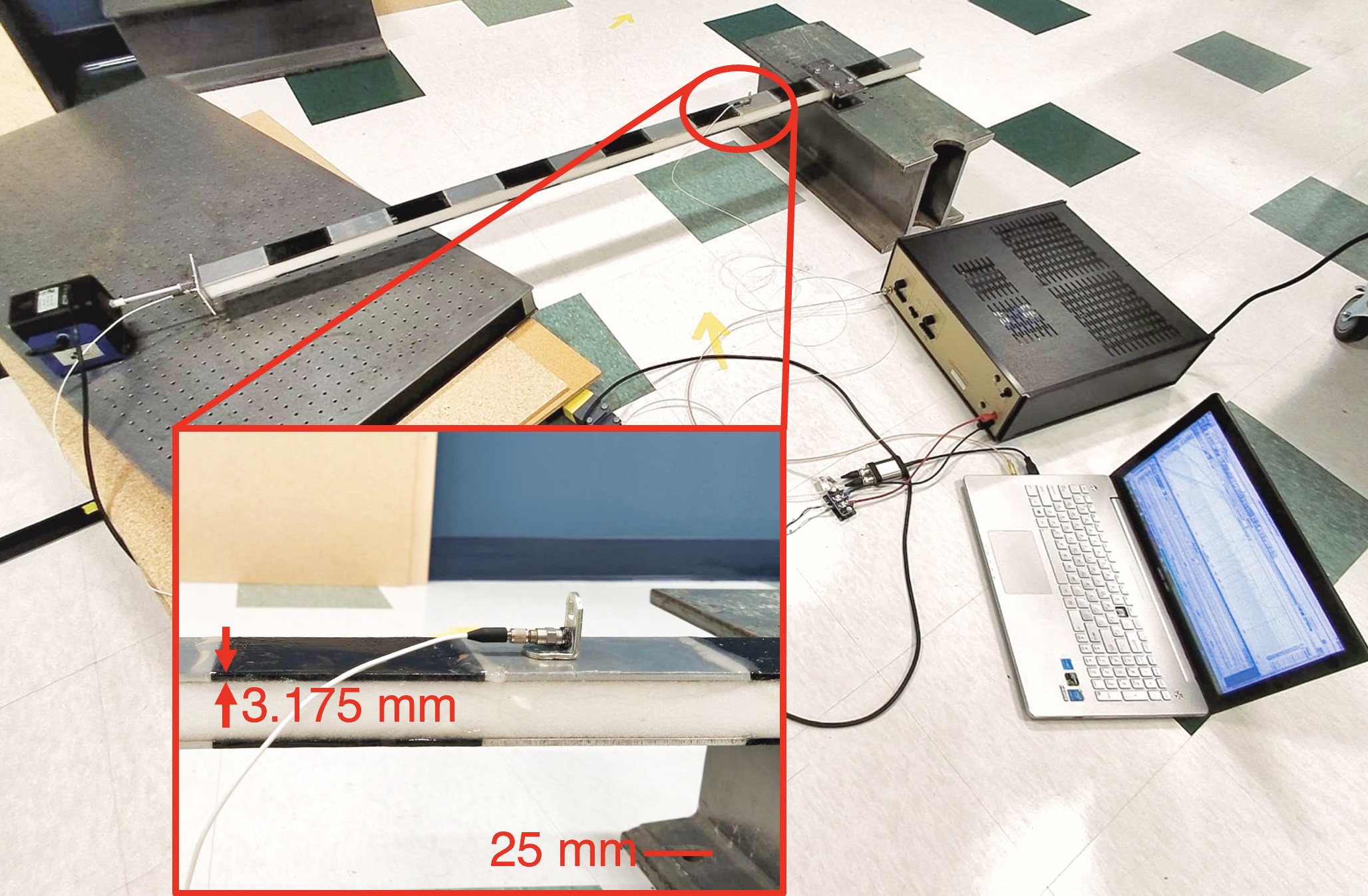}\label{fig:a4}}\:\:\:\
\subfloat[]{\includegraphics[width=7cm,height=6.5cm]{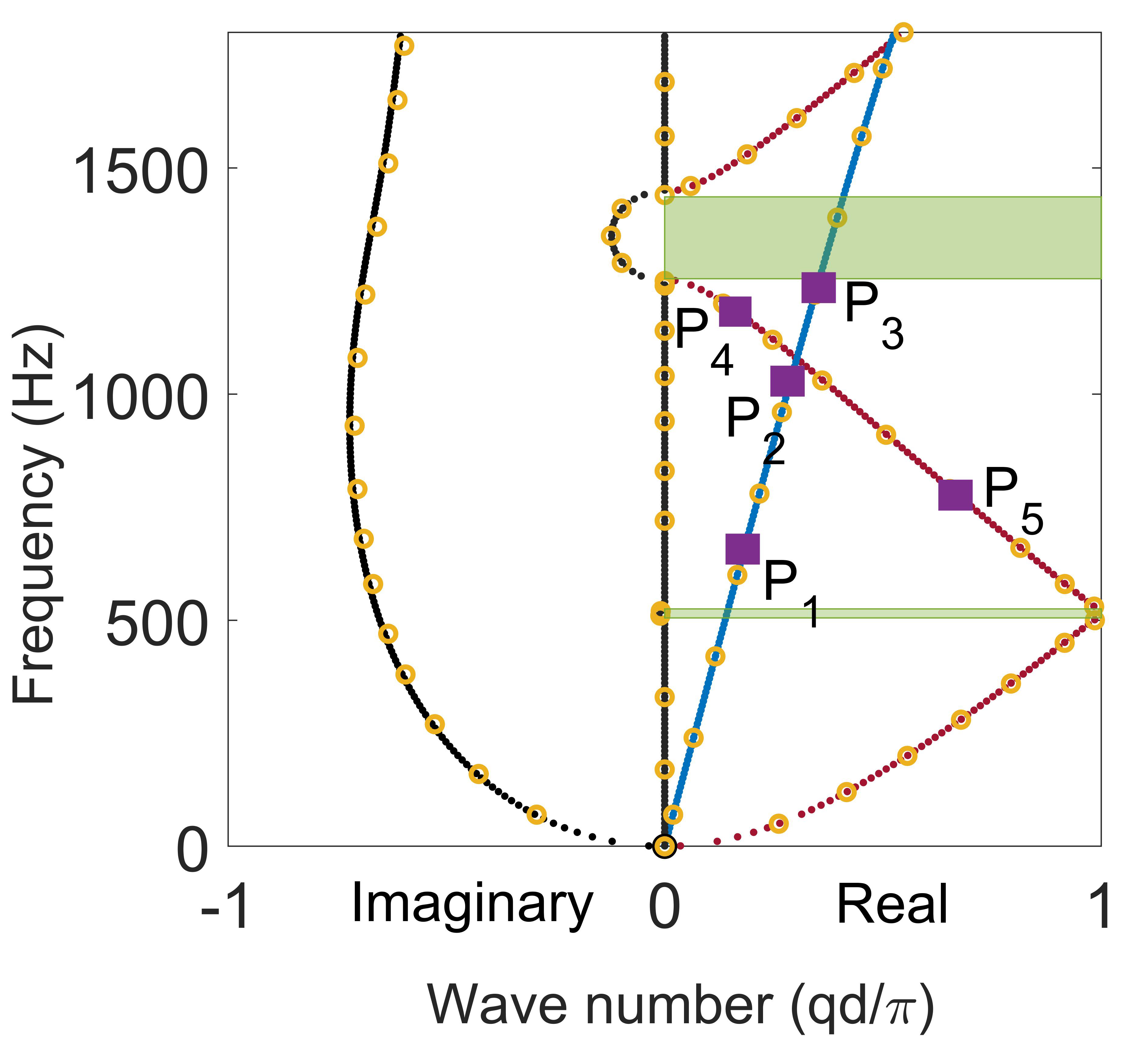}\label{fig:b4}}\hfil
\subfloat[]{\includegraphics[width=0.21\linewidth]{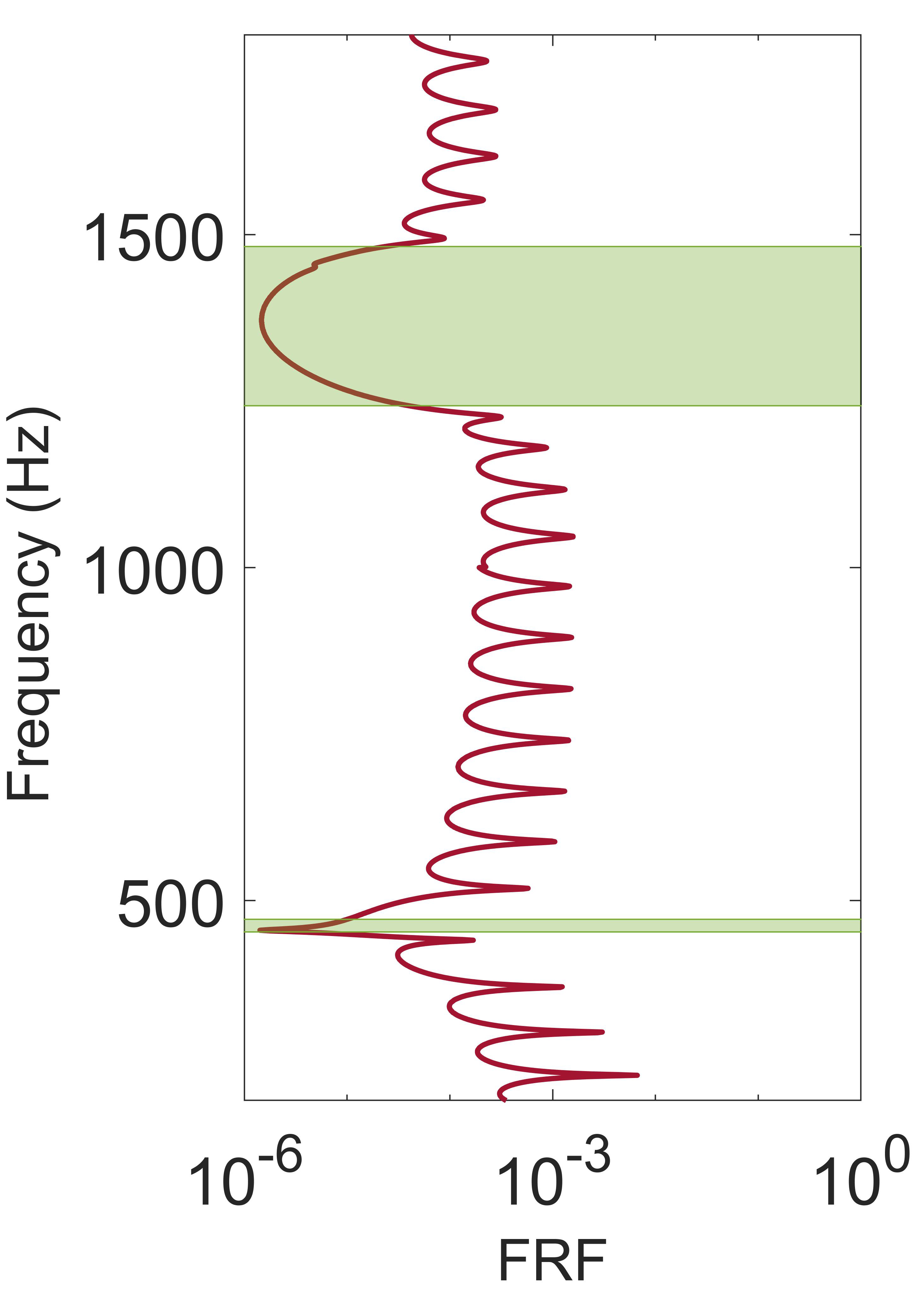}\label{fig:c4}}
\subfloat[]{\includegraphics[width=0.21\linewidth]{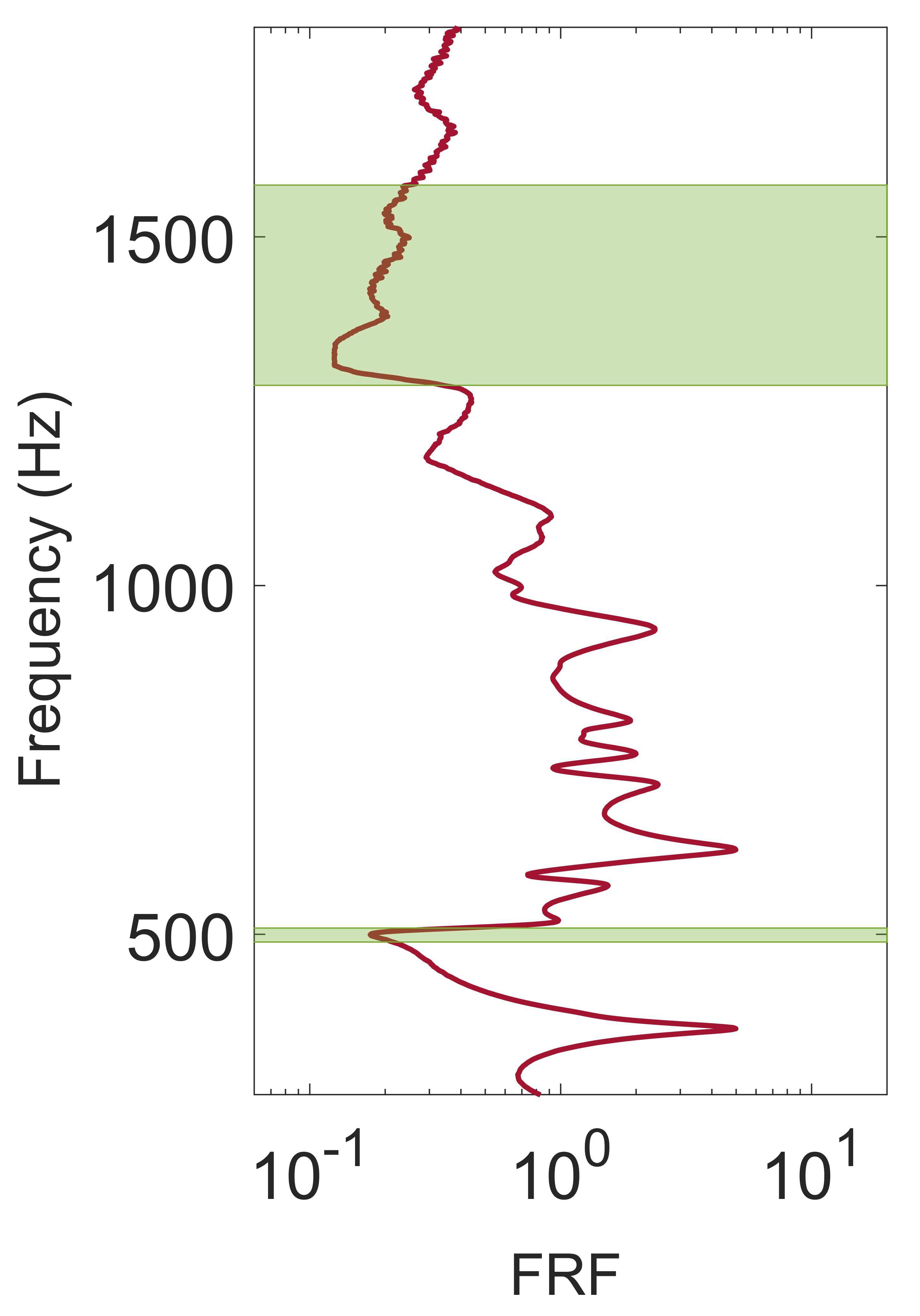}\label{fig:d4}}\hfil
\subfloat[]{\includegraphics[width=0.21\linewidth]{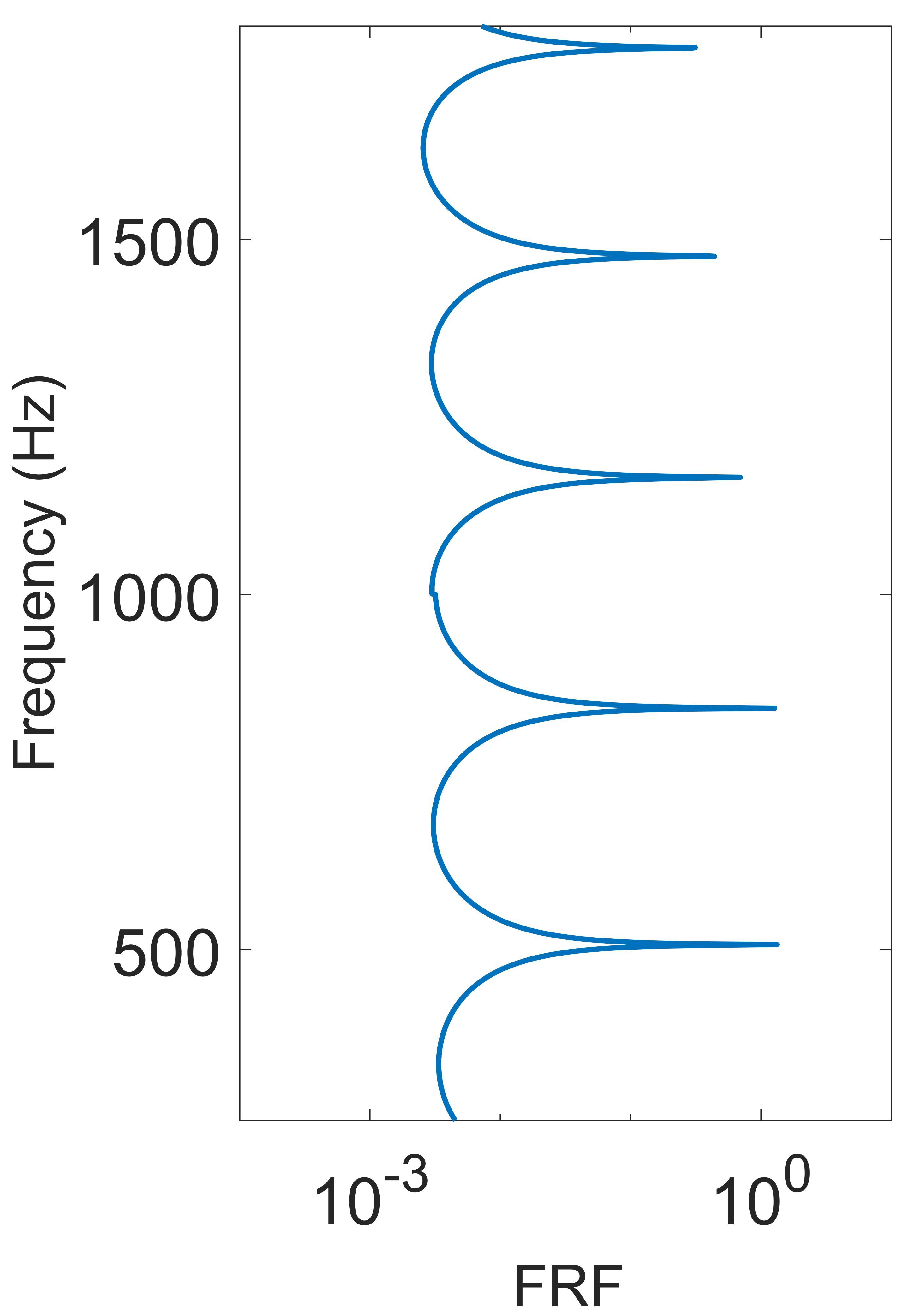}\label{fig:e4}}
\subfloat[]{\includegraphics[width=0.21\linewidth]{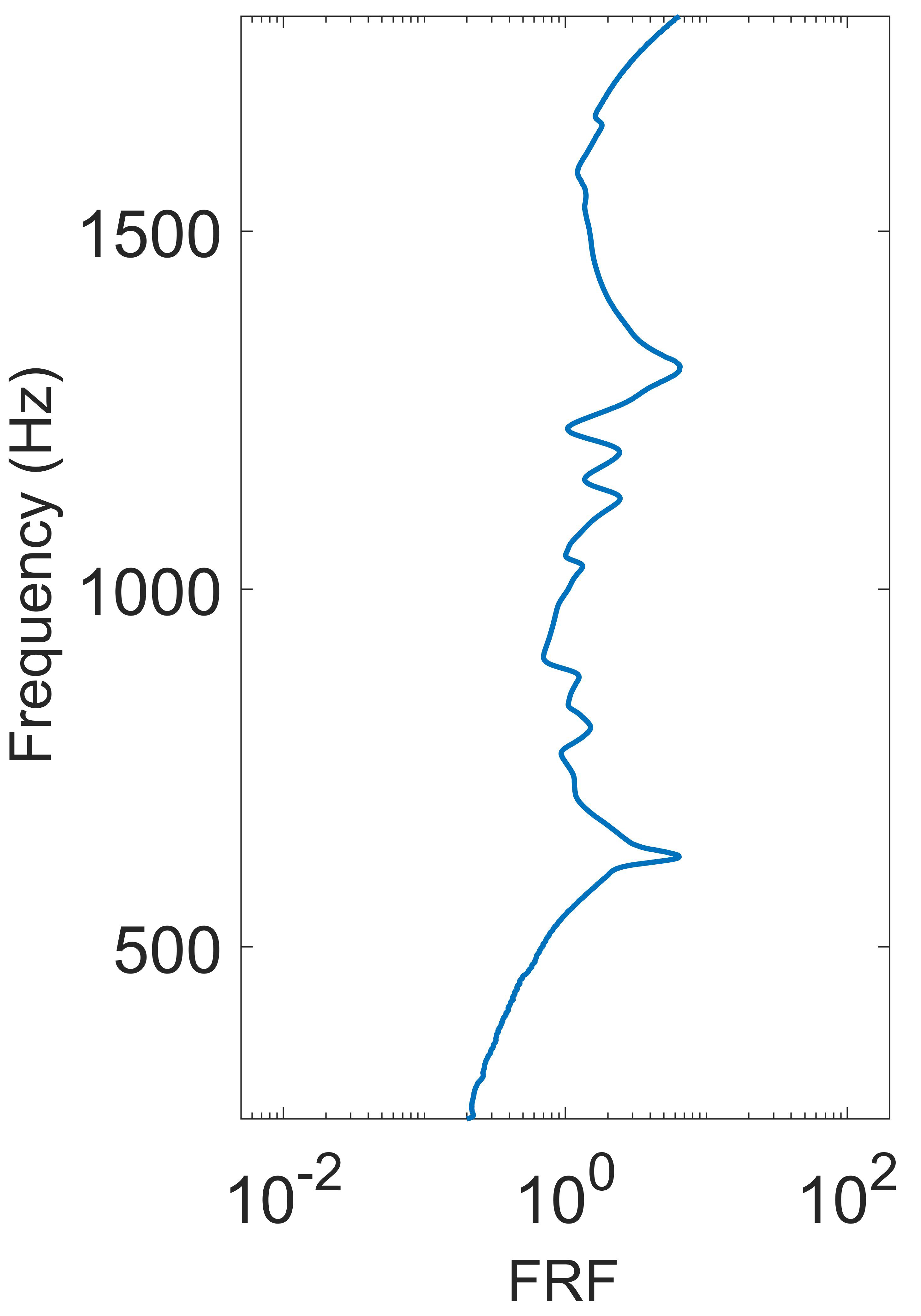}\label{fig:f4}}
\caption{(a) Experimental setup for FRF measurement of the sandwich beam with an MR symmetry. To avoid duplication, we are only presenting the setup for the longitudinal test. The one for the transverse test is similar to that shown in Fig.~\ref{F6}(a). (b) Band diagrams obtained using TMM (dots) and GDQM (circles) respectively. Transverse and longitudinal modes via TMM are presented in red and blue, respectively. (c) and (e): Simulated FRFs (with ten unit cells between the two measured points) of (c) transverse and (e) longitudinal vibration via FEM, respectively. (d) and (f): Experimentally measured FRFs (with eight unit cells between the two measured points) of (d) transverse and (f) longitudinal vibration, respectively. Green shades in (b)-(d) are transverse bandgaps.}
    \label{F4}
    \end{figure}

Band diagrams for sandwich beams with material periodicity shown in Fig.~\ref{F1}(a) (with an MR symmetry) and (b) (with a GR symmetry with $G=d/2$) calculated with TMM and GDQM are presented in Fig.~\ref{F4}(b) and Fig.~\ref{F6}(b), respectively. To confirm our theoretical results, frequency response functions obtained with FEM via ANSYS and experimental measurement for transverse and longitudinal vibrational modes are also presented in Figs.~\ref{F4} and~\ref{F6}, corresponding to the MR and GR (with $G=d/2$) symmetries, respectively. The experimental setup of the longitudinal and transverse vibration measurement along the MR and GR sandwich beams are presented in Fig.~\ref{F4}(a) and Fig.~\ref{F6}(a), respectively. Our sandwich structures are made of Rohacell foam as core, with a thickness of $h_c=25.4$ mm, and alternating aluminum (AL) and acrylonitrile butadiene styrene (ABS) plates as face sheets, with a thickness of $h_t(h_b)=3.175$ mm. Young's modulus $(E)$ and density $(\rho)$ of these three types of materials are listed in Table \ref{T1}. 

\begin{table}[h]
\small\sf\centering
\caption{Parameters of the sandwich beam with materially periodic face sheets\label{T1}}
\begin{tabular}{lllllllll}
\toprule
Property&Foam&AL&ABS&\\
\bottomrule\
$E$\:\ ($Pa$)&$75.4\times10^6$&$6.7\times10^{10}$&$2.14\times10^9$\\
\bottomrule\
$\rho \:\ $($kg/m^3$)&$75$&$2700$&$1040$\\
\bottomrule
\end{tabular}\\[10pt]
\end{table}

We clamp one end of the beam. The excitation is prescribed as a point force in the horizontal direction at the free end for longitudinal waves, as shown in Fig.~\ref{F4}(a), and vertically from the bottom near the free end for transverse waves, presented in Fig.~\ref{F6}(a). A Krohn-Hite model 7500 amplifier is connected to the shaker to supply the required power. Two 353B18 ICP accelerometers are placed on the sandwich beam face sheets, with one close to the shaker's excitation point and the other near the clamped end. The accelerometers are connected to a 2-channel 485B39 ICP sensor signal conditioner with a USB digital output through two 002C10 coaxial cables. SpectraPLUS-SC software is then used for recording and post-processing the received data from 485B39 accelerometers. 

\begin{figure}[H]\centering
\subfloat[]{\includegraphics[width=6cm,height=5.5cm]{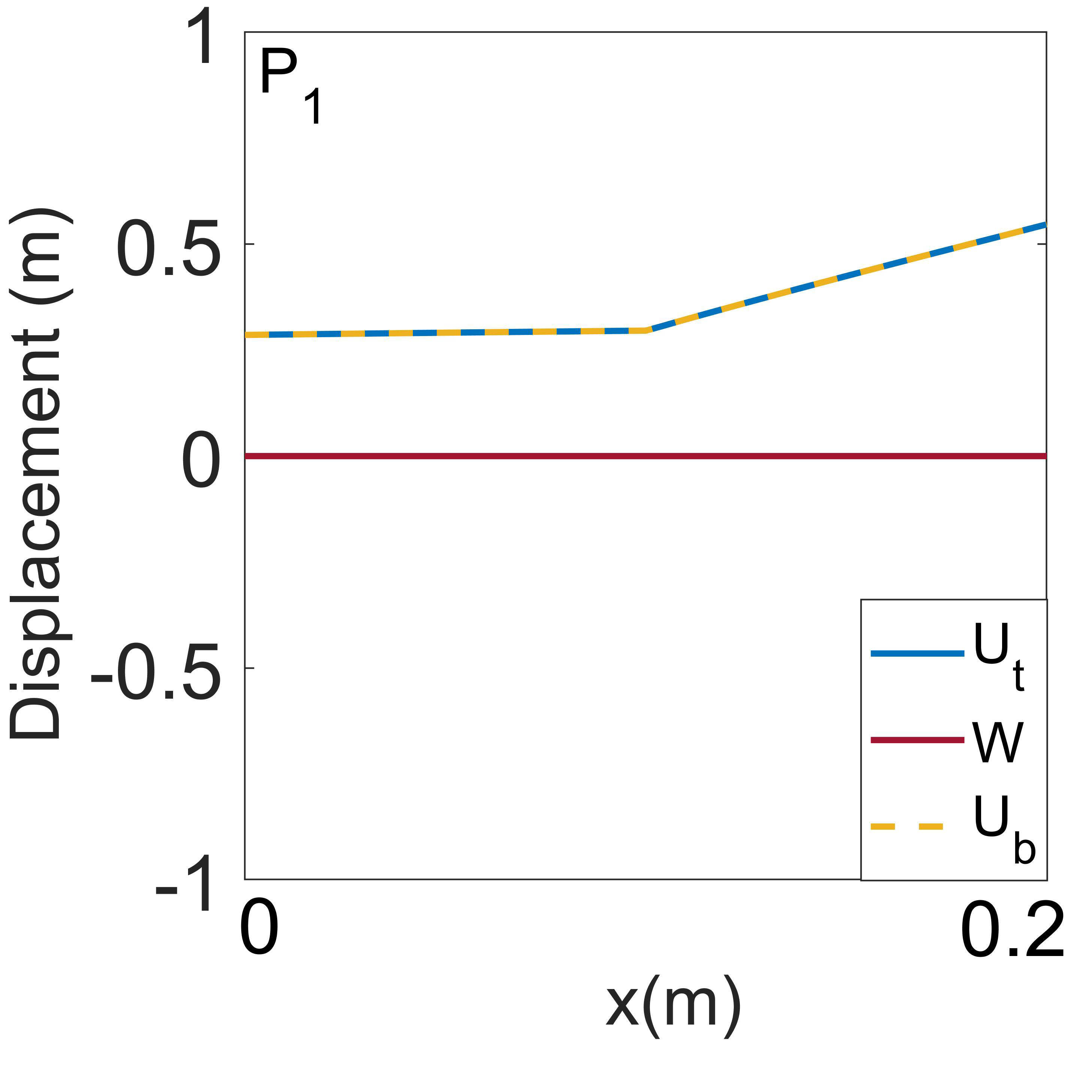}\label{fig:a5}}
\subfloat[]{\includegraphics[width=6cm,height=5.5cm]{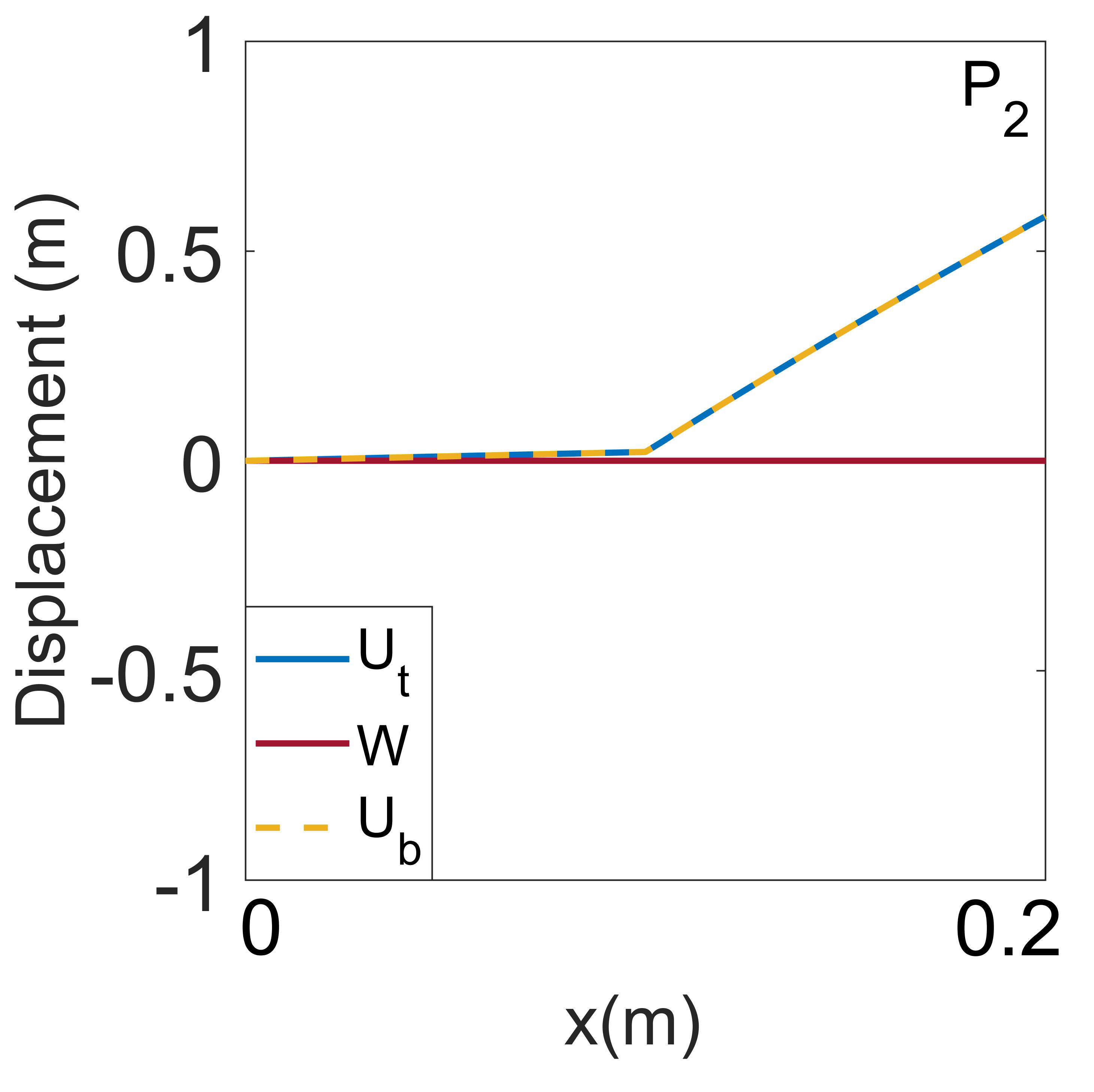}\label{fig:b5}}\hfil
\subfloat[]{\includegraphics[width=6cm,height=5.5cm]{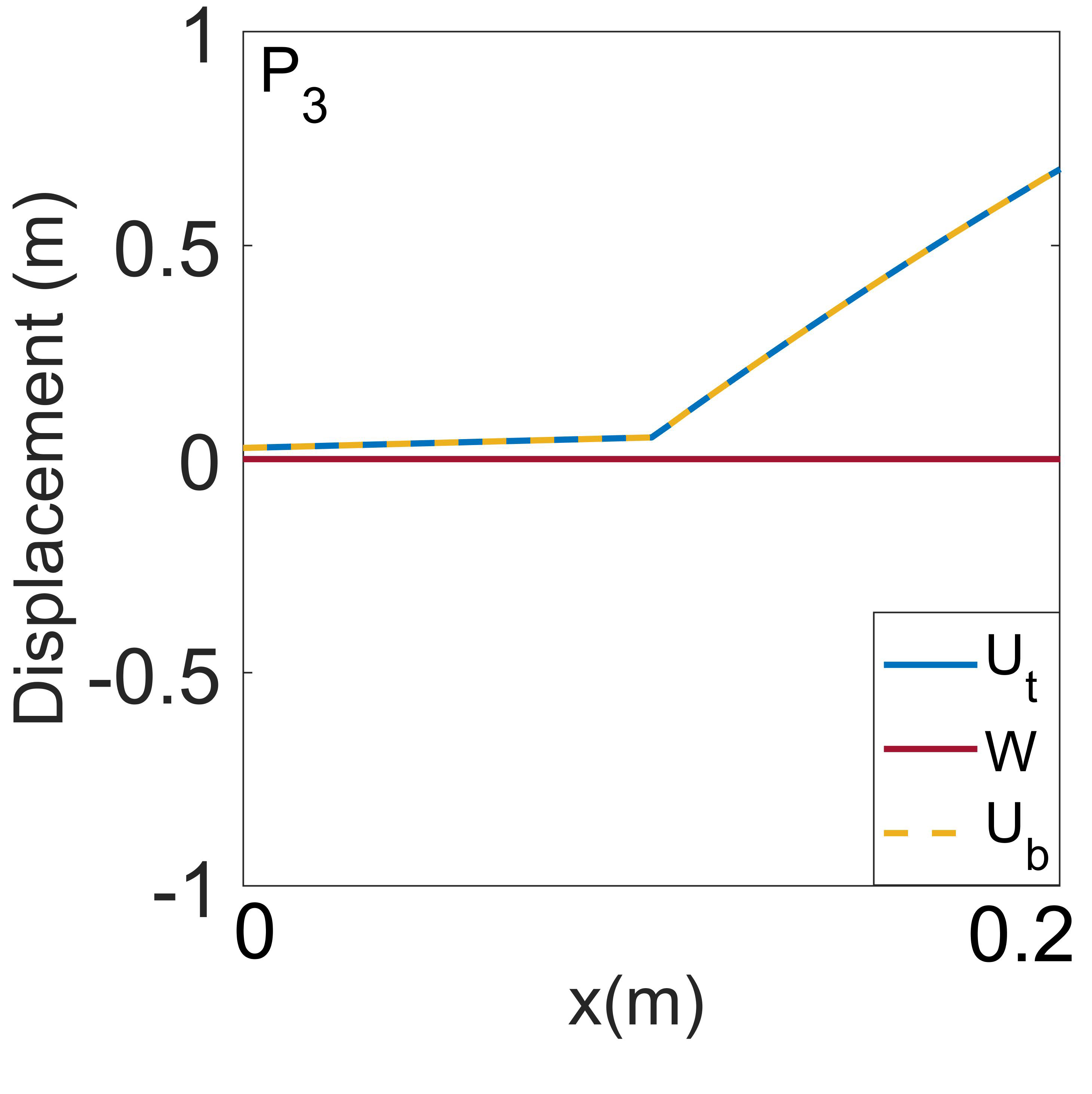}\label{fig:c5}}
\subfloat[]{\includegraphics[width=6cm,height=5.5cm]{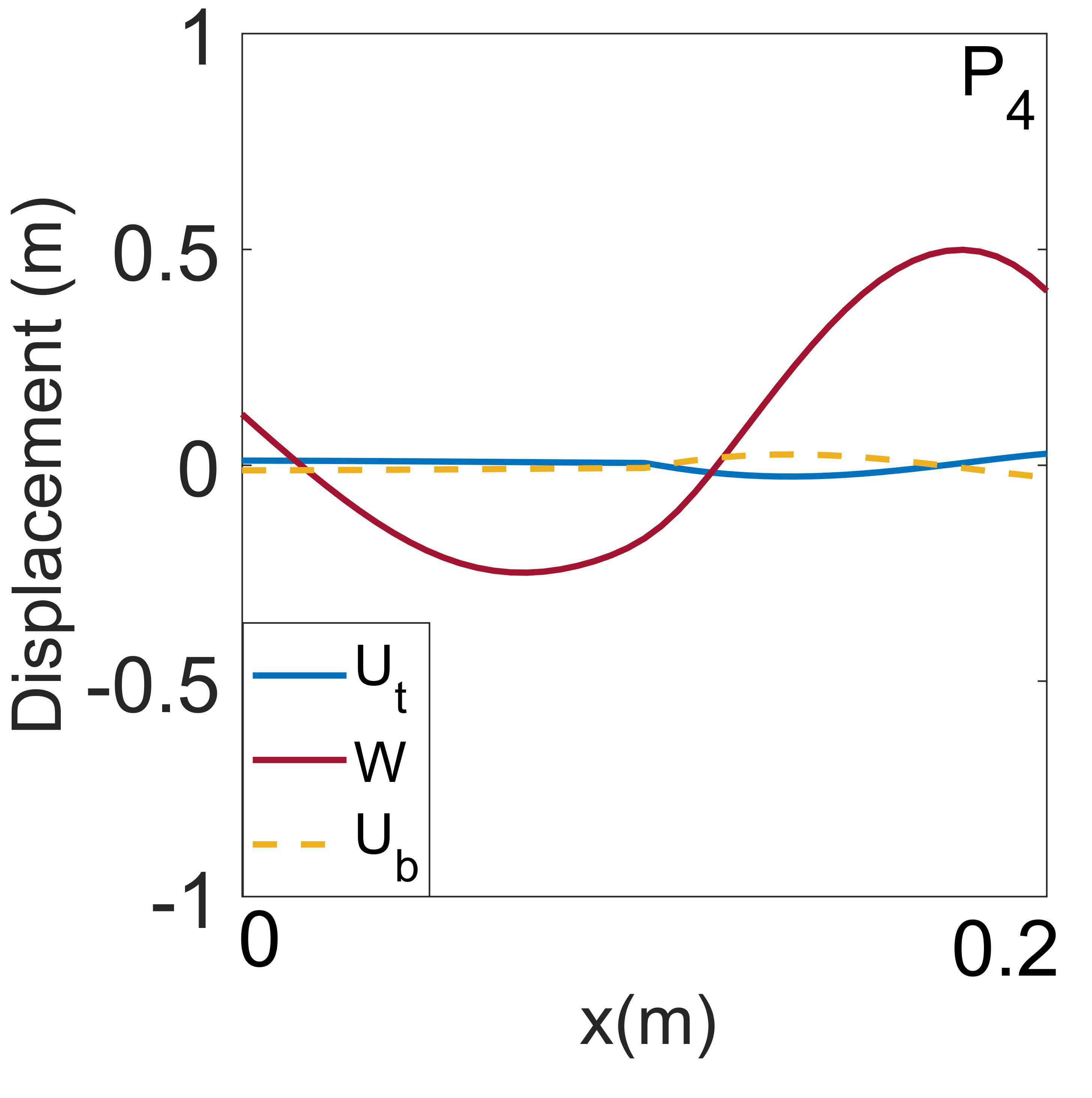}\label{fig:d5}}\\
\subfloat[]{\includegraphics[width=6cm,height=5.5cm]{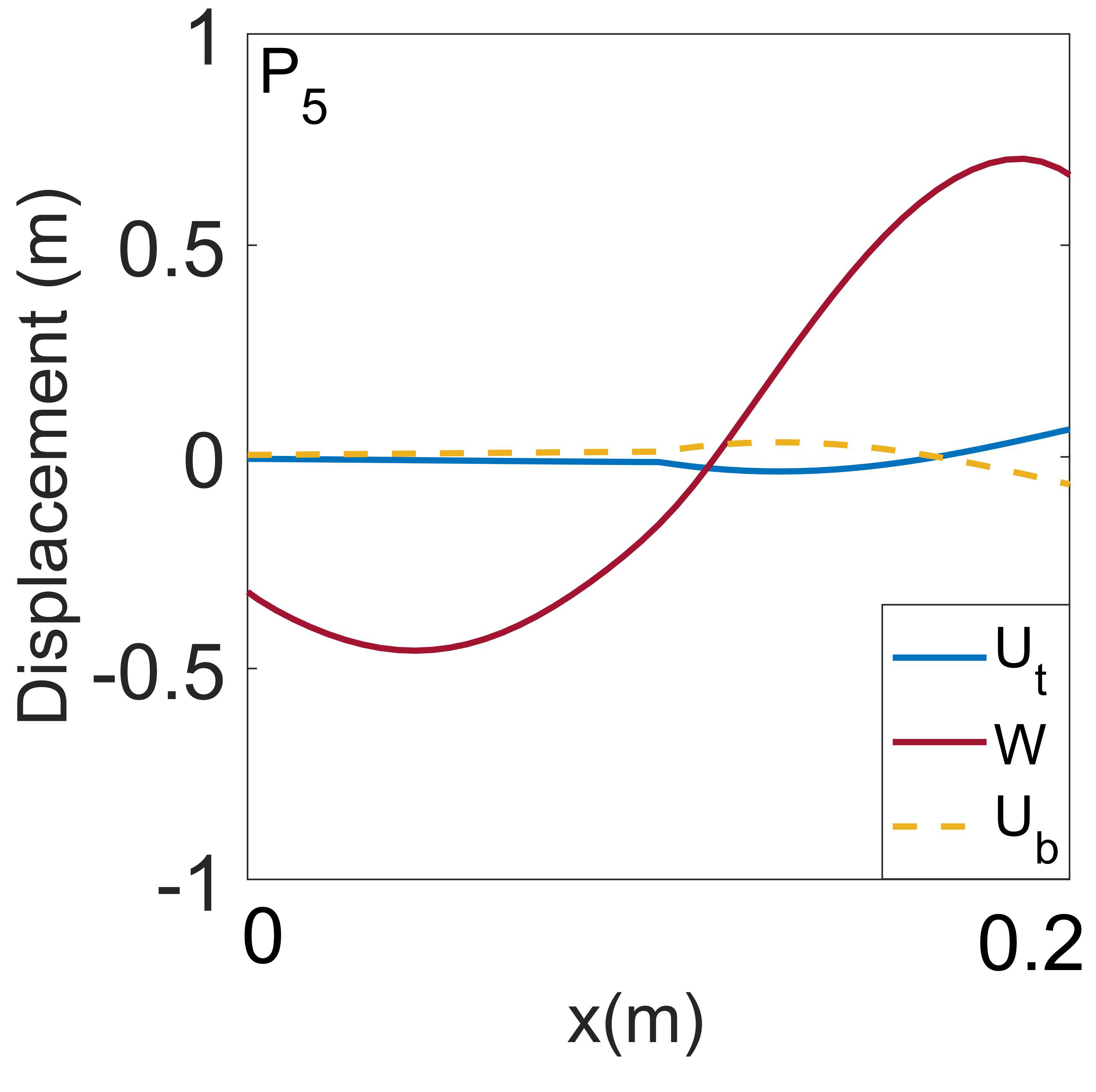}\label{fig:e5}}
\caption{Mode shapes of selected waves at points, (a) $P_1$,  (b) $P_2$, (c) $P_3$, (d) $P_4$, (e) $P_5$ specified as purple squares on Fig.~\ref{F4}(b) for the sandwich beam with periodic face sheets with an MR symmetry.}
    \label{F5}
    \end{figure}

\begin{figure}[H]\centering
\subfloat[]{\includegraphics[width=7cm,height=6cm]{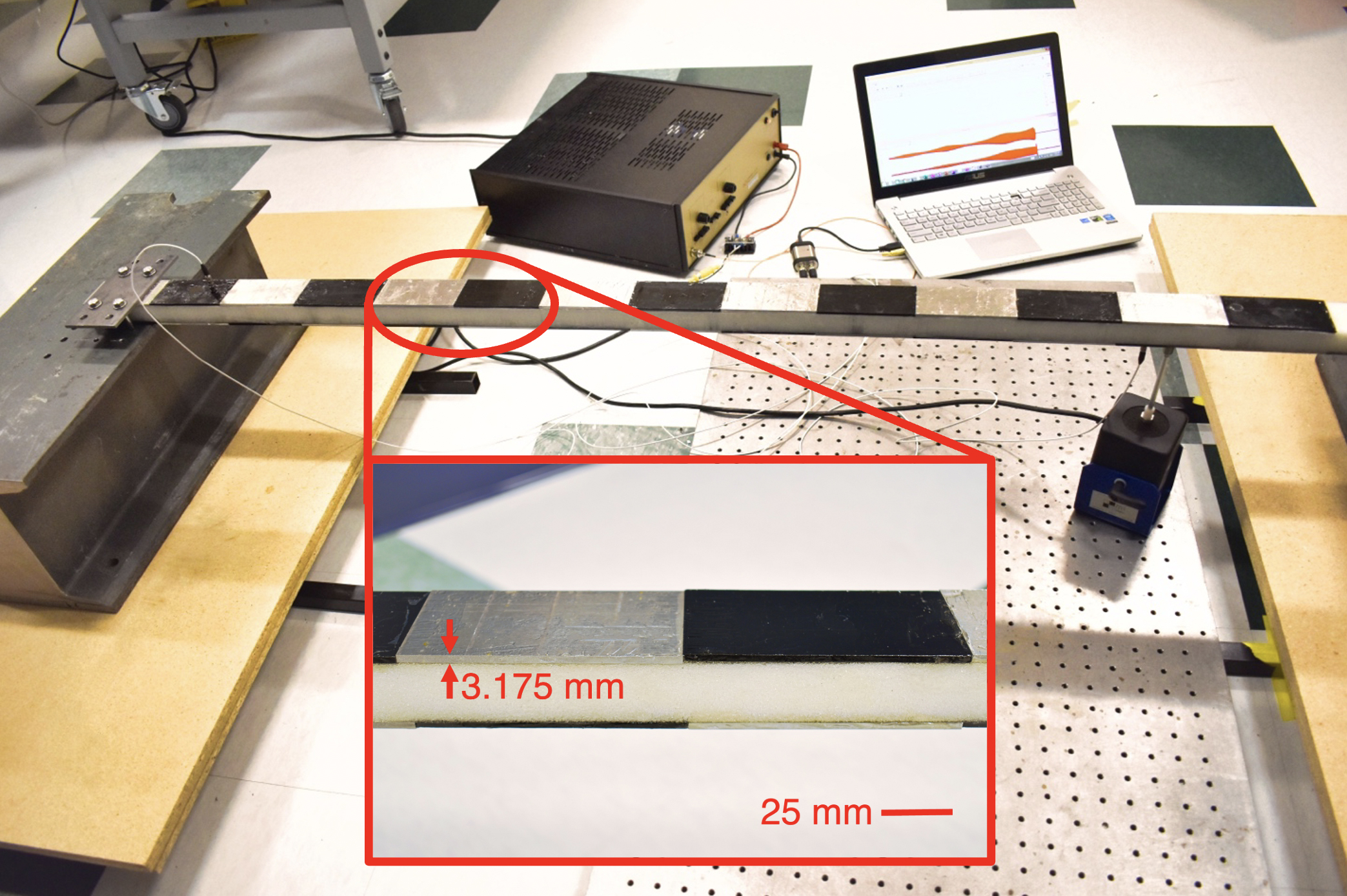}\label{fig:a6}}\:\:\
\subfloat[]{\includegraphics[width=7cm,height=6.5cm]{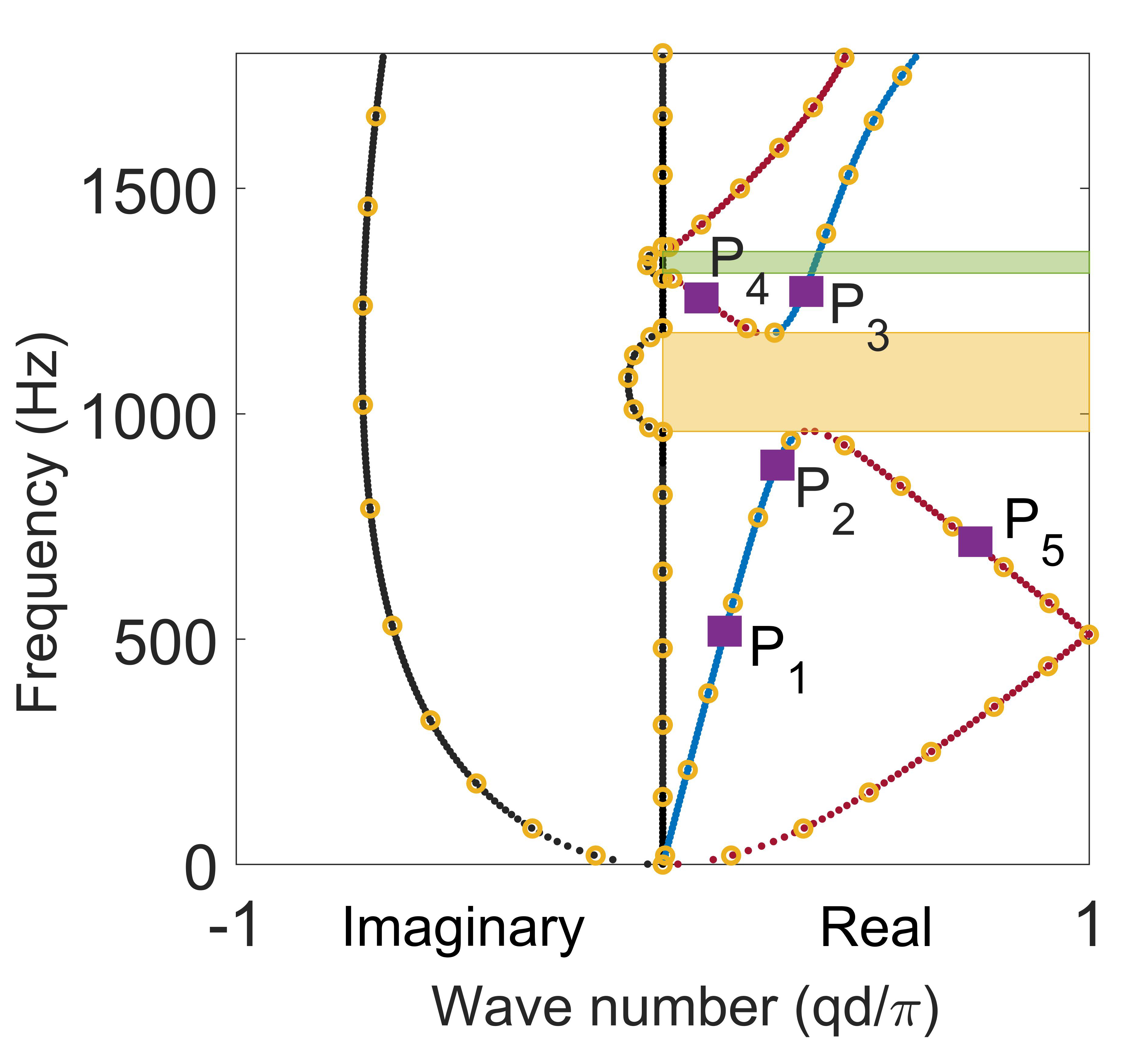}\label{fig:b6}}\hfil
\subfloat[]{\includegraphics[width=0.21\linewidth]{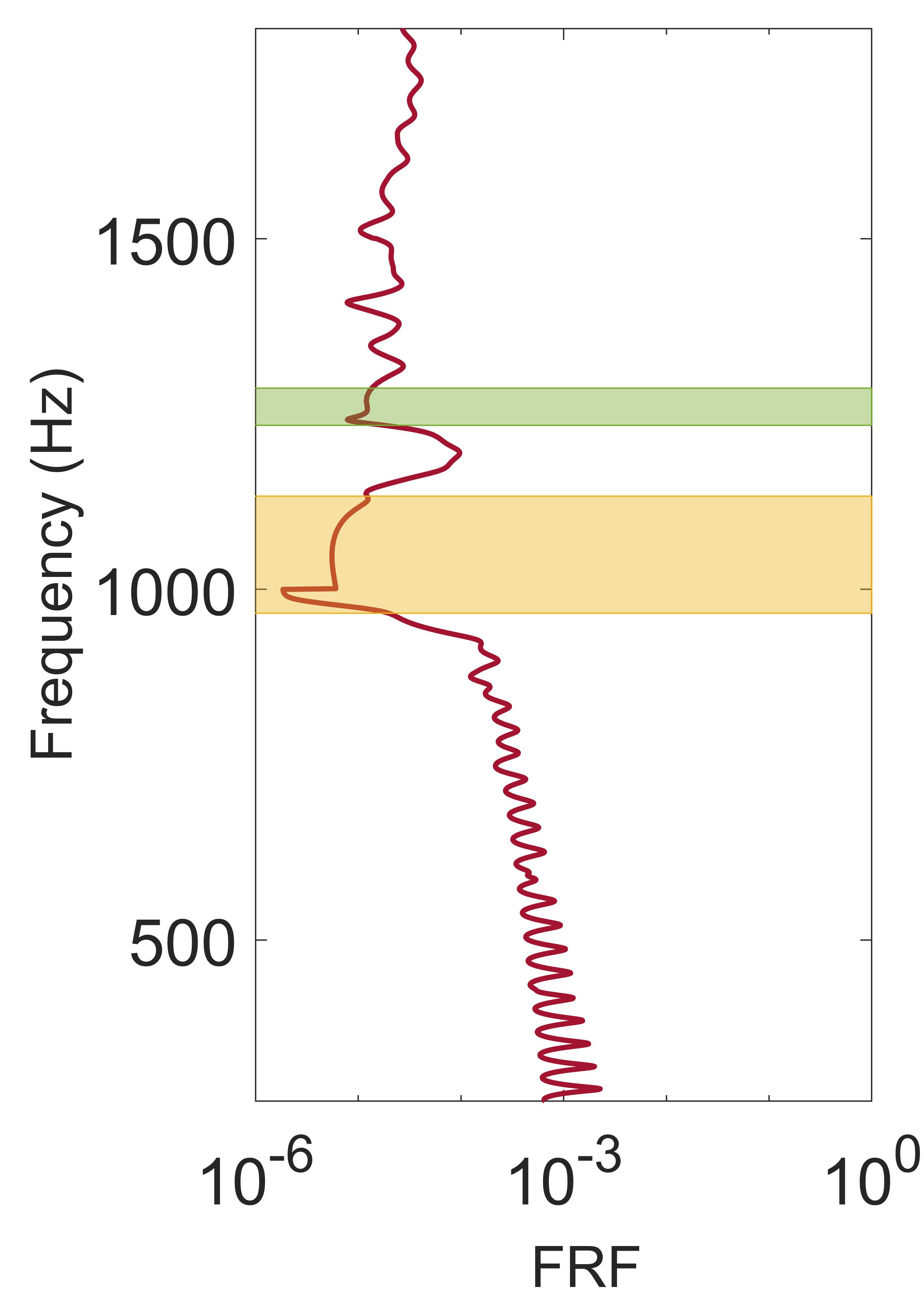}\label{fig:c6}}
\subfloat[]{\includegraphics[width=0.21\linewidth]{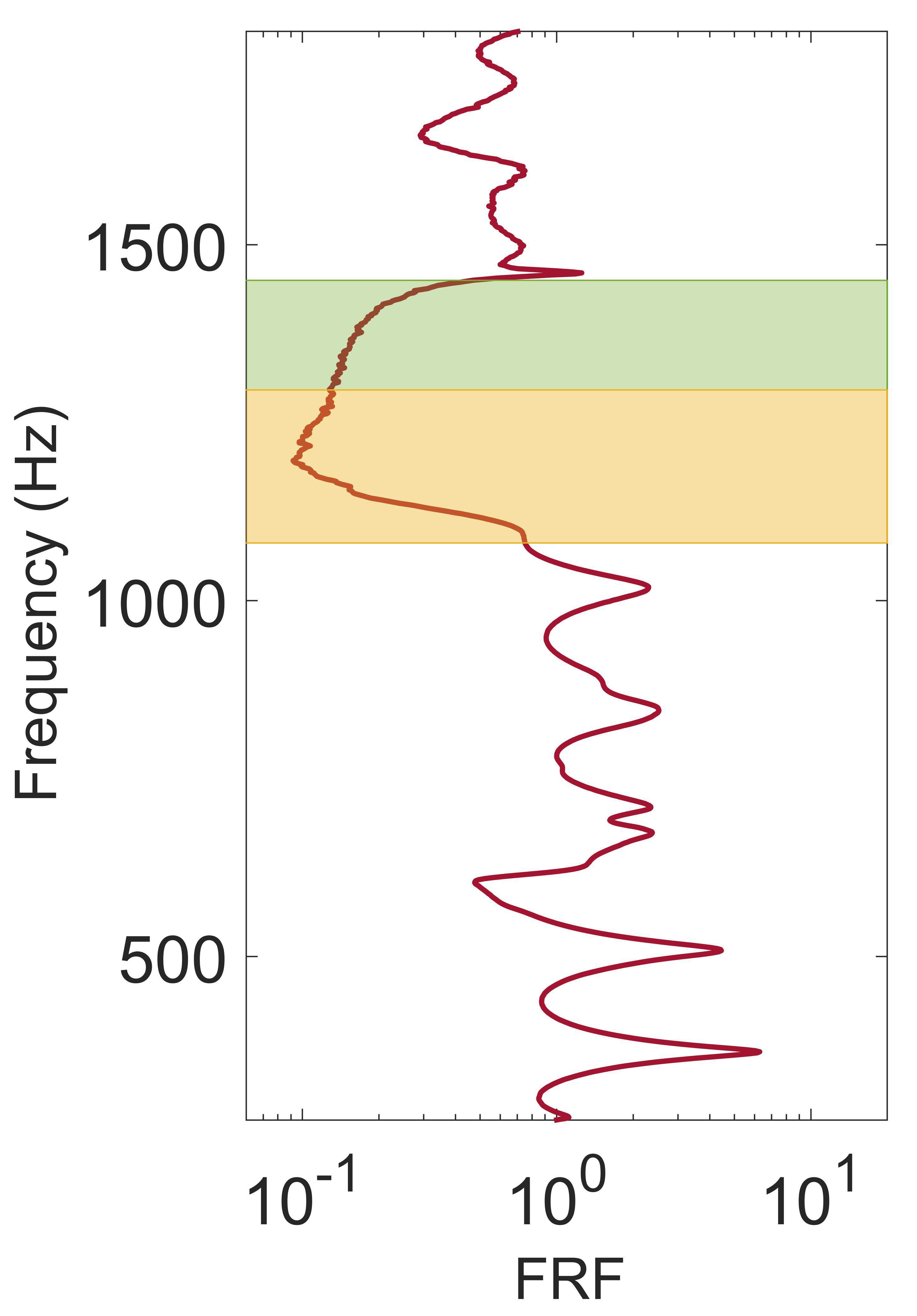}\label{fig:d6}}\hfil
\subfloat[]{\includegraphics[width=0.21\linewidth]{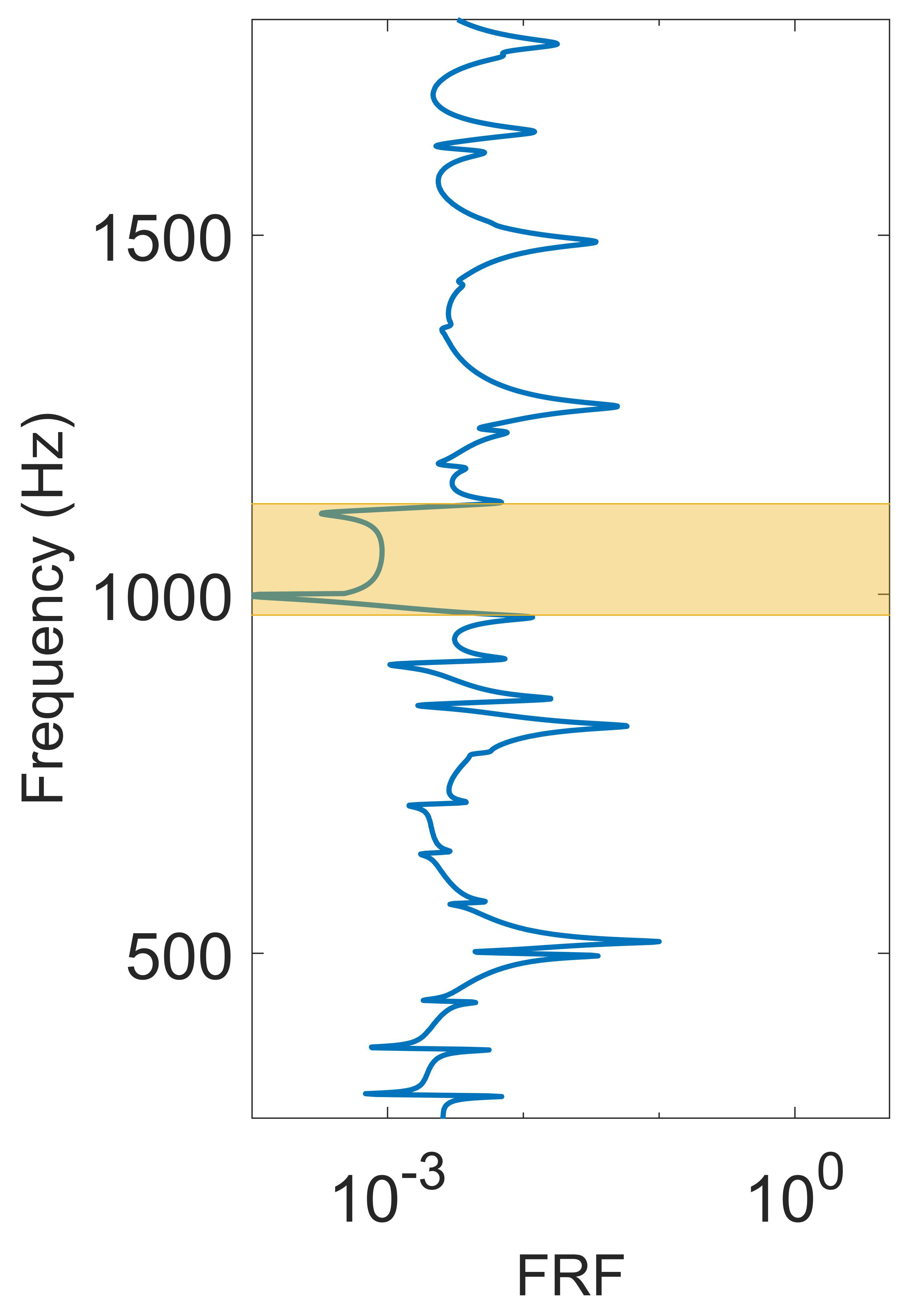}\label{fig:e6}}
\subfloat[]{\includegraphics[width=0.21\linewidth]{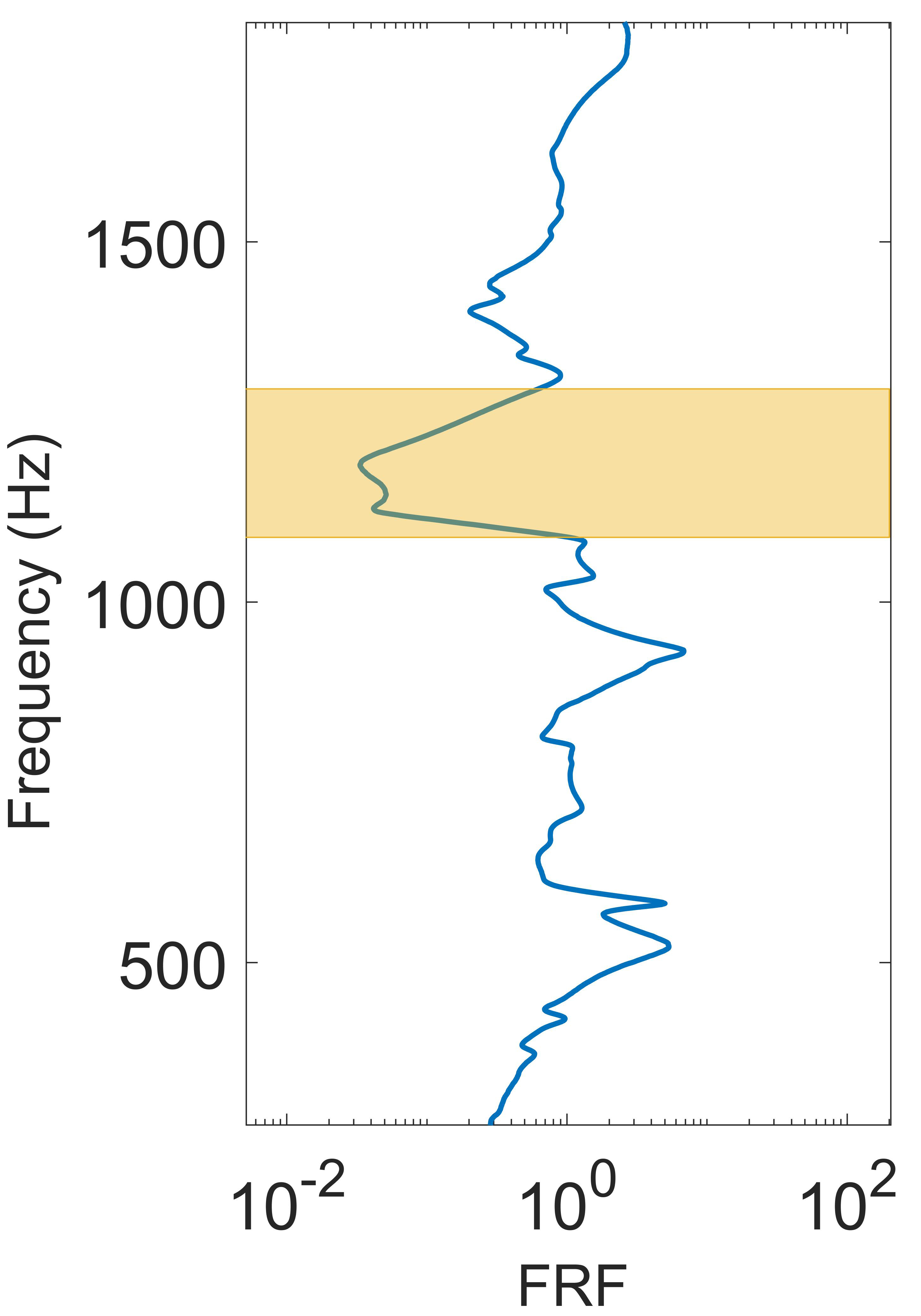}\label{fig:f6}}
\caption{(a) Experimental setup for FRF measurement of the sandwich beam with a GR symmetry. To avoid duplication, we are only presenting the setup for the transverse test. The one for the longitudinal test is similar to that shown in Fig.~\ref{F4}(a). (b) Band diagrams obtained using TMM (dots) and GDQM (circles) respectively. Transverse and longitudinal modes via TMM are presented in red and blue, respectively. (c) and (e): Simulated FRFs (with ten unit cells between the two measured points) of (c) transverse and (e) longitudinal vibration via FEM, respectively. (d) and (f): Experimentally measured FRFs (with eight unit cells between the two measured points) of (d) transverse and (f) longitudinal vibration, respectively. Yellow shades in (c)-(d) are $total$ bandgaps, and green shades in (c) and (d) are $partial$ bandgaps.}
    \label{F6}
    \end{figure}

A comparison of the band diagrams of sandwich beams with MR [Fig.~\ref{F4}(b)] and GR (with $G=d/2$) [Fig.~\ref{F6}(b)] symmetries reveal several important traits in their dynamic behaviors as a result of differences in their structural symmetry. In sandwich beams with an MR symmetry [Fig.~\ref{F4}(a)], bandgaps are open in either transverse or longitudinal (not shown due to high frequency) modes [in contrast to the band diagram for a uniform sandwich structure where there are no bandgaps, as shown in Fig. \ref{F2}(b)]. The crossing between the transverse and longitudinal bands makes them \textit{partial} bandgaps. For example, when exciting the beam at around 500 Hz, where the first bandgap is located, only the transverse phonon modes are prohibited from propagation, while the longitudinal mode can still propagate through. This is validated by the frequency response functions (FRFs) obtained via FEM and experiment, as shown in Fig.~\ref{F4}(c)-(f). As we can see, frequency ranges with low transmission [the two dips of FRFs marked with green shades in Fig.~\ref{F4}(c) and (d)] only exist in the transverse FRF, as shown in Fig.~\ref{F4}(c) and (d), corresponding to the two transverse bandgaps, presented as green shades in Fig.~\ref{F4}(b), while all longitudinal waves can propagate through under 1800 Hz, as shown in Fig.~\ref{F4}(e) and (f). It is also worth noting that the first bandgap opening between the acoustic and optic transverse modes occurs only at the boundary of the Brillouin zone, \textit{i.e.}, $qd/{\pi}=1$, indicating symmetry breaking only \textit{along} the repeating direction of the unit cell, \textit{i.e.}, the \textit{x}-direction, while the symmetry \textit{about} the \textit{x}-direction, \textit{i.e.}, \textit{along} the \textit{y}-direction, remains intact (which is what we have been referring to as mirror-reflection about the neutral axis of the sandwich beam). Hence, the bandgaps are decoupled due to symmetry broken only along the \textit{x}-direction. Mode shape analyses along the longitudinal ($P_1$, $P_2$, and $P_3$) and transverse ($P_4$ and $P_5$) branches also demonstrate the dissociation of the two modes, as shown in Fig.~\ref{F5}, in which longitudinal [Fig.~\ref{F5}(a)-(c)] and transverse [Fig.~\ref{F5}(d) and (e)] deformations dominate along their respective phonon branches. \

\begin{figure}[H]\centering
\subfloat[]{\includegraphics[width=6cm,height=5.5cm]{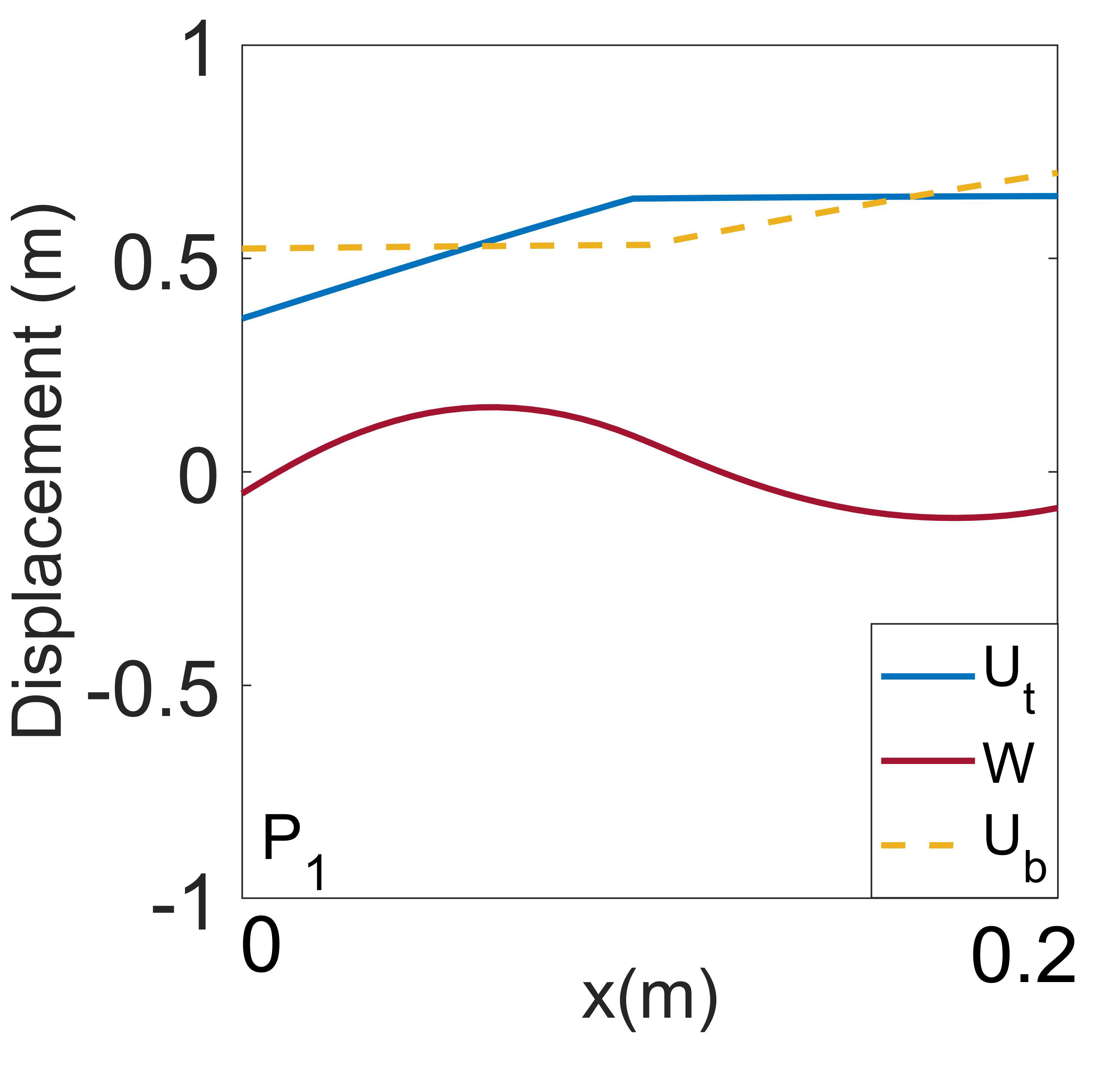}\label{fig:a7}}
\subfloat[]{\includegraphics[width=6cm,height=5.5cm]{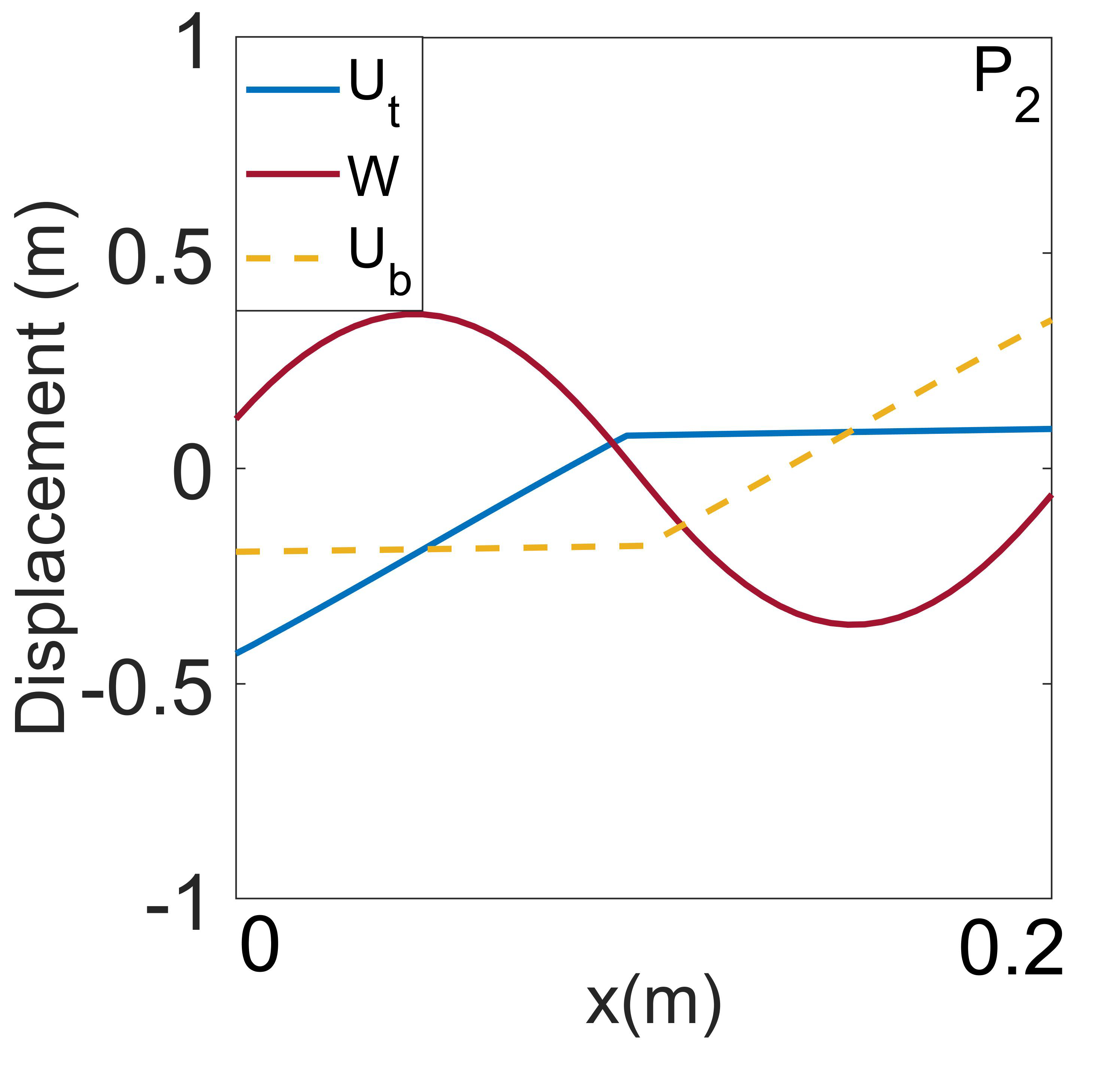}\label{fig:b7}}\hfil
\subfloat[]{\includegraphics[width=6cm,height=5.5cm]{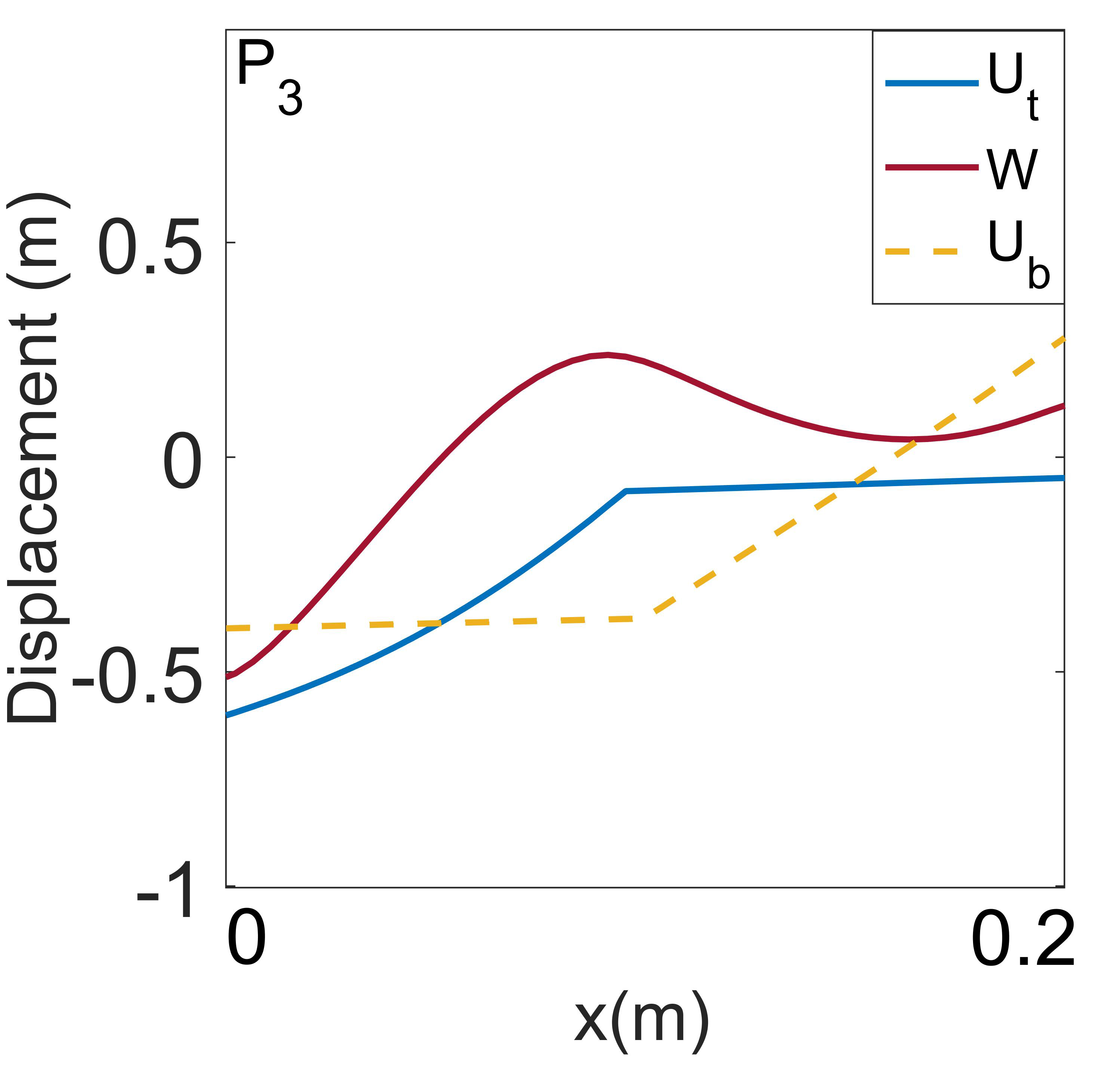}\label{fig:c7}}
\subfloat[]{\includegraphics[width=6cm,height=5.5cm]{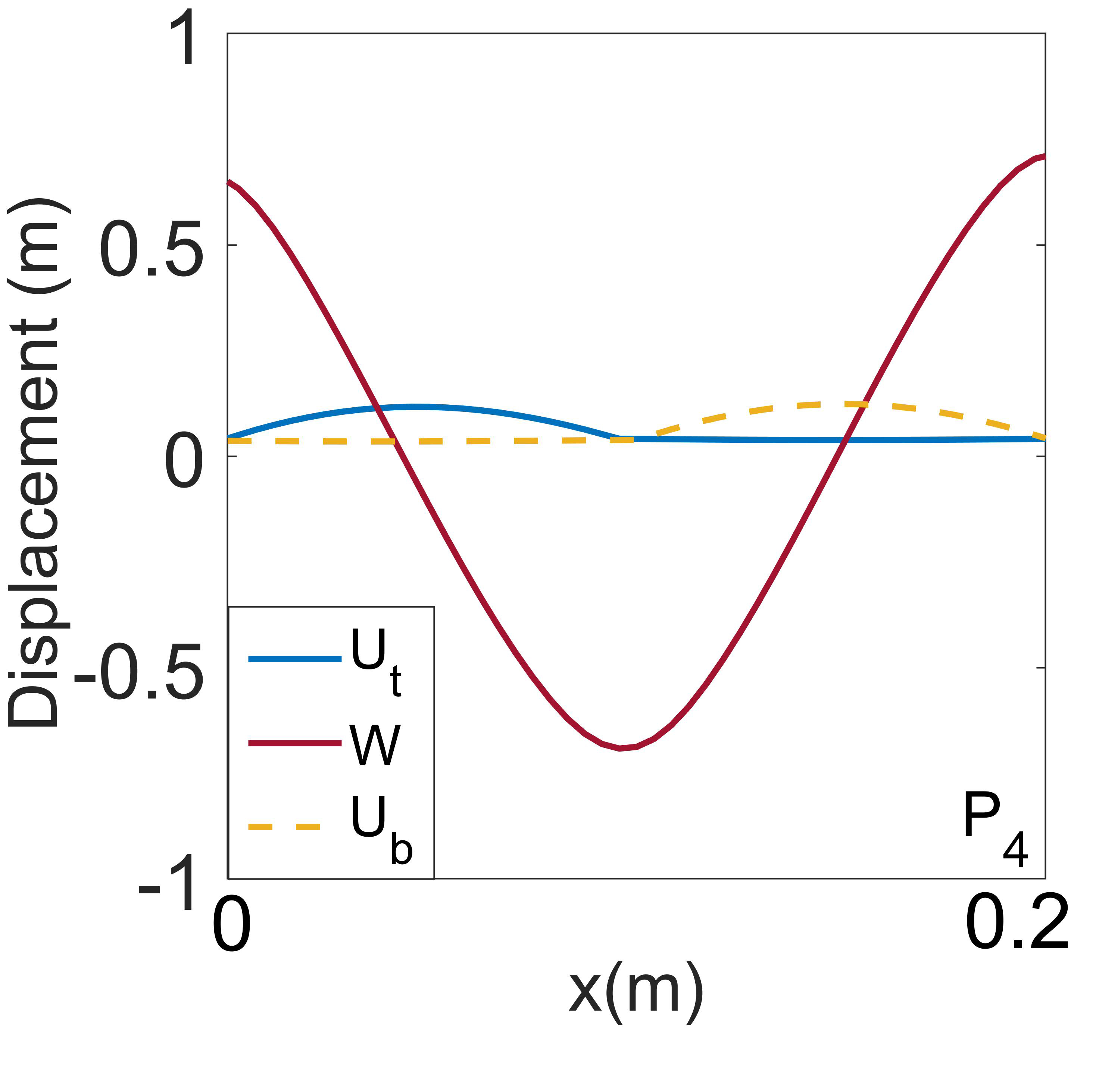}\label{fig:d7}}\\
\subfloat[]{\includegraphics[width=6cm,height=5.5cm]{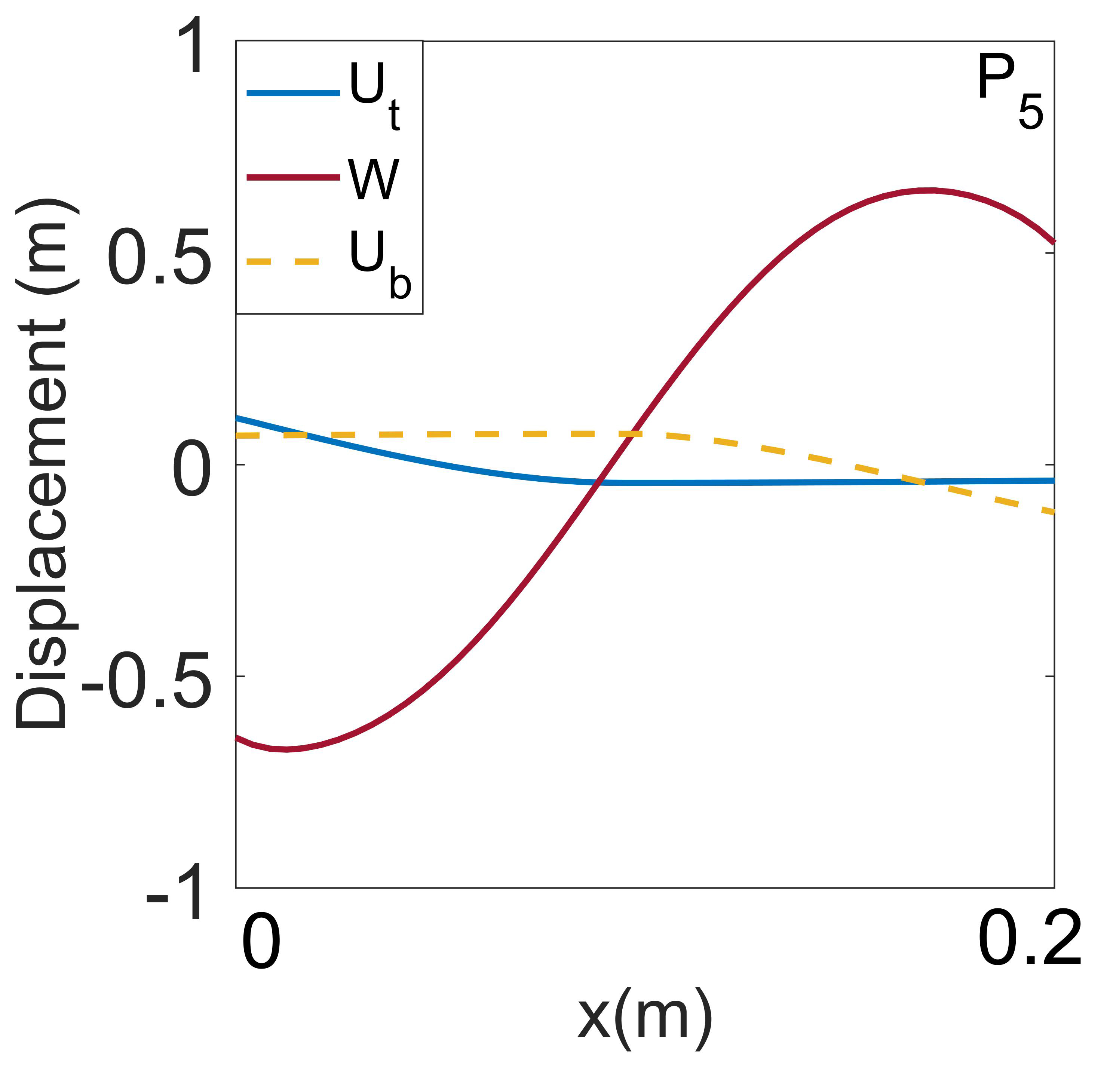}\label{fig:e7}}
\caption{Mode shapes of selected waves at points, (a) $P_1$,  (b) $P_2$, (c) $P_3$, (d) $P_4$ shown as purple squares in Fig. 6(b) for a sandwich beam with periodic face sheets with a GR symmetry.}
    \label{F7}
    \end{figure}

On the other hand, for sandwich beams with a GR symmetry, \textit{i.e.}, ${\textbf{\textit{T}}}_{G,0}$${\circ}$${\textbf{\textit{r}}}_{y=0}$, with $G=d/2$ [Fig.~\ref{F6}(a)], the $partial$ acoustic-optic transverse bandgap occurring at $qd/\pi=1$ diminishes (in this case, disappears), instead, a $total$ bandgap that is blocking both the transverse and the longitudinal waves emerges at the intersection of the two phonon branches, as presented with a yellow shade in Fig.~\ref{F6}(b). Our FEM and experimental FRF measurements also confirm the existence of such bandgaps with similar frequency ranges within which wave propagation in both transverse and longitudinal directions are effectively impeded, as shown in Fig.~\ref{F6}(c)-(e). Such a $total$ bandgap has also been observed in undulated beams and plates \cite{trainiti2015wave}, with the explanation that curvature of the beam couples the transverse and longitudinal behaviors. However, in our GR sandwich beam, we do not have such curved components to couple the two modes. Thus, such an explanation does not apply to our GR sandwich beams. Instead, we attribute such a \textit{total} bandgap to the simultaneous symmetry breaking along both $x-$ and $y-$directions in our GR $(G=d/2)$ structure, leading to the coupling of both transverse and longitudinal phonon modes within the bandgap. Our explanation applies to more general quasi-one-dimensional structures with GR symmetry, including the one studied in Ref. \cite{trainiti2015wave}. Our calculated mode shapes along the longitudinal and transverse branches also agree with such an explanation. As we can see, longitudinal and transverse displacements are only coupled near the \textit{total} bandgap, while remaining mostly decoupled elsewhere, as shown in Fig. \ref{F7}.

\begin{figure}[H]\centering
\subfloat[]{\includegraphics[width=7cm,height=6.4cm]{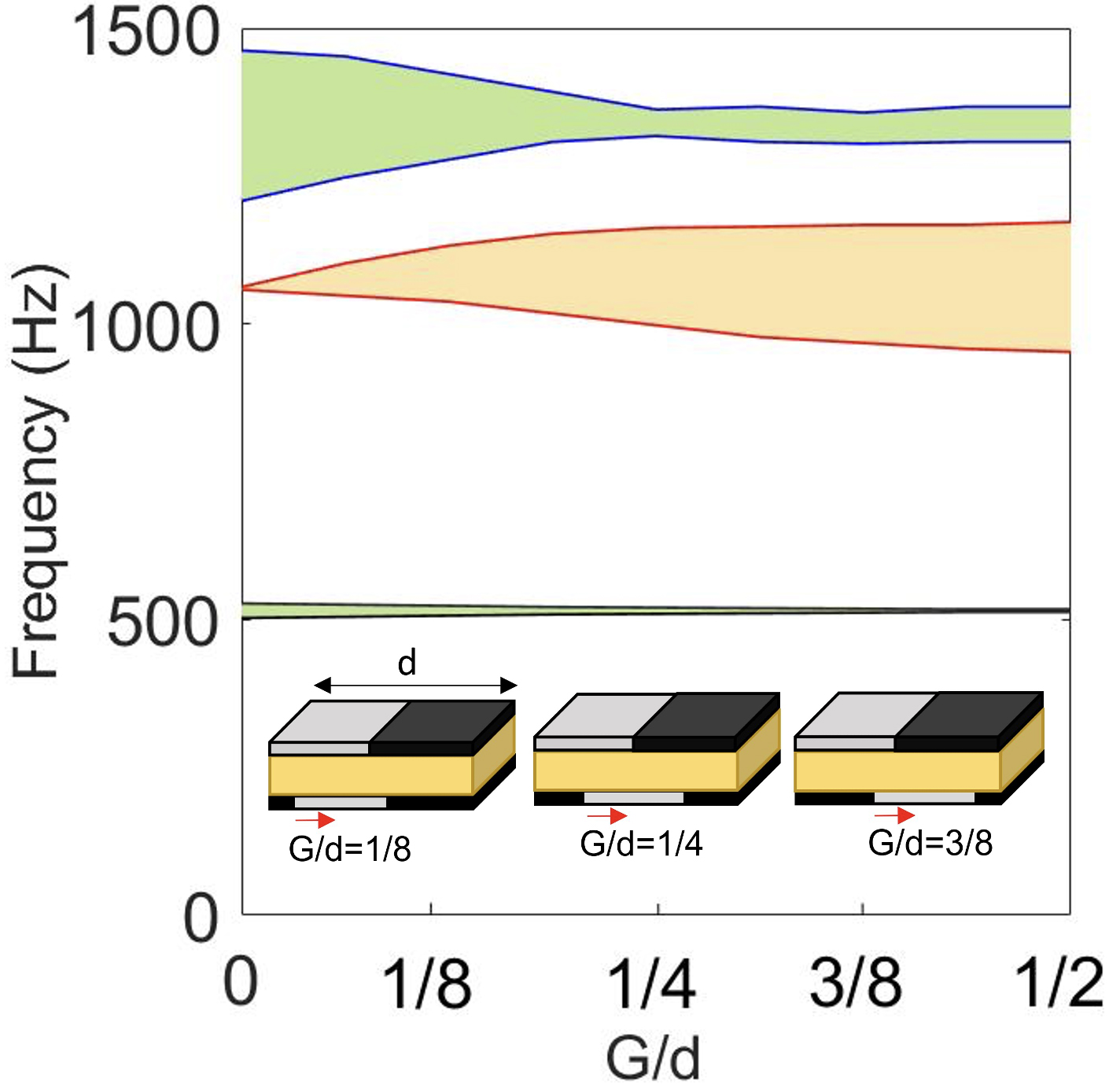}\label{fig:a8}}
\subfloat[]{\includegraphics[width=7cm,height=6.4cm]{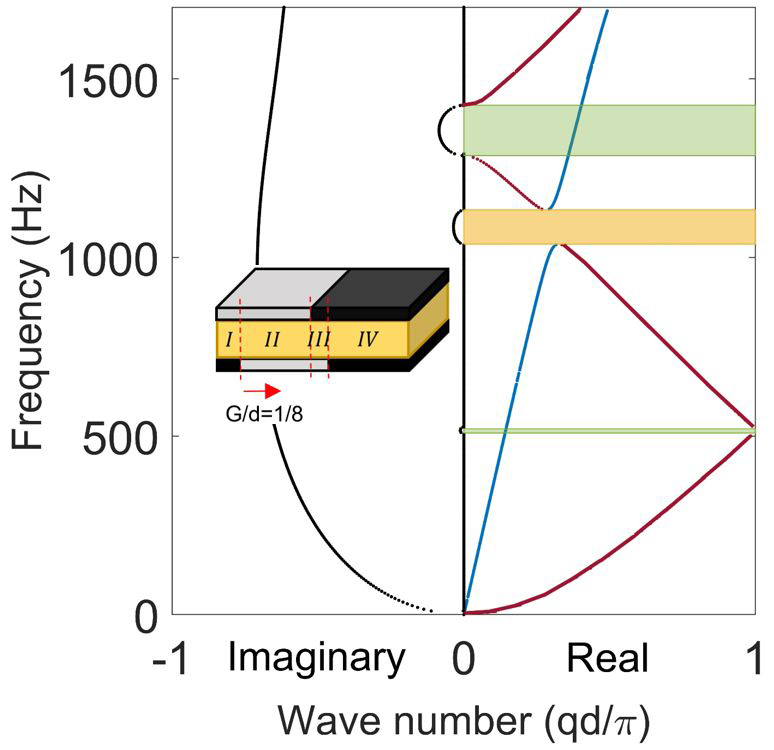}\label{fig:b8}}\hfil
\caption{(a) Effect of face sheets as a function of the glide magnitude, $G$, \textit{i.e.}, MR to GR symmetry, on the size of the bandgaps. (b) Band diagram for $G/d=1/8$. The unit cell is further divided into four subunits, \textit{I-IV}, for the convenience of our theoretical analysis.\label{F8}}
    \end{figure}

Our discussion above compares the differences between sandwich beams with MR (\textit{i.e.}, ${\textbf{\textit{T}}}_{G,0}$${\circ}$${\textbf{\textit{r}}}_{y=0}$, with $G=0$) and GR $(G=d/2)$ symmetries. One might wonder what happens if $0<G<d/2$. Our parametric study reveals that both decoupled (\textit{partial}) and coupled (\textit{total}) bandgaps are open when $G$ is within this range, as is shown in Fig. \ref{F8}. This is because when $0<G<d/2$, we can further divide the unit cell into four subunit cells [an example of $G=d/8$ is shown as the inset of Fig. \ref{F8}(b)], among which subunit cells $I$ and $III$ are asymmetric about the axial direction of the beam, while $II$ and $IV$ are symmetric (\textit{i.e.}, mirror-reflected). Such a symmetry difference within the same unit cell leads to the coexistence of both decoupled (\textit{partial}) and coupled (\textit{total}) bandgaps. The widths of these two bandgaps depend on the proportion of the contribution from their respective parts, \textit{i.e}., the larger the MR portion is, the wider the \textit{partial} bandgaps are and the narrower the \textit{total} bandgap is, and vice versa. As we can see, within 0 to $d/2$, the more we glide the bottom face sheet (\textit{i.e}., the larger the $G$ is), the less the MR portion is in a unit cell, and the narrower/wider the $partial/total$ bandgap(s) the structure creates. 

\begin{figure}[H]
\centering
\begin{minipage}[c]{\textwidth}
\centering
    \includegraphics[width=3in]{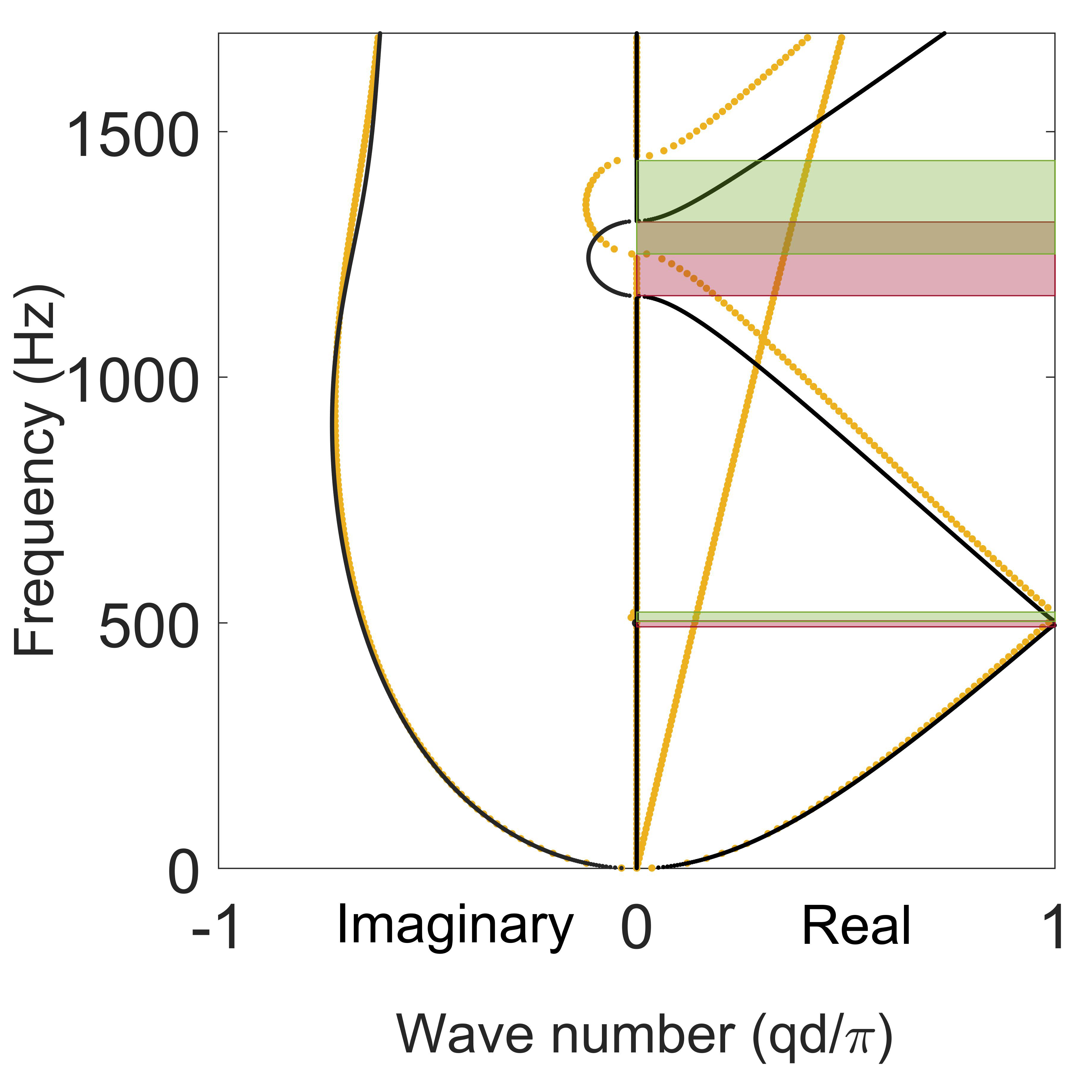}
\caption {Band diagrams of the sandwich beam with materially periodic face sheets with an MR symmetry. Yellow and black curves are obtained via CPT and Timoshenko beam theory with equivalent properties, respectively.\label{F9}}
\end{minipage}
\end{figure}

\indent We should also note that even though sandwich beams with an MR symmetry appear to be also approachable via a more commonly used Timoshenko beam theory with equivalent properties (as presented in Table~\ref{T2}) due to its unaltered neutral axis throughout the beam, our calculation reveals that such an equivalent Timoshenko beam cannot provide accurate results as our analysis does with CPT. The discrepancy is more pronounced in the optic phonon branches, as shown in Fig. \ref{F9}. This is because the equivalent Timoshenko beam theory fails to account for incoherent motions of the core and top and bottom face sheets as considered in our analysis with separate displacement fields. Moreover, equivalent Timoshenko beam cannot provide the longitudinal mode for the sandwich beam with periodic face sheets. The calculation of equivalent properties is presented in Appendix G.

\begin{table}[h]
\small\sf\centering
\caption{Equivalent properties of the sandwich beam with Rohacell foam as the core and periodic face sheets (made of AL and ABS materials) with an MR symmetry.\label{T2}}
\begin{tabular}{lllllllll}
\toprule
Face sheet material&$EI$($Pa{\cdot}m^4$)&$GA$($Pa{\cdot}m^2$)&$\rho A$($kg/m$)&$\rho I$($kg{\cdot}m$)\\
\bottomrule\
AL &$1629.7$&$17494$&$0.3619$&$6.7622\times10^{-5}$\\
\bottomrule\
ABS &$52.05$&$17494$&$0.1617$&$2.7243\times10^{-5}$\\
\bottomrule
\end{tabular}\\[10pt]
\end{table}

\begin{figure}[H]\centering
\subfloat[]{\includegraphics[width=6cm,height=5.5cm]{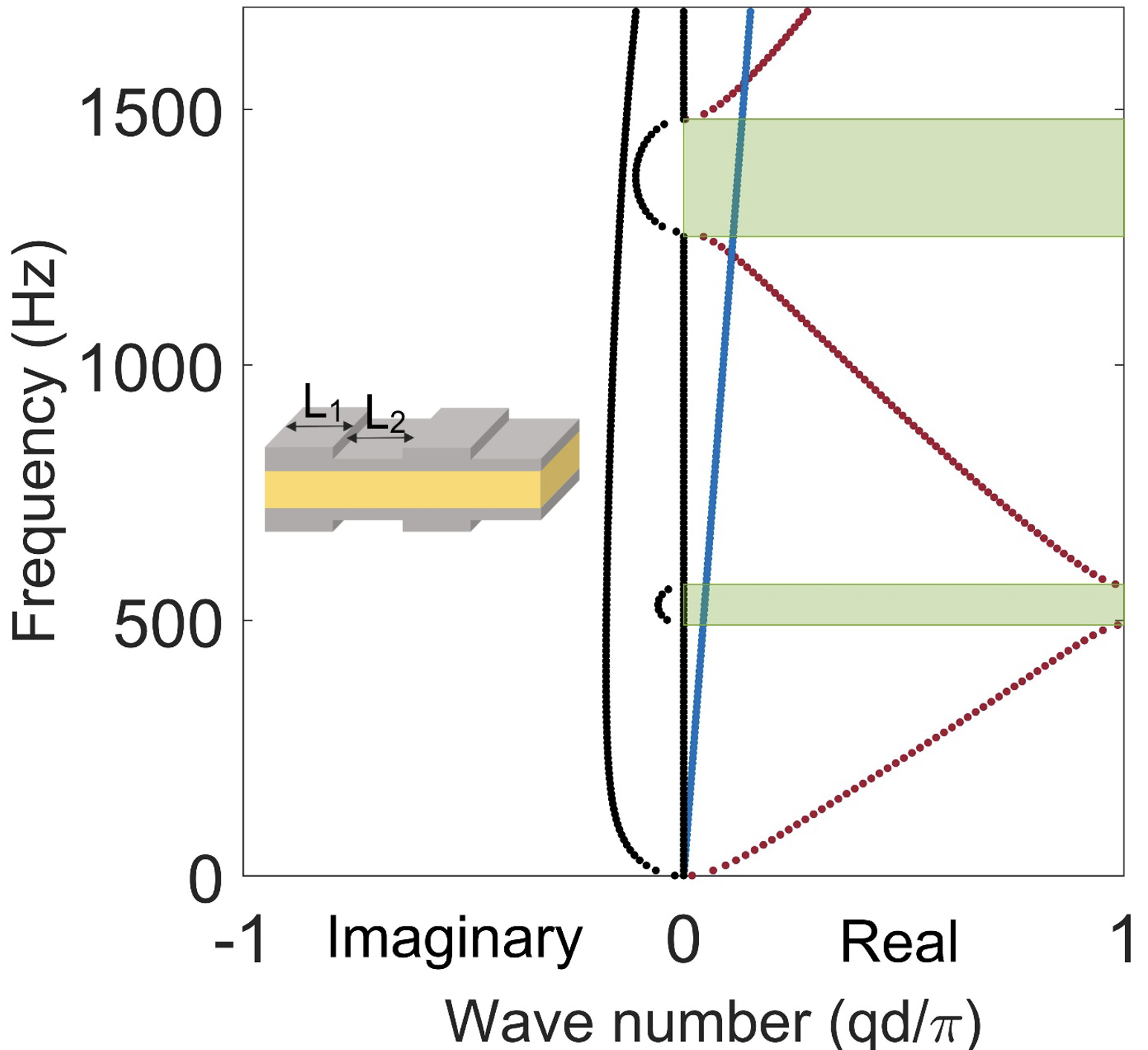}\label{fig:a10}}
\subfloat[]{\includegraphics[width=6cm,height=5.5cm]{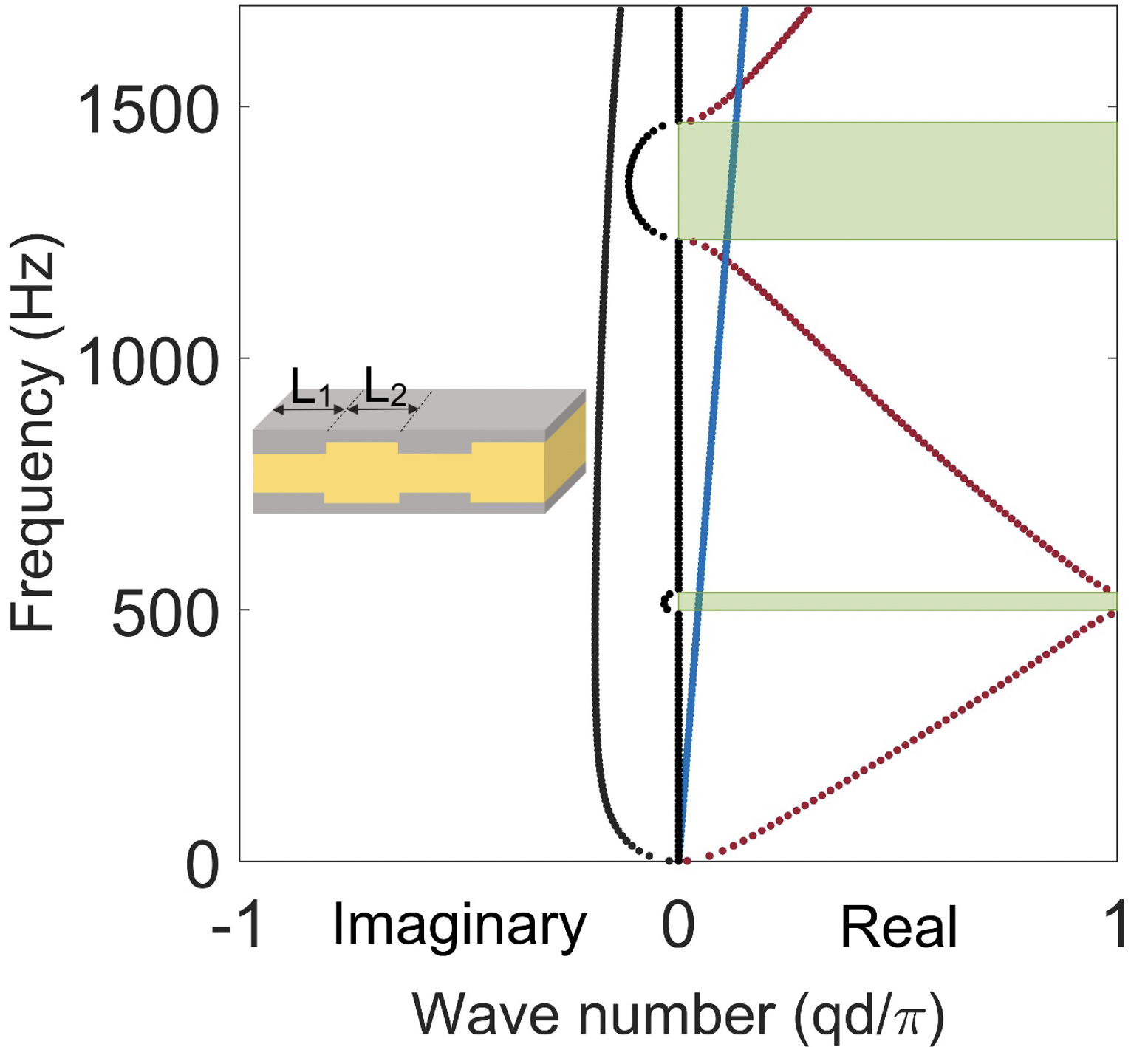}\label{fig:b10}}\hfil
\subfloat[]{\includegraphics[width=6cm,height=5.5cm]{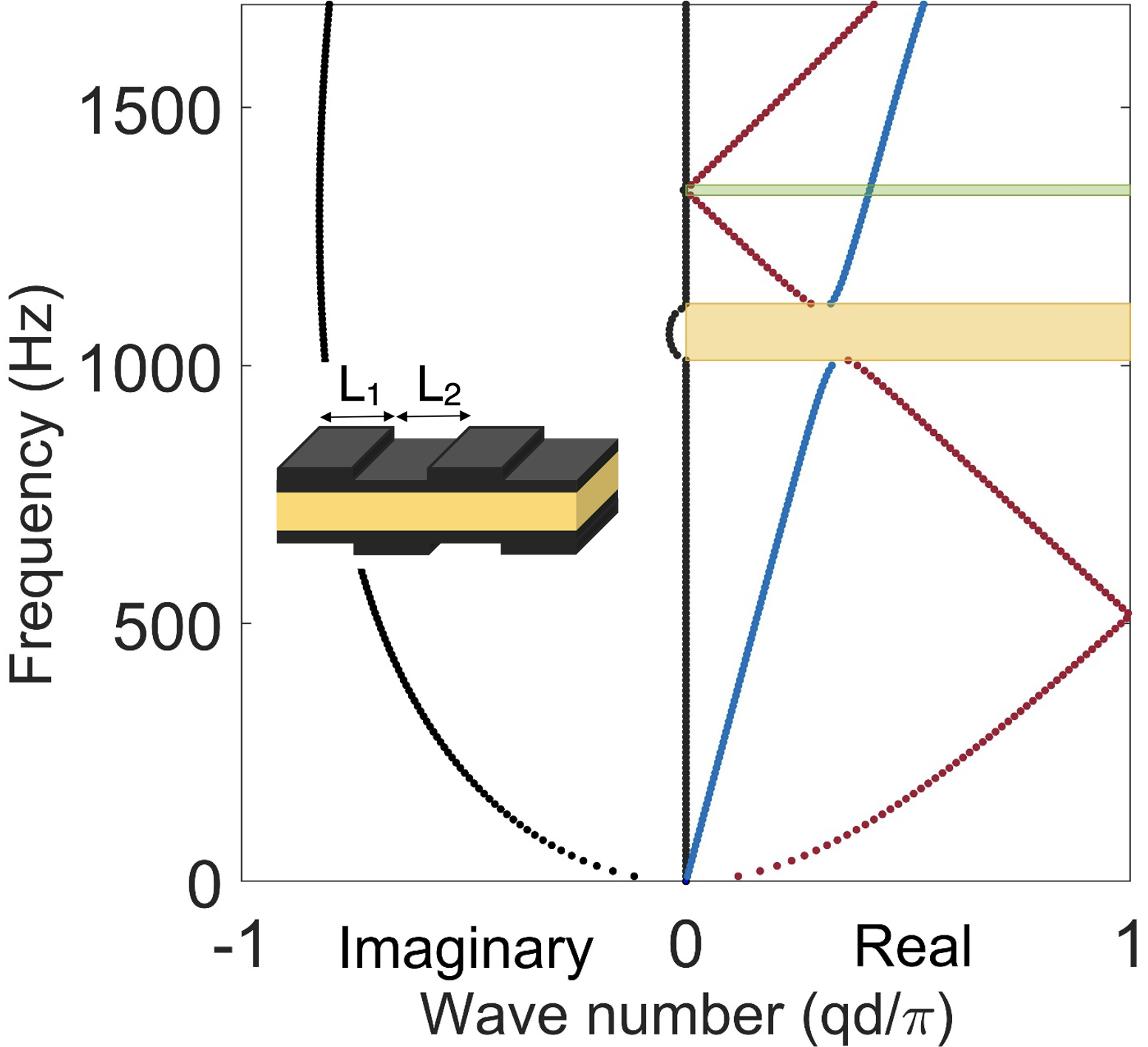}\label{fig:c10}}
\subfloat[]{\includegraphics[width=6cm,height=5.5cm]{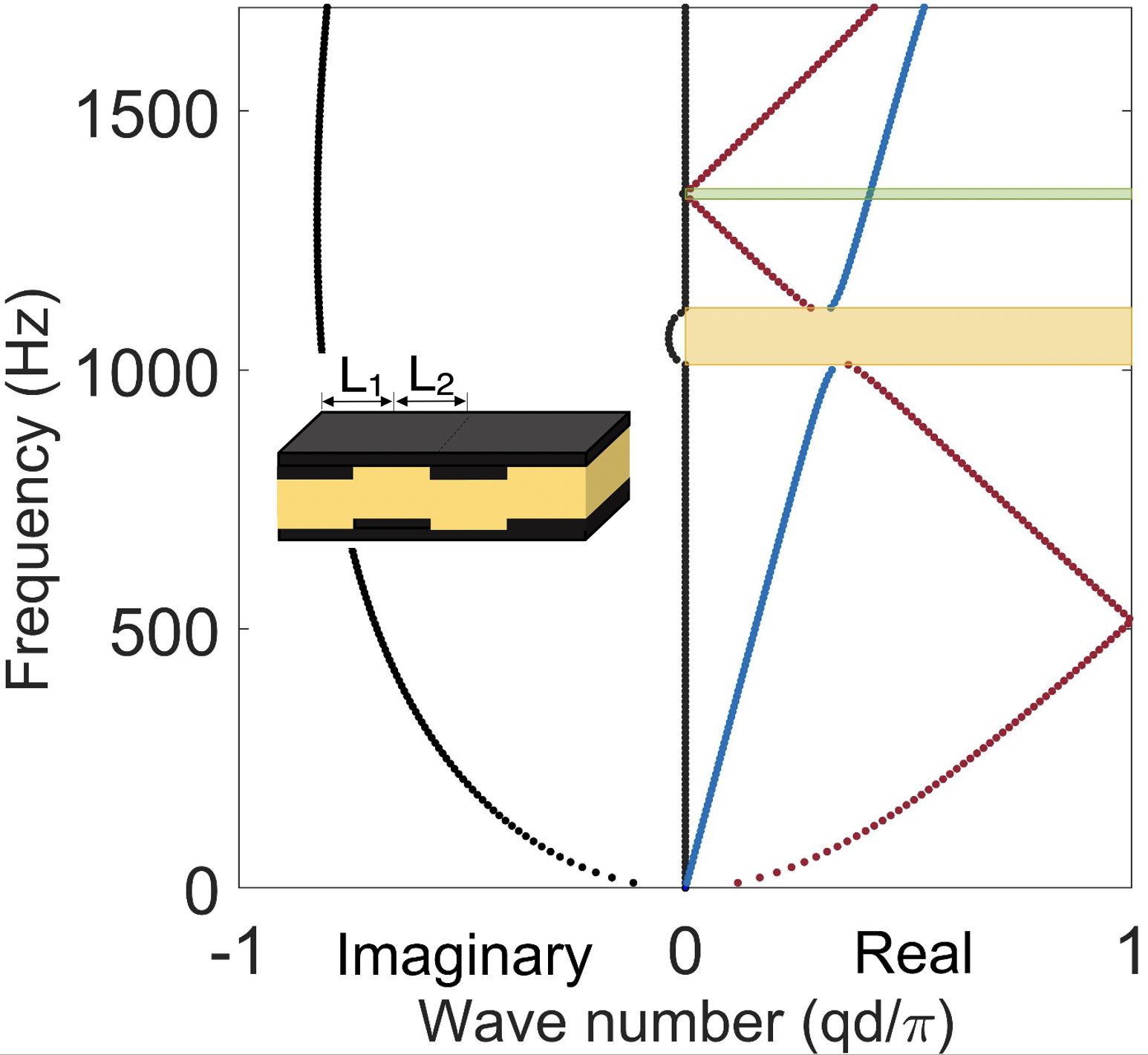}\label{fig:d10}}\hfil
\caption{Band diagram of sandwich beams with geometric periodicity in face sheets made of (a,b) AL with an MR symmetry and (c,d) ABS with a GR symmetry and subunit cells with length $L_1$=$L_2$=101.6 mm. (a) and (b): MR symmetry with (a) outward and (b) inward geometric variation with face sheet thicknesses $h_{t_1}(h_{b_1})$=6.35 mm and $h_{t_2}(h_{b_2})$=3.175 mm, and (a) uniform core thickness $h_{c_1}(h_{c_2})$= 25.4 mm, (b) varying core thicknesses $h_{c_1}=19.05$ mm and $h_{c_2}$=25.4 mm. The geometric parameters in (c) and (d) use ABS as face sheets with GR $(G=d/2)$ symmetry. The geometric parameters, \textit{i.e.}, $h_{t_1}$, $h_{t_2}$, $h_{b_1}$, $h_{b_2}$, $h_{c_1}$, and $h_{c_2}$ are consistent with the ones described in (a) and (b), respectively.
}
    \label{F10}
    \end{figure}

\begin{figure}[H]\centering
\subfloat[]{\includegraphics[width=0.37\linewidth]{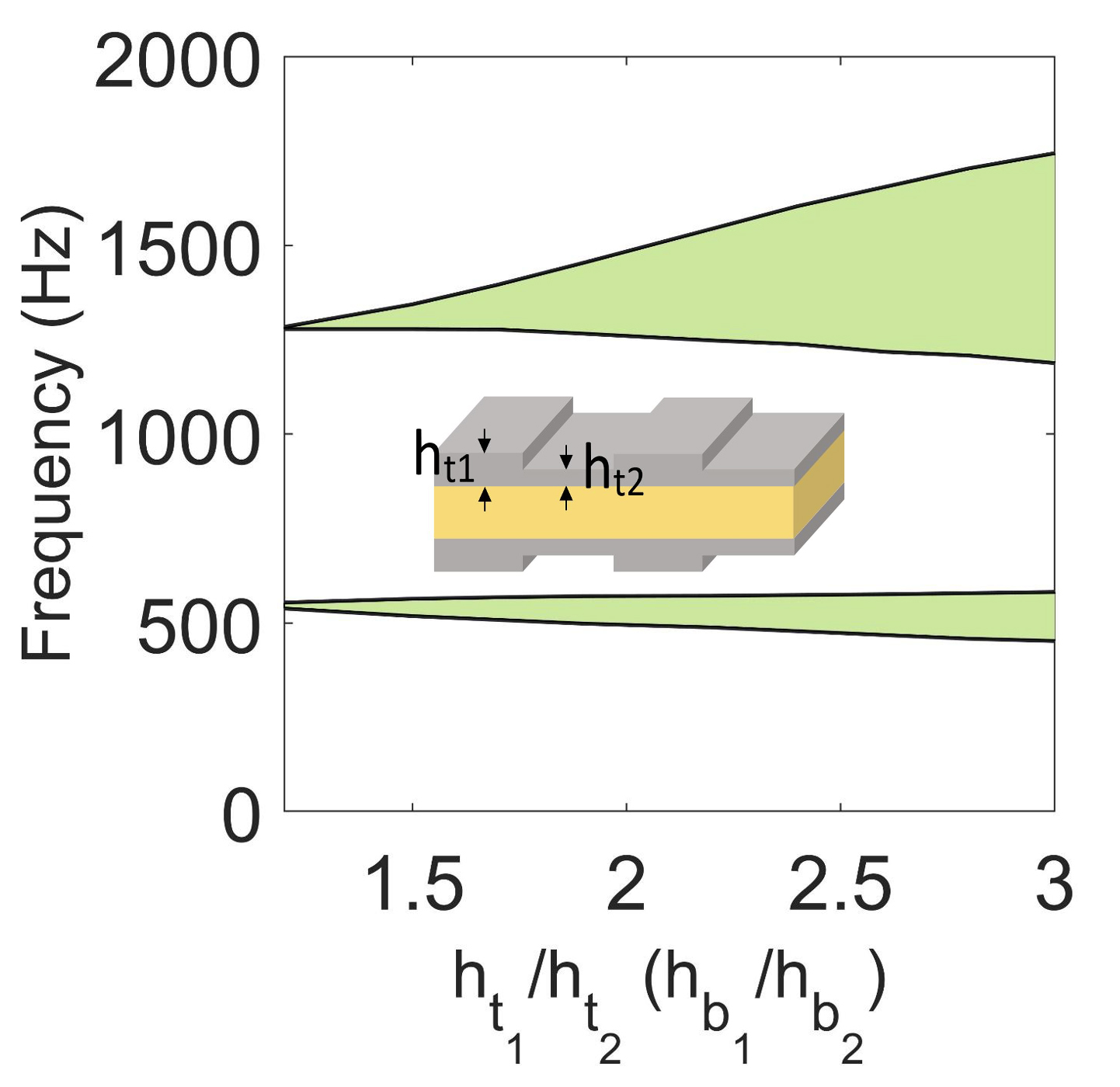}\label{fig:a11}}
\subfloat[]{\includegraphics[width=0.37\linewidth]{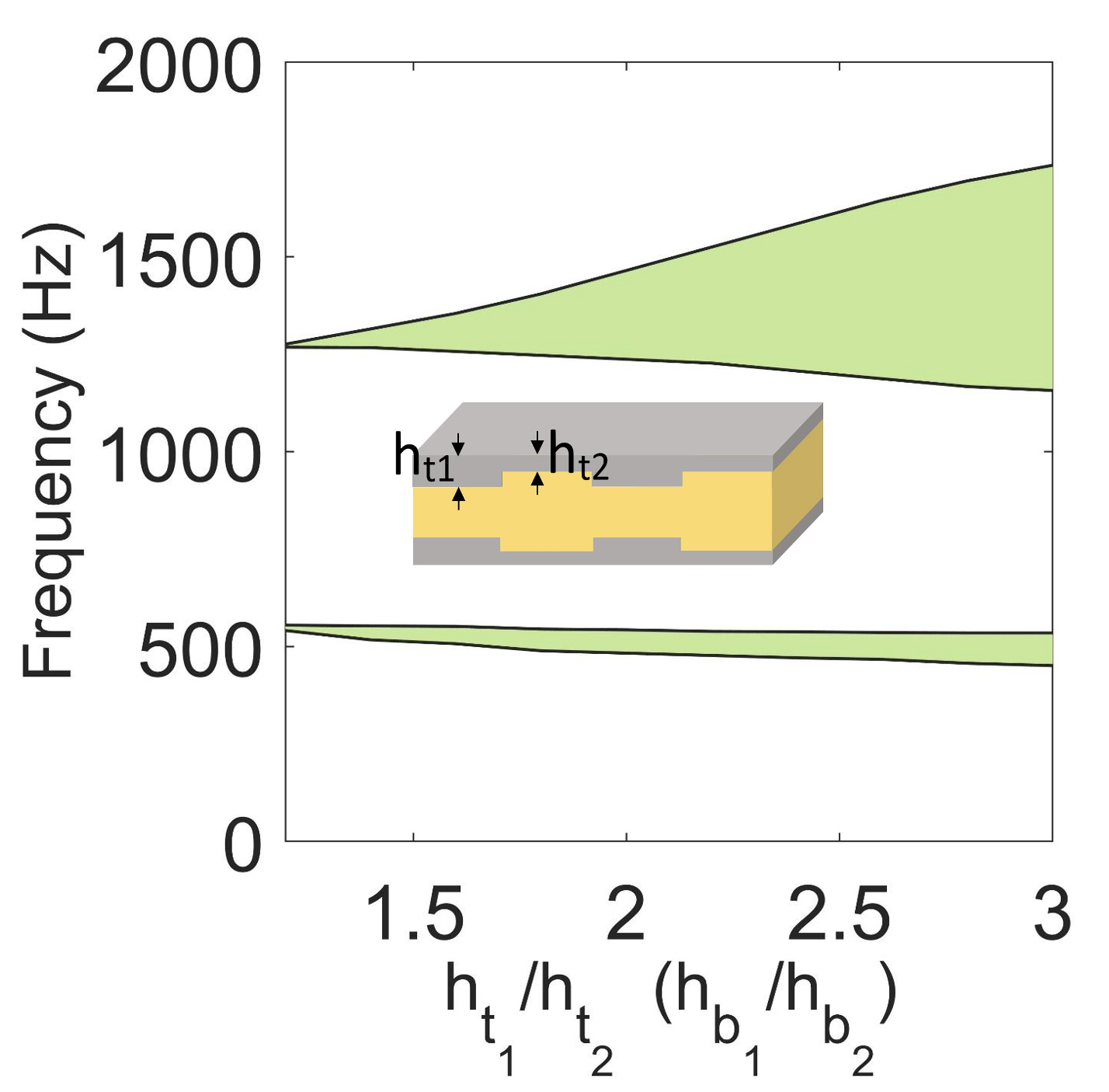}\label{fig:b11}}\hfil
\subfloat[]{\includegraphics[width=0.37\linewidth]{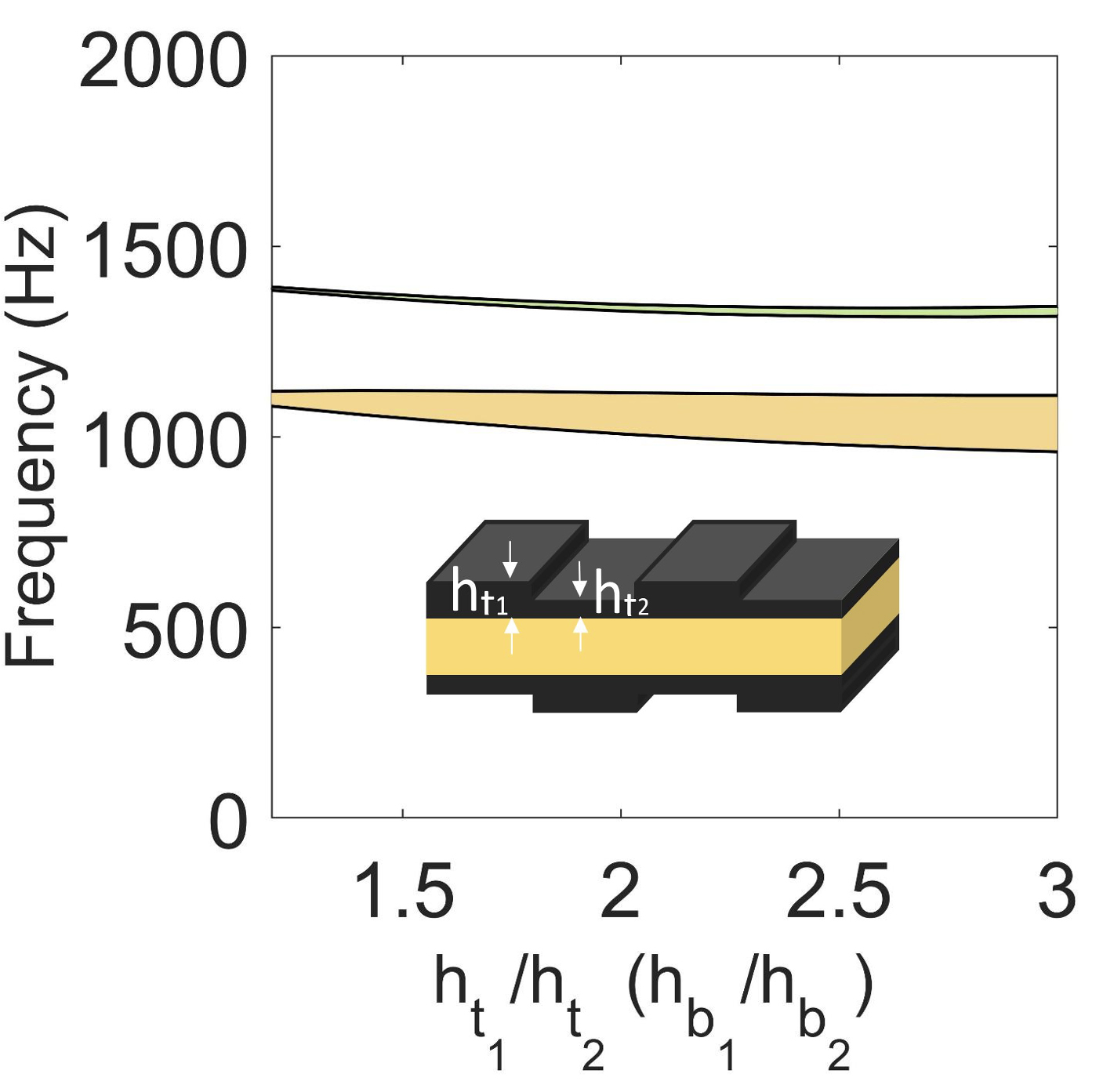}\label{fig:c11}}
\caption{Effect of face sheets thickness ratio on the size and location of $partial$ (green) and $total$ (yellow) bandgaps of sandwich beams with geometric periodicity in face sheets made of (a,b) AL with an MR symmetry and (c) ABS with a GR symmetry.}
    \label{F11}
    \end{figure}

To ensure the generosity of the symmetry effect, we also investigate sandwich beams with geometric periodicity, as are shown in Fig.~\ref{F1}(c)-(f). Similar to the two cases (MR and GR) with material periodicity, we observe $partial$ bandgaps only and both $total$ and $partial$ bandgaps in MR and GR (with $G=d/2$) sandwich beams, respectively, regardless of whether the geometry variation is inward or outward, as is shown in Fig.~\ref{F10}. Thus, the symmetry-bandgap relationship we alluded to in materially periodic sandwich beams still applies to the geometrically periodic scenarios. Contrary to their material counterpart, more geometric parameters can be tuned in these geometrically periodic sandwich beams to emphasize the differences in periodic face sheets. For example, a larger contrast in thicknesses of face sheets ($h_{t_1}/h_{t_2}$ and $h_{b_1}/h_{b_2}$) will lead to more widely opened bandgaps for both MR and GR sandwich beams, as shown in Fig.~\ref{F11}.

\section{Conclusion}
In this work, we investigate dynamic behaviors of sandwich beams with materially and geometrically periodic face sheets with MR and GR symmetries using CPT, which is more advantageous compared to the traditionally applied equivalent Timoshenko beam models. Numerical simulation and experimental measurements are performed to validate the accuracy of our theoretical results. We find that the sandwich beams with an MR symmetry can only open \textit{partial} bandgaps with decoupled transverse or longitudinal phonon modes. On the other hand, we obtain \textit{total} bandgaps that can effectively stop both transverse and longitudinal phonons from propagation in sandwich beams with a GR symmetry, which can be achieved without using a curved beam. We attribute this \textit{total} bandgap opening mechanism to the symmetry of structures, which offers a more general explanation of the dynamic behaviors of structures with similar symmetry profiles. Moreover, parametric studies suggest that both the existence and the magnitude of the \textit{total} (as well as partial) bandgaps can be further tuned by adjusting the glide magnitude and/or differences in materials/geometry of the periodic face sheets, which provides more interesting approaches to wave manipulation mechanisms at large.

\:\:\:\:\:\:\:\:\:\:\

\section{Data availability}
All data is stored on a secured hard drive at the Department of Mechanical Engineering, University of Vermont. The code, simulation and experimental data required to reproduce these findings are available from the corresponding author on reasonable request. 

\:\:\:\:\:\:\:\:\:\:\

\noindent\textbf{Acknowledgement}\\\\
\indent We are thankful to EVONIK Company for supporting our research by providing foam for our experiments. J. M. would like to acknowledge the University of Vermont Early EXtra Promotion of REsearch and Scholarly Success Grant Award for the support of this research.

\:\:\:\:\:\:\:\:\:\:\

\noindent\textbf{Appendix A}\\

The longitudinal and transverse displacements for the core $(U_c$ and $W_c)$, and top $(U_t$ and $W_t)$ and bottom $(U_b$ and $W_b)$ layers under compatibility conditions between the layers are expressed as 

\begin{equation}
Core\:\:\:\:\:\:\:\:\:\:\:\:\:\
\begin{array}{ll}
\end{array} \quad\left\{\! \begin{array}{ll}
U_c(x,y,\tau)= \displaystyle \frac{u_t+u_b}{2}+\frac{(h_t-h_b)}{4}w'+y(\frac{u_t-u_b}{h_c}+\frac{(h_t+h_b)}{2h_c}w')\\
W_c(x,y,\tau)= \displaystyle w(x,\tau)\end{array}\right.\nonumber\:\:\:\:\:\:\:\:\:\:\:\:\:\:\:\:\:\:\:\:\:\:\:\:\:\:\:\:\:\:\:\:\:\:\:\:\:\:\ (A1)
\end{equation}
\begin{equation}
Top\:\ face\:\ sheet
\begin{array}{ll}
\end{array} \quad\left\{\! \begin{array}{ll}
U_t(x,y,\tau)=u_t(x,\tau)-yw'(x,\tau)\\
W_t(x,y,\tau)=w(x,\tau)\end{array}\right.\nonumber\:\:\:\:\:\:\:\:\:\:\:\:\:\:\:\:\:\:\:\:\:\:\:\:\:\:\:\:\:\:\:\:\:\:\:\:\:\:\:\:\:\:\:\:\:\:\:\:\:\:\:\:\:\:\:\:\:\:\:\:\:\:\:\:\:\:\:\:\:\:\:\:\:\:\:\:\:\ (A2)
\end{equation}
\begin{equation}
Bottom\:\ face\:\ sheet
\begin{array}{ll}
\end{array} \quad\left\{\! \begin{array}{ll}
U_b(x,y,\tau)=u_b(x,\tau)-yw'(x,\tau)\\
W_b(x,y,\tau)=w(x,\tau)\end{array}\right.\nonumber\:\:\:\:\:\:\:\:\:\:\:\:\:\:\:\:\:\:\:\:\:\:\:\:\:\:\:\:\:\:\:\:\:\:\:\:\:\:\:\:\:\:\:\:\:\:\:\:\:\:\:\:\:\:\:\:\:\:\:\:\:\:\:\:\:\:\:\:\:\:\:\:\:\:\:\:\:\:\:\:\:\:\ (A3)
\end{equation}

The strain-displacement relationships for each layer are presented as follows
\begin{equation}
\varepsilon_{xx}^c=\frac{u_t'+u_b'}{2}+\frac{(h_t-h_b)}{4}w''+y(\frac{u_t'-u_b'}{h_c}+\frac{\left(h_t+h_b\right)}{2h_c}w'')\nonumber\:\:\:\:\:\:\:\:\:\:\:\:\:\:\:\:\:\:\:\:\:\:\:\:\:\:\:\:\:\:\:\:\:\:\:\:\:\:\:\:\:\:\:\:\:\:\:\:\:\:\:\:\:\:\:\:\:\:\:\:\:\:\:\:\:\:\:\:\:\:\:\:\:\:\:\:\:\:\:\:\:\:\ (A4)
\end{equation}
\begin{equation}
\gamma_{xy}^c=\frac{1}{h_c}(u_t-u_b+\frac{\left(h_t+h_b+2h_c\right)}{2}w')\nonumber\:\:\:\:\:\:\:\:\:\:\:\:\:\:\:\:\:\:\:\:\:\:\:\:\:\:\:\:\:\:\:\:\:\:\:\:\:\:\:\:\:\:\:\:\:\:\:\:\:\:\:\:\:\:\:\:\:\:\:\:\:\:\:\:\:\:\:\:\:\:\:\:\:\:\:\:\:\:\:\:\:\:\:\:\:\:\:\:\:\:\:\:\:\:\:\:\:\:\:\:\:\:\:\:\:\:\:\:\:\:\:\:\:\ (A5)
\end{equation}
\begin{equation}
{\varepsilon }_{xx}^t=u_t'-yw''\nonumber\:\:\:\:\:\:\:\:\:\:\:\:\:\:\:\:\:\:\:\:\:\:\:\:\:\:\:\:\:\:\:\:\:\:\:\:\:\:\:\:\:\:\:\:\:\:\:\:\:\:\:\:\:\:\:\:\:\:\:\:\:\:\:\:\:\:\:\:\:\:\:\:\:\:\:\:\:\:\:\:\:\:\:\:\:\:\:\:\:\:\:\:\:\:\:\:\:\:\:\:\:\:\:\:\:\:\:\:\:\:\:\:\:\:\:\:\:\:\:\:\:\:\:\:\:\:\:\:\:\:\:\:\:\:\:\:\:\:\:\:\:\:\:\:\:\:\:\:\:\:\:\:\:\:\:\:\:\:\:\:\:\:\:\:\ (A6)
\end{equation}
\begin{equation}
{\varepsilon }_{xx}^b=u_b'-yw''\nonumber\:\:\:\:\:\:\:\:\:\:\:\:\:\:\:\:\:\:\:\:\:\:\:\:\:\:\:\:\:\:\:\:\:\:\:\:\:\:\:\:\:\:\:\:\:\:\:\:\:\:\:\:\:\:\:\:\:\:\:\:\:\:\:\:\:\:\:\:\:\:\:\:\:\:\:\:\:\:\:\:\:\:\:\:\:\:\:\:\:\:\:\:\:\:\:\:\:\:\:\:\:\:\:\:\:\:\:\:\:\:\:\:\:\:\:\:\:\:\:\:\:\:\:\:\:\:\:\:\:\:\:\:\:\:\:\:\:\:\:\:\:\:\:\:\:\:\:\:\:\:\:\:\:\:\:\ (A7)
\end{equation}

\noindent where ${\varepsilon }_{xx}^c$, ${\varepsilon }_{xx}^t$, and ${\varepsilon }_{xx}^b$  are the normal strains in $x$-direction, respectively, and ${\gamma }_{xy}^c$ is the shear strain. 

The normal strain energy at the core and top and  bottom face sheets are defined as $V_{cN}$, $V_{tN}$, and $V_{bN}$ respectively as 
\begin{equation}
\begin{aligned}
\nonumber\\&\mathit{V_{cN}} = \displaystyle{\frac{1}{2}}E_cb\int_{0}^{L}\int^{\frac{h_c}{2}}_{-\ \frac{h_c}{2}}(\varepsilon_{xx}^c)^2dydx=\displaystyle{\frac{1}{6}}E_cA_c\int_{0}^{L}[u_t'^2+u_b'^2+u_t'u_b'+\left(h_t-\frac{h_b}{2}\right)u_t'w''\nonumber\\&+\left(\frac{h_t}{2}-h_b\right)u_b'w''+\frac{1}{4}\left(h_t^2-h_th_b+h_b^2\right)w''^2]dx\nonumber\\
\end{aligned}\nonumber\:\:\:\:\:\:\:\:\:\:\:\:\:\:\:\:\:\:\:\:\:\:\:\:\:\:\:\:\:\:\:\:\:\:\:\:\:\:\:\:\:\:\:\:\:\:\:\:\:\:\:\:\:\:\ (A8)
\end{equation}

\begin{equation}
\begin{aligned}
V_{tN} = \displaystyle{\frac{1}{2}}E_tb\int_{0}^{L}\int^{\frac{h_t}{2}}_{-\ \frac{h_t}{2}}(\varepsilon_{xx}^t)^2dydx=\displaystyle{\frac{1}{2}}E_tA_t\int_{0}^{L}u_t'^2dx+\displaystyle{\frac{1}{2}}E_tI_t\int_{0}^{L}w''^2dx
\end{aligned}\nonumber\:\:\:\:\:\:\:\:\:\:\:\:\:\:\:\:\:\:\:\:\:\:\:\:\:\:\:\:\:\:\:\:\:\:\:\:\:\:\:\:\:\:\:\:\:\:\:\:\:\:\:\:\:\:\ (A9)
\end{equation}

\begin{equation}
\begin{aligned}
&V_{bN} = \displaystyle{\frac{1}{2}}E_bb\int_{0}^{L}\int^{\frac{h_b}{2}}_{-\ \frac{h_b}{2}}(\varepsilon_{xx}^b)^2dydx=\displaystyle{\frac{1}{2}}E_bA_b\int_{0}^{L}u_b'^2dx+\displaystyle{\frac{1}{2}}E_bI_b\int_{0}^{L}w''^2dx
\end{aligned}\nonumber\:\:\:\:\:\:\:\:\:\:\:\:\:\:\:\:\:\:\:\:\:\:\:\:\:\:\:\:\:\:\:\:\:\:\:\:\:\:\:\:\:\:\:\:\:\:\:\:\:\:\:\:\:\:\ (A10)
\end{equation}

The shear strain energy at the core is defined as follows
\begin{equation}
\begin{aligned}
&&\mathit{V_{cS}} =\displaystyle{\frac{1}{2}}G_cb\int_{0}^{L}\int^{\frac{h_c}{2}}_{-\ \frac{h_c}{2}}(\gamma_{xy}^c)^2dydx=\displaystyle{\frac{1}{2}}G_cA_c\int_{0}^{L}\frac{[u_t-u_b+\frac{h_t+h_b+2h_c}{2}w']^2}{h_c^2}dx\nonumber\:\:\:\:\:\:\:\:\:\:\:\:\:\:\:\:\:\:\:\:\:\:\:\:\:\:\:\:\:\:\:\:\:\:\:\:\:\ (A11)
\end{aligned}
\end{equation}

\noindent where $E_c$, $E_t$, and $E_b$ are Young's modulus of the core and top and bottom face sheets, respectively, $G_c$ is the shear modulus of the core, $A_c$, $A_t$, and $A_b$ are the cross-sectional area of the core and top and bottom face sheets, respectively, $I_t$ and $I_b$ are the polar moment of inertia of the top and bottom face sheets, respectively, $m_c$, $m_t$, and $m_b$ are the mass per unit length of the core and top and and bottom face sheets, respectively, $b$ is the width of the sandwich beam and $L$ is the length of each subunit cell. 

kinetic energy at core and top and bottom layers, $T_{cN}$, $T_{tN}$, and $T_{bN}$, respectively, are obtained as follows

\begin{equation}
\begin{aligned}
&T_c = \displaystyle{\frac{1}{2}\rho_cb\int^L_0}{\int^{\frac{h_c}{2}}_{-\frac{h_c}{2}}{\left[{\left[\frac{\dot{u}_t+\dot{u}_b}{2}+\frac{\left(h_t-h_b\right)}{4}\dot{w}'+y\left(\frac{\dot{u}_t-\dot{u}_b}{h_c}+\frac{\left(h_t+h_b\right)}{2h_c}\dot{w}'\right)\right]}^2+{\dot{w}}^2\right]dydx}}=\nonumber\:\:\:\:\:\:\:\:\:\:\:\:\:\:\:\:\:\:\ (A12)\\
&\frac{1}{6}m_c\int^L_0{\left[\dot{u}^2_t+\dot{u}^2_b+\dot{u}_t\dot{u}_b+\left(h_t-\frac{h_b}{2}\right)\dot{u}_t\dot{w}'+\left(\frac{h_t}{2}-h_b\right)\dot{u}_b\dot{w}'+\frac{1}{4}\left(h^2_t-h_th_b+h^2_b\right)\dot{w}'^2\right]}dx+\frac{1}{6}m_c\int^L_0{\dot{w}^2}dx\nonumber
\end{aligned}
\end{equation}
\begin{equation}
\begin{aligned}
&T_t = \displaystyle{\frac{1}{2}\rho_tb\int^L_0}{\int^{\frac{h_t}{2}}_{-\frac{h_t}{2}}{\left[{\left(\dot{u}_t-y\dot{w}'\right)}^2+{\dot{w}}^2\right]}}dydx=\frac{1}{2}\int^L_0{(m_t{\dot{u}}^2_t}+{\rho }_tI_t{\dot{w}}^{'2})dx+\frac{1}{2}m_t\int^L_0{{\dot{w}}^2dx} \nonumber\:\:\:\:\:\:\:\:\:\:\:\:\:\:\:\:\:\:\:\:\:\:\:\:\:\ (A13)
\end{aligned}
\end{equation}

\begin{eqnarray}
\begin{aligned}
&T_b = \displaystyle{\frac{1}{2}\rho_bb\int^L_0}{\int^{\frac{h_b}{2}}_{-\frac{h_b}{2}}{\left[{\left(\dot{u}_b-y\dot{w}'\right)}^2+{\dot{w}}^2\right]}}dydx=\frac{1}{2}\int^L_0{(m_b{\dot{u}}^2_b}+{\rho }_bI_b\dot{w}^{'2})dx+\frac{1}{2}m_b\int^L_0{{\dot{w}}^2dx}\nonumber\:\:\:\:\:\:\:\:\:\:\:\:\:\:\:\:\:\:\:\:\:\:\:\ (A14)
\end{aligned}
\end{eqnarray}

\noindent where $m_c$, $m_t$, and $m_b$ are the mass per unit length at core and top and bottom layers, respectively.\\\\




\noindent\textbf{Appendix B}\\

Coefficients associated with equations of motions [Eqs. (7-9)] are as follows
\begin{eqnarray}
&&{A_1}^{\alpha}=-{m_t}^{\alpha}-\frac{{m_c}^{\alpha}}{3},\ \ \ \ \ {A_2}^{\alpha}=-\frac{{m_c}^{\alpha}}{6},\ \ \ \ \  {A_3}^{\alpha}=-\frac{{m_c}^{\alpha}}{6}\left({h_t}^{\alpha}-\frac{{h_b}^{\alpha}}{2}\right),\ \ \ \ \ {A_4}^{\alpha}=-{m_b}^{\alpha}-\frac{{m_c}^{\alpha}}{3}\ \nonumber\\ 
&& {A_5}^{\alpha}=-\frac{{m_c}^{\alpha}}{6}\left(\frac{{h_t}^{\alpha}}{2}-{h_b}^{\alpha}\right),\ \ \ \ \ {A_6}^{\alpha}={{\rho }_t}^{\alpha}{I_t}^{\alpha}+{{\rho }_b}^{\alpha}{I_b}^{\alpha}+\frac{{m_c}^{\alpha}}{12}\left({{h}^{\alpha}_t}^2-{h}^{\alpha}_t{h}^{\alpha}_b+{{h}^{\alpha}_b}^2\right),    
\ \ \ \ \  {B_1}^{\alpha}={E_t}^{\alpha}{A_t}^{\alpha}+\frac{{E_c}^{\alpha}{A_c}^{\alpha}}{3} \ \ \ \ \ \nonumber\\ 
&& {B_2}^{\alpha}=-\frac{{G_c}^{\alpha}{A_c}^{\alpha}}{{{h}^{\alpha}_c}^2},\ \ \ \ \ {B_3}^{\alpha}=\frac{{E_c}^{\alpha}{A_c}^{\alpha}}{6},\ \ \ \ \ {B_4}^{\alpha}=\frac{{G_c}^{\alpha}{A_c}^{\alpha}}{{{h}^{\alpha}_c}^2}\frac{\left({h_t}^{\alpha}+{h_b}^{\alpha}+2{h_c}^{\alpha}\right)}{2},\ \ \ \ \  {B_5}^{\alpha}=\frac{{E_c}^{\alpha}{A_c}^{\alpha}}{6}\left({h_t}^{\alpha}-\frac{{h_b}^{\alpha}}{2}\right)\ \ \ \ \ \nonumber\\
&& {B_6}^{\alpha}={E_b}^{\alpha}{A_b}^{\alpha}+\frac{{E_c}^{\alpha}{A_c}^{\alpha}}{3},\ \  {B_7}^{\alpha}=\frac{{E_c}^{\alpha}{A_c}^{\alpha}}{6}\left(\frac{{h_t}^{\alpha}}{2}-{h_b}^{\alpha}\right),\ \  {B_8}^{\alpha}={E_t}^{\alpha}{I_t}^{\alpha}+{E_b}^{\alpha}{I_b}^{\alpha}+\frac{{E_c}^{\alpha}{A_c}^{\alpha}}{12}\left({{h}^{\alpha}_t}^2-{h}^{\alpha}_t{h}^{\alpha}_b+{{h}^{\alpha}_b}^2\right) \nonumber\\
&& {B_9}^{\alpha}=\frac{{G_c}^{\alpha}{A_c}^{\alpha}}{{{h}^{\alpha}_c}^2}{\left(\frac{{h_t}^{\alpha}+{h_b}^{\alpha}+2{h_c}^{\alpha}}{2}\right)}^2 ,\ \ \ \ \ {M_T}^{\alpha}={m_t}^{\alpha}+{m_b}^{\alpha}+{m_c}^{\alpha} \nonumber\:\:\:\:\:\:\:\:\:\:\:\:\:\:\:\:\:\:\:\:\:\:\:\:\:\:\:\:\:\:\:\:\:\:\:\:\:\:\:\:\:\:\:\:\:\:\:\:\:\:\:\:\:\:\:\:\:\:\:\      (B1)
\end{eqnarray}

\noindent\textbf{Appendix C}\\

Superscript $\alpha$ in Eq. (B1) denotes the subunit cell $\alpha=I, II,..., \beta$, which is omitted from the coefficients in the following equations for the sake of simplicity.\\
\indent Substituting Eq. (14) into Eqs. (7-9), we obtain three sets of polynomial equations in $q$ as follows

\small \begin{eqnarray}
&&\left(-A_1{\omega }^2-B_1q^2+B_2\right)C+\left(-A_2{\omega }^2-B_3q^2-B_2\right)D+\left(-A_3{\omega }^2-B_5q^2-B_4\right)iqE=0 \nonumber\\ 
&&\left(-A_2{\omega }^2-B_3q^2-B_2\right)C+\left(-A_4{\omega }^2-B_6q^2+B_2\right)D+\left(-A_5{\omega }^2-B_7q^2+B_4\right)iqE=0 \nonumber\\ 
&& \left(A_3{\omega }^2+B_5q^2+B_4\right)iqC+\left(A_5{\omega }^2+B_7q^2-B_4\right)iqD+\left(A_6{\omega }^2q^2-B_8q^4-B_9q^2+M_T{\omega }^2\right)E=0\nonumber\:\:\:\:\:\:\:\:\:\:\:\:\:\:\:\:\:\  (C1)
\end{eqnarray}

Rewriting Eqs. $(C1)$ in a matrix form 

\begin{eqnarray}
&&\left[ \begin{array}{ccc}
-A_1{\omega }^2-B_1q^2+B_2 & -A_2{\omega }^2-B_3q^2-B_2& \left(-A_3{\omega }^2-B_5q^2-B_4\right)iq\\ 
-A_2{\omega }^2-B_3q^2-B_2& -A_4{\omega }^2-B_6q^2+B_2& \left(-A_3{\omega }^2-B_5q^2-B_4\right)iq\\ 
\left(A_3{\omega }^2+B_5q^2+B_4\right)iq& \left(A_5{\omega }^2+B_7q^2-B_4\right)iq& A_6{\omega }^2q^2-B_8q^4-B_9q^2+M_T{\omega }^2 \end{array}
\right]\left[ \begin{array}{c}
C \\ 
D \\ 
E \end{array}
\right]=0\nonumber\
\:\:\:\:\:\:\:\:\ (C2)
\end{eqnarray}

 Setting the determinant of the matrix equal to zero gives an 8th-order polynomial equation in $q$ as
 
 \begin{eqnarray}
&&{\gamma}_1q^8+\gamma _2q^6+{\gamma }_3q^4+{\gamma }_4q^2+{\gamma}_5=0 \nonumber\:\:\:\:\:\:\:\:\:\:\:\:\:\:\:\:\:\:\:\:\:\:\:\:\:\:\:\:\:\:\:\:\:\:\:\:\:\:\:\:\:\:\:\:\:\:\:\:\:\:\:\:\:\:\:\:\:\:\:\:\:\:\:\:\:\:\:\:\:\:\:\:\:\:\:\:\:\:\:\:\:\:\:\:\:\:\:\:\:\:\:\:\:\:\:\:\:\:\:\:\:\:\:\:\:\:\:\:\:\:\:\:\:\:\:\:\:\:\:\:\:\:\:\:\:\:\:\:\:\:\:\:\:\:\:\:\:\:\ (C3)
\end{eqnarray}
\nonumber\
\noindent where
 \begin{eqnarray}
&&{\gamma}_1=B_8B^2_3-2B_3B_5B_7+B_6B^2_5+B_1B^2_7-B_1B_6B_8\nonumber\:\:\:\:\:\:\:\:\:\:\:\:\:\:\:\:\:\:\:\:\:\:\:\:\:\:\:\:\:\:\:\:\:\:\:\:\:\:\:\:\:\:\:\:\:\:\:\:\:\:\:\:\:\:\:\:\:\:\:\:\:\:\:\:\:\:\:\:\:\:\:\:\:\:\:\:\:\:\:\:\:\:\:\:\:\:\:\:\:\:\:\:\:\:\:\:\:\:\:\:\:\:\:\:\:\:\:\:\ (C4)
\end{eqnarray}
\begin{eqnarray}
&&{\gamma }_2=B^2_3B_9- B_2B^2_7 -B_2B^2_5 + B_1B_2B_8 - 2B_1B_4B_7 + 2B_3B_4B_5+2B_2B_3B_8-2B_2B_5B_7-\nonumber\\ &&2B_3B_4B_7+2B_4B_5B_6-B_1B_6B_9+B_2B_6B_8+(A_1B^2_7+A_4B^2_5-A_6B^2_3+ 2A_2B_3B_8-2A_3B_3B_7-\nonumber\\ 
&&A_4B_1B_8+2A_5B_1B_7-2A_5B_3B_5+A_6B_1B_6- 2A_2B_5B_7+2A_3B_5B_6-A_1B_6B_8){\omega }^2\nonumber\:\:\:\:\:\:\:\:\:\:\:\:\:\:\:\:\:\:\:\:\:\:\:\:\:\:\:\:\:\:\:\:\:\:\:\:\:\:\:\:\:\:\:\:\:\ (C5)
\end{eqnarray}
\begin{eqnarray}
&&{\gamma }_3=(A^2_5B_1+A^2_3B_6+A^2_2B_8-2A_2A_6B_3- 2A_3A_5B_3+A_4A_6B_1-2A_2A_3B_7-2A_2A_5B_5+2A_3A_4B_5\nonumber\\
&&-A_1A_4B_8+ 2A_1A_5B_7+A_1A_6B_6){\omega }^4+(-A_6B_1B_2-2A_3B_2B_5+2A_3B_3B_4-2A_5B_1B_4+A_1B_2B_8+\nonumber\\&&2A_2B_4B_5-2A_6B_2B_3-2A_1B_4B_7+2A_2B_2B_8-2A_3B_2B_7-2A_5B_2B_5-2A_5B_3B_4-2A_2B_4B_7+\nonumber\\
&&2A_3B_4B_6+2A_4B_4B_5+2A_2B_3B_9-A_4B_1B_9+A_4B_2B_8-2A_5B_2B_7-A_6B_2B_6-A_1B_6B_9+B_1B_6M_T-\nonumber\\
&&B^2_3M_T){\omega }^2+B_1B^2_4+2B_3B^2_4+B^2_4B_6+B_1B_2B_9+2B_2B_3B_9+B_2B_6B_9\nonumber\:\:\:\:\:\:\:\:\:\:\:\:\:\:\:\:\:\:\:\:\:\:\:\:\:\:\:\:\:\:\:\:\:\:\:\:\:\:\:\:\:\:\:\:\:\:\:\:\:\:\:\:\:\:\:\:\:\:\:\:\:\:\:\:\:\:\:\:\ (C6)
\end{eqnarray}
\begin{eqnarray}
&&{\gamma }_4=\left(A_1A^2_5+A^2_3A_4-A^2_2A_6-2A_2A_3A_5+ A_1A_4A_6\right){\omega }^6+(-A^2_3B_2-A^2_5B_2+A^2_2B_9- A_1A_6B_2\nonumber\\
&&+2A_2A_3B_4-2A_1A_5B_4 -2A_2A_6B_2-2A_3A_5B_2- 2A_2A_5B_4+2A_3A_4B_4-A_4A_6B_2-A_1A_4B_9-\nonumber\\
&&2A_2B_3M_T+A_4B_1M_T+ A_1B_6M_T){\omega }^4+(A_4B^2_4+A_1B_2B_9+2A_2B_2B_9+A_4B_2B_9+A_1B^2_4+2A_2B^2_4-\nonumber\\
&&2B_2B_3M_T-B_2B_6M_T- B_1B_2M_T){\omega }^2 \nonumber\:\:\:\:\:\:\:\:\:\:\:\:\:\:\:\:\:\:\:\:\:\:\:\:\:\:\:\:\:\:\:\:\:\:\:\:\:\:\:\:\:\:\:\:\:\:\:\:\:\:\:\:\:\:\:\:\:\:\:\:\:\:\:\:\:\:\:\:\:\:\:\:\:\:\:\:\:\:\:\:\:\:\:\:\:\:\:\:\:\:\:\:\:\:\:\:\:\:\:\:\:\:\:\:\:\:\:\:\:\:\:\:\:\:\:\:\:\:\:\:\:\:\:\:\:\:\:\:\:\:\:\:\ (C7)
\end{eqnarray}
\begin{eqnarray}
&&{\gamma }_5=M_T\left[\left(A_1A_4-A^2_2\right){\omega }^6-\left(A_1B_2+2A_2B_2+A_4B_2\right){\omega }^4\right]\nonumber\:\:\:\:\:\:\:\:\:\:\:\:\:\:\:\:\:\:\:\:\:\:\:\:\:\:\:\:\:\:\:\:\:\:\:\:\:\:\:\:\:\:\:\:\:\:\:\:\:\:\:\:\:\:\:\:\:\:\:\:\:\:\:\:\:\:\:\:\:\:\:\:\:\:\:\:\:\:\:\:\:\:\:\:\:\:\:\:\:\:\:\:\ (C8)
\end{eqnarray}\\

\noindent\textbf{Appendix D}\\

Relationships between coefficients $C^{\left(\alpha\right)}_l,\ D^{\left(\alpha\right)}_l,\ and\ E^{\left(\alpha\right)}_l$ in which $l=1,2,\dots,8\ $, can be obtained by setting Eq. (15) into Eqs. (7-9) as 

\begin{eqnarray}
&&{\mathit{\Lambda }}^{(\alpha)}_l=\frac{{\mu }^{\left(\alpha\right)}_l}{{\beta }^{(\alpha)}_l} \nonumber\\
&&{\mathit{Z}}^{(\alpha)}_l=\frac{{\zeta }^{\left(\alpha\right)}_l}{{\eta }^{(\alpha)}_l} \nonumber\:\:\:\:\:\:\:\:\:\:\:\:\:\:\:\:\:\:\:\:\:\:\:\:\:\:\:\:\:\:\:\:\:\:\:\:\:\:\:\:\:\:\:\:\:\:\:\:\:\:\:\:\:\:\:\:\:\:\:\:\:\:\:\:\:\:\:\:\:\:\:\:\:\:\:\:\:\:\:\:\:\:\:\:\:\:\:\:\:\:\:\:\:\:\:\:\:\:\:\:\:\:\:\:\:\:\:\:\:\:\:\:\:\:\:\:\:\:\:\:\:\:\:\:\:\:\:\:\:\:\:\:\:\:\:\:\:\:\:\:\:\:\:\:\:\:\:\:\:\:\:\:\:\:\:\:\:\:\:\:\:\:\:\:\:\:\:\:\:\:\:\:\:\:\:\:\:\:\:\:\:\:\             (D1)
\end{eqnarray} 

Here ${\mathit{\Lambda }}^{(\alpha)}_l$ denotes the relation between top face sheet's longitudinal displacement $\overline{u}_t$ and transverse displacement $\overline{w}$ associated with subunit cell $\alpha$. ${\mathit{Z}}^{(\alpha)}_l$ also denotes the relation between bottom face sheet's longitudinal displacement $\overline{u}_b$ and transverse displacement $\overline{w}$.

\begin{eqnarray}
&&{\mu}_l=(B_3B_8\ -\ B_5B_7)q^6\ +[B_4B_5\ +\ B_2B_8\ -\ B_4B_7\ +\ B_3B_9+(-\ A_6B_3+A_2B_8-A_3B_7-A_5B_5){\omega }^2]q^4+\nonumber\\
&&[B^2_4+B_2B_9-(\ A_2A_6+\ A_3A_5){\omega }^4+(\ A_3B_4-A_6B_2-\ A_5B_4+A_2B_9-B_3M_T){\omega }^2]q^2-(A_2{\omega }^2-B_2){M_T\omega }^2\nonumber\\
&&{\beta }_l=[(A_2A_3-A_1A_5){\omega }^4+(B_1B_4+B_2B_5+B_3B_4+B_2B_7+(A_3B_3-A_5B_1+A_2B_5-A_1B_7+A_1B_4+A_3B_2+\nonumber\\
&&A_2B_4+A_5B_2){\omega }^2)q^2+\left(-B_1B_7+B_3B_5\right)q^4]qi
\nonumber\\
&&{\zeta }_l=[B_1B_9-2B_4B_5-B_2B_8+\left(-A_6B_1-2A_3B_5+A_1B_8\right){\omega }^2]q^4+(-B^2_5+B_1B_8)q^6+[(- A^2_3-A_1A_6){\omega }^4+\nonumber\\
&&(-2A_3B_4+A_6B_2+A_1B_9-B_1M){\omega }^2-\ B_2B_9-\ B^2_4]q^2+(\ -A_1{\omega }^2+\ B_2){\omega }^2 
\nonumber\\
&&{\eta }_l=-[[(A_2A_3-\ A_1A_5){\omega }^2+(A_1B_4+\ A_3B_2+\ A_2B_4+\ A_5B_2)]{\omega }^2+(B_1B_4+\ B_2B_5+\ B_3B_4+\ B_2B_7+\nonumber\\
&&A_3B_3{\omega }^2-A_5B_1{\omega }^2+A_2B_5{\omega }^2- A_1B_7{\omega }^2)q^2+(-B_1B_7+B_3B_5)q^4]qi\nonumber\:\:\:\:\:\:\:\:\:\:\:\:\:\:\:\:\:\:\:\:\:\:\:\:\:\:\:\:\:\:\:\:\:\:\:\:\:\:\:\:\:\:\:\:\:\:\:\:\:\:\:\:\:\:\:\:\:\:\:\:\:\:\:\:\:\:\:\:\:\:\:\:\:\:\    (D2)
\end{eqnarray} 

Here superscript $\alpha$ and subscript $l$ are omitted from the coefficients in Eq. $(D2)$ for the sake of simplicity. \\
\indent The relation between top axial force $\overline{P}_t$, bottom axial force $\overline{P}_b$, shear force $\overline{S}$, moment force $\overline{M}$, and transverse displacement $\overline{w}$ can be obtained by substitution of the eight obtained roots for $q$ together with the relation between displacements $\overline{u}_t$, $\overline{u}_b$, and $\overline{w}$ [Eq. $(D1)$] into Eqs. (10-13) as follows:

\begin{equation} 
{\chi }^{(\alpha)}_l=\frac{{\kappa }^{\left(\alpha\right)}_l}{{\tau }^{\left(\alpha\right)}_l}\nonumber\:\:\:\:\:\:\:\:\:\:\:\:\:\:\:\:\:\:\:\:\:\:\:\:\:\:\:\:\:\:\:\:\:\:\:\:\:\:\:\:\:\:\:\:\:\:\:\:\:\:\:\:\:\:\:\:\:\:\:\:\:\:\:\:\:\:\:\:\:\:\:\:\:\:\:\:\:\:\:\:\:\:\:\:\:\:\:\:\:\:\:\:\:\:\:\:\:\:\:\:\:\:\:\:\:\:\:\:\:\:\:\:\:\:\:\:\:\:\:\:\:\:\:\:\:\:\:\:\:\:\:\:\:\:\:\:\:\:\:\:\:\:\:\:\:\:\:\:\:\:\:\:\:\:\:\:\:\:\:\:\:\:\:\:\:\:\:\:\:\:\:\:\:\:\:\:\:\    (D3)
\end{equation} 
where  
\begin{eqnarray} 
&&{\kappa }_l=[B_2B^2_5+A_2B^2_5{\omega }^2-B_1B_2B_8+B_1B_4B_7- B_3B_4B_5-B_2B_3B_8+B_2B_5B_7-A_2B_1B_8{\omega}^2+\nonumber\\
&& A_3B_1B_7{\omega }^2-A_3B_3B_5{\omega }^2+A_1B_3B_8{\omega }^2-A_1B_5B_7{\omega }^2]q^4+[(-A^2_3B_3+A_2A_6B_1+A_3A_5B_1-\nonumber\\
&&A_1A_6B_3+A_2A_3B_5-A_1A_5B_5){\omega }^4+(- A_3B_1B_4+ A_6B_1B_2+A_1B_4B_5+ A_3B_2B_5- 2A_3B_3B_4+\nonumber\\
&&A_5B_1B_4+A_2B_4B_5+A_6B_2B_3 - A_2B_1B_9+A_5B_2B_5+A_1B_3B_9){\omega }^2-B_3B^2_4-B_1B^2_4-B_1B_2B_9-\nonumber\\
&&B_2B_3B_9]q^2 +A_2B_1M_T{\omega }^4-A_1B_3M_T{\omega }^4+B_1B_2M_T{\omega }^2+B_2B_3M_T{\omega }^2
\nonumber\\
&&{\tau}_l=(B_1B_7-B_3B_5)q^4+[-B_1B_4-B_2B_5-B_3B_4-B_2B_7(-A_3B_3+A_5B_1-A_2B_5+A_1B_7){\omega }^2]q^2\nonumber\\
&&+(A_1A_5-A_2A_3){\omega }^4+(-A_1B_4-A_3B_2-A_2B_4-A_5B_2){\omega }^2\nonumber\:\:\:\:\:\:\:\:\:\:\:\:\:\:\:\:\:\:\:\:\:\:\:\:\:\:\:\:\:\:\:\:\:\:\:\:\:\:\:\:\:\:\:\:\:\:\:\:\:\:\:\:\:\:\:\:\:\:\:\:\:\:\:\:\:\:\:\:\:\:\:\:\:\:\:\:\:\:\:\:\:\:\:\:\:\:\:\   (D4)
\end{eqnarray}

\indent Here ${\chi }^{(\alpha)}_l$ denotes the relation between top axial force $\overline{P}_t$ and transverse displacement $\overline{w}$. Superscript $\alpha$ and subscript $l$ are omitted from the coefficients in Eq. $(D4)$ for the sake of simplicity.

\begin{equation} 
{\mathit{\Phi }}^{(\alpha)}_l=\frac{{\varrho }^{\left(\alpha\right)}_l}{{\varsigma }^{\left(\alpha\right)}_l}\nonumber\:\:\:\:\:\:\:\:\:\:\:\:\:\:\:\:\:\:\:\:\:\:\:\:\:\:\:\:\:\:\:\:\:\:\:\:\:\:\:\:\:\:\:\:\:\:\:\:\:\:\:\:\:\:\:\:\:\:\:\:\:\:\:\:\:\:\:\:\:\:\:\:\:\:\:\:\:\:\:\:\:\:\:\:\:\:\:\:\:\:\:\:\:\:\:\:\:\:\:\:\:\:\:\:\:\:\:\:\:\:\:\:\:\:\:\:\:\:\:\:\:\:\:\:\:\:\:\:\:\:\:\:\:\:\:\:\:\:\:\:\:\:\:\:\:\:\:\:\:\:\:\:\:\:\:\:\:\:\:\:\:\:\:\:\:\:\:\:\:\:\:\:\:\:\:\    (D5)
\end{equation}

\noindent where 
\begin{eqnarray} 
&&{\varrho }_l=(-B_1B^2_7 -B^2_3B_8-B^2_5B_6+B_1B_6B_8+2B_3B_5B_7)q^6+[B_2B^2_7 -B^2_3B_9+B_1B_4B_7-B_3B_4B_5-\nonumber\\
&&B_2B_3B_8+B_2B_5B_7 + 2B_3B_4B_7-2B_4B_5B_6+B_1B_6B_9- B_2B_6B_8+(A_6B^2_3-A_1B^2_7- A_2B_3B_8+\nonumber\\ &&2A_3B_3B_7-A_5B_1B_7+A_5B_3B_5-A_6B_1B_6+A_2B_5B_7- 2A_3B_5B_6+A_1B_6B_8){\omega }^2]q^4+[-B^2_4B_6 -B_3B^2_4-\nonumber\\ &&B_2B_3B_9-B_2B_6B_9+(-A_3B_3B_4+A_6B_2B_3+A_1B_4B_7+A_3B_2B_7+A_5B_3B_4+A_2B_4B_7-2A_3B_4B_6-\nonumber\\ 
&&A_2B_3B_9+ A_5B_2B_7+A_6B_2B_6+A_1B_6B_9){\omega }^2+(-A^2_3B_6+A_2A_6B_3+A_3A_5B_3+A_2A_3B_7-A_1A_5B_7-\nonumber\\&& A_1A_6B_6){\omega }^4]q^2+A_2B_3M_T{\omega }^4- A_1B_6M_T{\omega }^4+B_2B_3M_T{\omega }^2+B_2B_6M_T{\omega }^2\nonumber\\ &&
{\varsigma }_l=(B_1B_7-B_3B_5)q^4+[-B_1B_4-B_2B_5-B_3B_4-B_2B_7+(-A_3B_3+A_5B_1-A_2B_5+A_1B_7){\omega }^2]q^2+\nonumber\\
&&(A_1A_5-A_2A_3){\omega}^4+(-A_1B_4-A_3B_2-A_2B_4-A_5B_2){\omega }^2\nonumber\:\:\:\:\:\:\:\:\:\:\:\:\:\:\:\:\:\:\:\:\:\:\:\:\:\:\:\:\:\:\:\:\:\:\:\:\:\:\:\:\:\:\:\:\:\:\:\:\:\:\:\:\:\:\:\:\:\:\:\:\:\:\:\:\:\:\:\:\:\:\:\:\:\:\:\:\:\:\:\:\:\:\:\:\:\  (D6)
\end{eqnarray} 

Here ${\mathit{\Phi }}^{(\alpha)}_l$ denotes the relation between bottom axial force $\overline{P}_b$ and transverse displacement $\overline{w}$. superscript $\alpha$ and subscript $l$ are omitted from the coefficients in Eq. $(D6)$ for the sake of simplicity.

\begin{equation} 
{\mathit{H}}=B_8q^2-B_5qi-B_7qi
\nonumber\:\:\:\:\:\:\:\:\:\:\:\:\:\:\:\:\:\:\:\:\:\:\:\:\:\:\:\:\:\:\:\:\:\:\:\:\:\:\:\:\:\:\:\:\:\:\:\:\:\:\:\:\:\:\:\:\:\:\:\:\:\:\:\:\:\:\:\:\:\:\:\:\:\:\:\:\:\:\:\:\:\:\:\:\:\:\:\:\:\:\:\:\:\:\:\:\:\:\:\:\:\:\:\:\:\:\:\:\:\:\:\:\:\:\:\:\:\:\:\:\:\:\:\:\:\:\:\:\:\:\:\:\:\:\:\:\:\:\:\:\:\:\:\:\:\:\:\   (D7)
\end{equation} 

\noindent where,  ${\mathit{H}}$ denotes the relation between moment force $\overline{M}$ and transverse displacement $\overline{w}$. 

\begin{equation} 
{\mathit{\Gamma }}=\frac{M_T{\omega }^2}{qi}\nonumber\:\:\:\:\:\:\:\:\:\:\:\:\:\:\:\:\:\:\:\:\:\:\:\:\:\:\:\:\:\:\:\:\:\:\:\:\:\:\:\:\:\:\:\:\:\:\:\:\:\:\:\:\:\:\:\:\:\:\:\:\:\:\:\:\:\:\:\:\:\:\:\:\:\:\:\:\:\:\:\:\:\:\:\:\:\:\:\:\:\:\:\:\:\:\:\:\:\:\:\:\:\:\:\:\:\:\:\:\:\:\:\:\:\:\:\:\:\:\:\:\:\:\:\:\:\:\:\:\:\:\:\:\:\:\:\:\:\:\:\:\:\:\:\:\:\:\:\:\:\:\:\:\:\:\:\:\:\:\:\:\:\:\:\:\:\:\:\:\:\:\:\:\:\:\    (D8)
\end{equation} 

\noindent where, ${\mathit{\Gamma }}$ denotes the relation between shear force $\overline{S}$ and transverse displacement $\overline{w}$. 
\\\\\\\

\noindent\textbf{Appendix E}\\

\begin{eqnarray} 
\left[\mathit{\Upsilon }\right]{\mathit{\Psi }}^{(I)}_{n+1}=\left[\mathit{\Pi }\right]{\mathit{\Psi }}^{(II)}_n\nonumber\:\:\:\:\:\:\:\:\:\:\:\:\:\:\:\:\:\:\:\:\:\:\:\:\:\:\:\:\:\:\:\:\:\:\:\:\:\:\:\:\:\:\:\:\:\:\:\:\:\:\:\:\:\:\:\:\:\:\:\:\:\:\:\:\:\:\:\:\:\:\:\:\:\:\:\:\:\:\:\:\:\:\:\:\:\:\:\:\:\:\:\:\:\:\:\:\:\:\:\:\:\:\:\:\:\:\:\:\:\:\:\:\:\:\:\:\:\:\:\:\:\:\:\:\:\:\:\:\:\:\:\:\:\:\:\:\:\:\:\:\:\:\:\:\:\:\:\:\:\:\:\:\:\:\:\:\:\:\:\:\:\:\:\:\:\:\:\:\:\:\:\:\      (E1)
\end{eqnarray} 

where
\begin{eqnarray} 
{\mathit{\Psi }}^{(I)}_{n+1}=\left\{ \begin{array}{c}
E^{\left(I\right)}_1e^{iq^{\left(I\right)}_1(nd)} \\ 
E^{\left(I\right)}_2e^{iq^{\left(I\right)}_2(nd)} \\ 
E^{\left(I\right)}_3e^{iq^{\left(I\right)}_3(nd)} \\ 
E^{\left(I\right)}_4e^{iq^{\left(I\right)}_4(nd)} \\ 
E^{\left(I\right)}_5e^{iq^{\left(I\right)}_5(nd)} \\ 
E^{\left(I\right)}_6e^{iq^{\left(I\right)}_6(nd)} \\ 
E^{\left(I\right)}_7e^{iq^{\left(I\right)}_7(nd)} \\ 
E^{\left(I\right)}_8e^{iq^{\left(I\right)}_8(nd)} \end{array}
\right\},\  {\mathit{\Psi }}^{(II)}_n=\left\{ \begin{array}{c}
E^{\left(II\right)}_1e^{iq^{\left(II\right)}_1(n-1)d} \\ 
E^{\left(II\right)}_2e^{iq^{\left(II\right)}_2(n-1)d} \\ 
E^{\left(II\right)}_3e^{iq^{\left(II\right)}_3(n-1)d} \\ 
E^{\left(II\right)}_4e^{iq^{\left(II\right)}_4(n-1)d} \\ 
E^{\left(II\right)}_5e^{iq^{\left(II\right)}_5(n-1)d} \\ 
E^{\left(II\right)}_6e^{iq^{\left(II\right)}_6(n-1)d} \\ 
E^{\left(II\right)}_7e^{iq^{\left(II\right)}_7(n-1)d} \\ 
E^{\left(II\right)}_8e^{iq^{\left(II\right)}_8(n-1)d} \end{array}
\right\}\nonumber\:\:\:\:\:\:\:\:\:\:\:\:\:\:\:\:\:\:\:\:\:\:\:\:\:\:\:\:\:\:\:\:\:\:\:\:\:\:\:\:\:\:\:\:\:\:\:\:\:\:\:\:\:\:\:\:\:\:\:\:\:\:\:\:\:\:\:\:\:\:\:\:\:\:\:\:\:\:\:\:\:\:\:\:\:\:\:\:\:\:\   (E2)\end{eqnarray}
\begin{eqnarray}  
\left[\mathit{\Upsilon }\right]=\left[ \begin{array}{cccccccc}
1 & 1 & 1 & 1 & 1 & 1 & 1 & 1 \\ 
iq^{(I)}_1 & iq^{(I)}_2 & iq^{(I)}_3 & iq^{(I)}_4 & iq^{(I)}_5 & iq^{(I)}_6 & iq^{(I)}_7 & iq^{(I)}_8 \\ 
{\mathit{\Lambda }}^{(I)}_1 & {\mathit{\Lambda }}^{(I)}_2 & {\mathit{\Lambda }}^{(I)}_3 & {\mathit{\Lambda }}^{(I)}_4 & {\mathit{\Lambda }}^{(I)}_5 & {\mathit{\Lambda }}^{(I)}_6 & {\mathit{\Lambda }}^{(I)}_7 & {\mathit{\Lambda }}^{(I)}_8 \\ 
{\mathit{Z}}^{(I)}_1 & {\mathit{Z}}^{(I)}_2 & {\mathit{Z}}^{(I)}_3 & {\mathit{Z}}^{(I)}_4 & {\mathit{Z}}^{(I)}_5 & {\mathit{Z}}^{(I)}_6 & {\mathit{Z}}^{(I)}_7 & {\mathit{Z}}^{(I)}_8 \\ 
{\chi }^{(I)}_1 & {\chi }^{(I)}_2 & {\chi }^{(I)}_3 & {\chi }^{(I)}_4 & {\chi }^{(I)}_5 & {\chi }^{(I)}_6 & {\chi }^{(I)}_7 & {\chi }^{(I)}_8 \\ 
{\mathit{\Phi }}^{(I)}_1 & {\mathit{\Phi }}^{(I)}_2 & {\mathit{\Phi }}^{(I)}_3 & {\mathit{\Phi }}^{(I)}_4 & {\mathit{\Phi }}^{(I)}_5 & {\mathit{\Phi }}^{(I)}_6 & {\mathit{\Phi }}^{(I)}_7 & {\mathit{\Phi }}^{(I)}_8 \\ 
{\mathit{H}}^{(I)}_1 & {\mathit{H}}^{(I)}_2 & {\mathit{H}}^{(I)}_3 & {\mathit{H}}^{(I)}_4 & {\mathit{H}}^{(I)}_5 & {\mathit{H}}^{(I)}_6 & {\mathit{H}}^{(I)}_7 & {\mathit{H}}^{(I)}_8 \\ 
{\mathit{\Gamma }}^{(I)}_1 & {\mathit{\Gamma }}^{(I)}_2 & {\mathit{\Gamma }}^{(I)}_3 & {\mathit{\Gamma }}^{(I)}_4 & {\mathit{\Gamma }}^{(I)}_5 & {\mathit{\Gamma }}^{(I)}_6 & {\mathit{\Gamma }}^{(I)}_7 & {\mathit{\Gamma }}^{(I)}_8 \end{array}
\right]\nonumber\:\:\:\:\:\:\:\:\:\:\:\:\:\:\:\:\:\:\:\:\:\:\:\:\:\:\:\:\:\:\:\:\:\:\:\:\:\:\:\:\:\:\:\:\:\:\:\:\:\:\:\:\:\:\:\:\:\:\:\:\:\:\:\:\:\:\:\:\:\:\:\:\:\:\:\:\:\:\:\:\:\:\:\:\  (E3)
\end{eqnarray} 
\begin{eqnarray} 
&&\left[\mathit{\Pi }\right]=\left[ \begin{array}{cccccccc}
1 & 1 & 1 & 1 & 1 & 1 & 1 & 1 \\ 
iq^{(II)}_1 & iq^{(II)}_2 & iq^{(II)}_3 & iq^{(II)}_4 & iq^{(II)}_5 & iq^{(II)}_6 & iq^{(II)}_7 & iq^{(II)}_8 \\ 
{\mathit{\Lambda }}^{(II)}_1 & {\mathit{\Lambda }}^{(II)}_2 & {\mathit{\Lambda }}^{(II)}_3 & {\mathit{\Lambda }}^{(II)}_4 & {\mathit{\Lambda }}^{(II)}_5 & {\mathit{\Lambda }}^{(II)}_6 & {\mathit{\Lambda }}^{(II)}_7 & {\mathit{\Lambda }}^{(II)}_8 \\ 
{\mathit{Z}}^{(II)}_1 & {\mathit{Z}}^{(II)}_2 & {\mathit{Z}}^{(II)}_3 & {\mathit{Z}}^{(I)}_4 & {\mathit{Z}}^{(II)}_5 & {\mathit{Z}}^{(II)}_6 & {\mathit{Z}}^{(II)}_7 & {\mathit{Z}}^{(II)}_8 \\ 
{\chi }^{(II)}_1 & {\chi }^{(II)}_2 & {\chi }^{(II)}_3 & {\chi }^{(I)}_4 & {\chi }^{(II)}_5 & {\chi }^{(II)}_6 & {\chi }^{(II)}_7 & {\chi }^{(II)}_8 \\ 
{\mathit{\Phi }}^{(II)}_1 & {\mathit{\Phi }}^{(II)}_2 & {\mathit{\Phi }}^{(II)}_3 & {\mathit{\Phi }}^{(I)}_4 & {\mathit{\Phi }}^{(II)}_5 & {\mathit{\Phi }}^{(II)}_6 & {\mathit{\Phi }}^{(II)}_7 & {\mathit{\Phi }}^{(II)}_8 \\ 
{\mathit{H}}^{(II)}_1 & {\mathit{H}}^{(II)}_2 & {\mathit{H}}^{(II)}_3 & {\mathit{H}}^{(I)}_4 & {\mathit{H}}^{(II)}_5 & {\mathit{H}}^{(II)}_6 & {\mathit{H}}^{(II)}_7 & {\mathit{H}}^{(II)}_8 \\ 
{\mathit{\Gamma }}^{(II)}_1 & {\mathit{\Gamma }}^{(II)}_2 & {\mathit{\Gamma }}^{(II)}_3 & {\mathit{\Gamma }}^{(I)}_4 & {\mathit{\Gamma }}^{(II)}_5 & {\mathit{\Gamma }}^{(II)}_6 & {\mathit{\Gamma }}^{(II)}_7 & {\mathit{\Gamma }}^{(II)}_8 \end{array}
\right]\boldsymbol{.} \nonumber\\
&&\left[ \begin{array}{cccccccc}
e^{iq^{(II)}_1d} & 0 & 0 & 0 & 0 & 0 & 0 & 0 \\ 
0 & e^{iq^{(II)}_2d} & 0 & 0 & 0 & 0 & 0 & 0 \\ 
0 & 0 & e^{iq^{(II)}_3d} & 0 & 0 & 0 & 0 & 0 \\ 
0 & 0 & 0 & e^{iq^{(II)}_4d} & 0 & 0 & 0 & 0 \\ 
0 & 0 & 0 & 0 & e^{iq^{(II)}_5d} & 0 & 0 & 0 \\ 
0 & 0 & 0 & 0 & 0 & e^{iq^{(II)}_6d} & 0 & 0 \\ 
0 & 0 & 0 & 0 & 0 & 0 & e^{iq^{(II)}_7d} & 0 \\ 
0 & 0 & 0 & 0 & 0 & 0 & 0 & e^{iq^{(II)}_8d} \end{array}
\right]\nonumber\:\:\:\:\:\:\:\:\:\:\:\:\:\:\:\:\:\:\:\:\:\:\:\:\:\:\:\:\:\:\:\:\:\:\:\:\:\:\:\  (E4) 
\end{eqnarray} 

\noindent where $d$ is the lattice constant. By applying continuity conditions at the interface between subunit cells $I$ and $II$ from unit cell $n+1$ (Fig. \ref{F3}) we obtain the following matrices. 

\begin{equation}  
\left[\mathit{\Theta }\right]{\mathit{\Psi }}^{(I)}_{n+1}=\left[\mathit{\Delta }\right]{\mathit{\Psi }}^{(II)}_{n+1}\nonumber\:\:\:\:\:\:\:\:\:\:\:\:\:\:\:\:\:\:\:\:\:\:\:\:\:\:\:\:\:\:\:\:\:\:\:\:\:\:\:\:\:\:\:\:\:\:\:\:\:\:\:\:\:\:\:\:\:\:\:\:\:\:\:\:\:\:\:\:\:\:\:\:\:\:\:\:\:\:\:\:\:\:\:\:\:\:\:\:\:\:\:\:\:\:\:\:\:\:\:\:\:\:\:\:\:\:\:\:\:\:\:\:\:\:\:\:\:\:\:\:\:\:\:\:\:\:\:\:\:\:\:\:\:\:\:\:\:\:\:\:\:\:\:\:\:\:\:\:\:\:\:\:\:\  (E5)  
\end{equation} 

\noindent where 

\begin{eqnarray} 
{\mathit{\Psi }}^{(I)}_{n+1}=\left\{ \begin{array}{c}
E^{\left(I\right)}_1e^{iq^{\left(I\right)}_1(nd)} \\ 
E^{\left(I\right)}_2e^{iq^{\left(I\right)}_2(nd)} \\ 
E^{\left(I\right)}_3e^{iq^{\left(I\right)}_3(nd)} \\ 
E^{\left(I\right)}_4e^{iq^{\left(I\right)}_4(nd)} \\ 
E^{\left(I\right)}_5e^{iq^{\left(I\right)}_5(nd)} \\ 
E^{\left(I\right)}_6e^{iq^{\left(I\right)}_6(nd)} \\ 
E^{\left(I\right)}_7e^{iq^{\left(I\right)}_7(nd)} \\ 
E^{\left(I\right)}_8e^{iq^{\left(I\right)}_8(nd)} \end{array}
\right\},\  {\mathit{\Psi }}^{(II)}_{n+1}=\left\{ \begin{array}{c}
E^{\left(II\right)}_1e^{iq^{\left(II\right)}_1(nd)} \\ 
E^{\left(II\right)}_2e^{iq^{\left(II\right)}_2(nd)} \\ 
E^{\left(II\right)}_3e^{iq^{\left(II\right)}_3(nd)} \\ 
E^{\left(II\right)}_4e^{iq^{\left(II\right)}_4(nd)} \\ 
E^{\left(II\right)}_5e^{iq^{\left(II\right)}_5(nd)} \\ 
E^{\left(II\right)}_6e^{iq^{\left(II\right)}_6(nd)} \\ 
E^{\left(II\right)}_7e^{iq^{\left(II\right)}_7(nd)} \\ 
E^{\left(II\right)}_8e^{iq^{\left(II\right)}_8(nd)} \end{array}
\right\}\nonumber\:\:\:\:\:\:\:\:\:\:\:\:\:\:\:\:\:\:\:\:\:\:\:\:\:\:\:\:\:\:\:\:\:\:\:\:\:\:\:\:\:\:\:\:\:\:\:\:\:\:\:\:\:\:\:\:\:\:\:\:\:\:\:\:\:\:\:\:\:\:\:\:\:\:\:\:\:\:\:\:\:\:\:\:\:\:\:\:\:\:\:\:\:\  (E6) 
\end{eqnarray} 

\begin{equation}
\begin{aligned}
\left[\mathit{\Theta }\right]=\left[ \begin{array}{cccccccc}
1 & 1 & 1 & 1 & 1 & 1 & 1 & 1 \\ 
iq^{(I)}_1 & iq^{(I)}_2 & iq^{(I)}_3 & iq^{(I)}_4 & iq^{(I)}_5 & iq^{(I)}_6 & iq^{(I)}_7 & iq^{(I)}_8 \\ 
{\mathit{\Lambda }}^{(I)}_1 & {\mathit{\Lambda }}^{(I)}_2 & {\mathit{\Lambda }}^{(I)}_3 & {\mathit{\Lambda }}^{(I)}_4 & {\mathit{\Lambda }}^{(I)}_5 & {\mathit{\Lambda }}^{(I)}_6 & {\mathit{\Lambda }}^{(I)}_7 & {\mathit{\Lambda }}^{(I)}_8 \\
{\mathit{Z}}^{(I)}_1 & {\mathit{Z}}^{(I)}_2 & {\mathit{Z}}^{(I)}_3 & {\mathit{Z}}^{(I)}_4 & {\mathit{Z}}^{(I)}_5 & {\mathit{Z}}^{(I)}_6 & {\mathit{Z}}^{(I)}_7 & {\mathit{Z}}^{(I)}_8 \\ 
{\chi }^{(I)}_1 & {\chi }^{(I)}_2 & {\chi }^{(I)}_3 & {\chi }^{(I)}_4 & {\chi }^{(I)}_5 & {\chi }^{(I)}_6 & {\chi }^{(I)}_7 & {\chi }^{(I)}_8 \\ 
{\mathit{\Phi }}^{(I)}_1 & {\mathit{\Phi }}^{(I)}_2 & {\mathit{\Phi }}^{(I)}_3 & {\mathit{\Phi }}^{(I)}_4 & {\mathit{\Phi }}^{(I)}_5 & {\mathit{\Phi }}^{(I)}_6 & {\mathit{\Phi }}^{(I)}_7 & {\mathit{\Phi }}^{(I)}_8 \\ 
{\mathit{H}}^{(I)}_1 & {\mathit{H}}^{(I)}_2 & {\mathit{H}}^{(I)}_3 & {\mathit{H}}^{(I)}_4 & {\mathit{H}}^{(I)}_5 & {\mathit{H}}^{(I)}_6 & {\mathit{H}}^{(I)}_7 & {\mathit{H}}^{(I)}_8 \\ 
{\mathit{\Gamma }}^{(I)}_1 & {\mathit{\Gamma }}^{(I)}_2 & {\mathit{\Gamma }}^{(I)}_3 & {\mathit{\Gamma }}^{(I)}_4 & {\mathit{\Gamma }}^{(I)}_5 & {\mathit{\Gamma }}^{(I)}_6 & {\mathit{\Gamma }}^{(I)}_7 & {\mathit{\Gamma }}^{(I)}_8 \end{array}
\right]\boldsymbol{.} \nonumber\\
\left[ \begin{array}{cccccccc}
e^{iq^{(I)}_1L} & 0 & 0 & 0 & 0 & 0 & 0 & 0 \\ 
0 & e^{iq^{(I)}_2L} & 0 & 0 & 0 & 0 & 0 & 0 \\ 
0 & 0 & e^{iq^{(I)}_3L} & 0 & 0 & 0 & 0 & 0 \\ 
0 & 0 & 0 & e^{iq^{(I)}_4L} & 0 & 0 & 0 & 0 \\ 
0 & 0 & 0 & 0 & e^{iq^{(I)}_5L} & 0 & 0 & 0 \\ 
0 & 0 & 0 & 0 & 0 & e^{iq^{(I)}_6L} & 0 & 0 \\ 
0 & 0 & 0 & 0 & 0 & 0 & e^{iq^{(I)}_7L} & 0 \\ 
0 & 0 & 0 & 0 & 0 & 0 & 0 & e^{iq^{(I)}_8L} \end{array}
\right]\nonumber
\end{aligned}
\:\:\:\:\:\:\:\:\:\:\:\:\:\:\:\:\:\:\:\:\:\:\:\:\:\:\:\:\:\:\:\:\:\:\:\:\:\:\:\:\:\:\:\:\:\:\:\:\:\:\:\:\  (E7)
\end{equation}

\begin{eqnarray}
\begin{aligned}
\left[\mathit{\Delta }\right]=\left[ \begin{array}{cccccccc}
1 & 1 & 1 & 1 & 1 & 1 & 1 & 1 \\ 
iq^{(II)}_1 & iq^{(II)}_2 & iq^{(II)}_3 & iq^{(II)}_4 & iq^{(II)}_5 & iq^{(II)}_6 & iq^{(II)}_7 & iq^{(II)}_8 \\ 
{\mathit{\Lambda }}^{(II)}_1 & {\mathit{\Lambda }}^{(II)}_2 & {\mathit{\Lambda }}^{(II)}_3 & {\mathit{\Lambda }}^{(II)}_4 & {\mathit{\Lambda }}^{(II)}_5 & {\mathit{\Lambda }}^{(II)}_6 & {\mathit{\Lambda }}^{(II)}_7 & {\mathit{\Lambda }}^{(II)}_8 \\ 
{\mathit{Z}}^{(II)}_1 & {\mathit{Z}}^{(II)}_2 & {\mathit{Z}}^{(II)}_3 & {\mathit{Z}}^{(II)}_4 & {\mathit{Z}}^{(II)}_5 & {\mathit{Z}}^{(II)}_6 & {\mathit{Z}}^{(II)}_7 & {\mathit{Z}}^{(II)}_8 \\ 
{\chi }^{(II)}_1 & {\chi }^{(II)}_2 & {\chi }^{(II)}_3 & {\chi }^{(II)}_4 & {\chi }^{(II)}_5 & {\chi }^{(II)}_6 & {\chi }^{(II)}_7 & {\chi }^{(II)}_8 \\ 
{\mathit{\Phi }}^{(II)}_1 & {\mathit{\Phi }}^{(II)}_2 & {\mathit{\Phi }}^{(II)}_3 & {\mathit{\Phi }}^{(II)}_4 & {\mathit{\Phi }}^{(II)}_5 & {\mathit{\Phi }}^{(II)}_6 & {\mathit{\Phi }}^{(II)}_7 & {\mathit{\Phi }}^{(II)}_8 \\ 
{\mathit{H}}^{(II)}_1 & {\mathit{H}}^{(II)}_2 & {\mathit{H}}^{(II)}_3 & {\mathit{H}}^{(II)}_4 & {\mathit{H}}^{(II)}_5 & {\mathit{H}}^{(II)}_6 & {\mathit{H}}^{(II)}_7 & {\mathit{H}}^{(II)}_8 \\ 
{\mathit{\Gamma }}^{(II)}_1 & {\mathit{\Gamma }}^{(II)}_2 & {\mathit{\Gamma }}^{(II)}_3 & {\mathit{\Gamma }}^{(II)}_4 & {\mathit{\Gamma }}^{(II)}_5 & {\mathit{\Gamma }}^{(II)}_6 & {\mathit{\Gamma }}^{(II)}_7 & {\mathit{\Gamma }}^{(II)}_8 \end{array}
\right]\boldsymbol{.} \nonumber\\
\left[ \begin{array}{cccccccc}
e^{iq^{(II)}_1L}& 0 & 0 & 0 & 0 & 0 & 0 & 0 \\ 
0 & e^{iq^{(II)}_2L} & 0 & 0 & 0 & 0 & 0 & 0 \\ 
0 & 0 & e^{iq^{(II)}_3L} & 0 & 0 & 0 & 0 & 0 \\ 
0 & 0 & 0 & e^{iq^{(II)}_4L} & 0 & 0 & 0 & 0 \\ 
0 & 0 & 0 & 0 & e^{iq^{(II)}_5L} & 0 & 0 & 0 \\ 
0 & 0 & 0 & 0 & 0 & e^{iq^{(II)}_6L} & 0 & 0 \\ 
0 & 0 & 0 & 0 & 0 & 0 & e^{iq^{(II)}_7L} & 0 \\ 
0 & 0 & 0 & 0 & 0 & 0 & 0 & e^{iq^{(II)}_8L} \end{array}
\right]\nonumber
\end{aligned}
\:\:\:\:\:\:\:\:\:\:\:\:\:\:\:\:\:\:\:\:\:\:\:\:\:\:\:\:\:\:\:\:\:\:\:\:\:\:\:\:\:\:\:\:\:\:\:\:\:\:\:\:\:\:\ (E8)
\end{eqnarray}
\\\\
where, $L$ is the length of each subunit cell ($L_1=L_2=L$). 
\\\\
\noindent\textbf{Appendix F}\\
\\\\
\indent The weighting coefficients, its derivative, and arrangements of grid points are presented determined as:

\begin{eqnarray}
&& \displaystyle C^{(1)}_{pq}=\frac{M^{(1)}({\overline {x}}_p)}{(\overline {x}_p-\overline {x}_q)M^{(1)}(\overline {x}_q)
},{q}\neq{p}\nonumber\\
&&\displaystyle C^{(1)}_{pq}=-\sum_{q=1,q\neq{p}}^{N}C^{(1)}_{pq}\nonumber\:\:\:\:\:\:\:\:\:\:\:\:\:\:\:\:\:\:\:\:\:\:\:\:\:\:\:\:\:\:\:\:\:\:\:\:\:\:\:\:\:\:\:\:\:\:\:\:\:\:\:\:\:\:\:\:\:\:\:\:\:\:\:\:\:\:\:\:\:\:\:\:\:\:\:\:\:\:\:\:\:\:\:\:\:\:\:\:\:\:\:\:\:\:\:\:\:\:\:\:\:\:\:\:\:\:\:\:\:\:\:\:\:\:\:\:\:\:\:\:\:\:\:\:\:\:\:\:\:\:\:\:\:\:\:\:\:\:\:\:\:\:\:\:\:\:\:\:\:\:\:\:\    (F1)
\end{eqnarray}
\noindent
where
\begin{eqnarray}
&&\displaystyle M^{(1)}(\overline {x}_p)=\prod_{k=1,k\neq{p}}^{N} (\overline {x}_p-\overline {x}_k)\nonumber\:\:\:\:\:\:\:\:\:\:\:\:\:\:\:\:\:\:\:\:\:\:\:\:\:\:\:\:\:\:\:\:\:\:\:\:\:\:\:\:\:\:\:\:\:\:\:\:\:\:\:\:\:\:\:\:\:\:\:\:\:\:\:\:\:\:\:\:\:\:\:\:\:\:\:\:\:\:\:\:\:\:\:\:\:\:\:\:\:\:\:\:\:\:\:\:\:\:\:\:\:\:\:\:\:\:\:\:\:\:\:\:\:\:\:\:\:\:\:\:\:\:\:\:\:\:\:\:\:\:\:\:\:\:\:\      (F2)
\end{eqnarray}
\indent The higher-order derivatives are obtained as follows: 
\begin{eqnarray}
&& \displaystyle C^{(r)}_{pq}=r\Big[c^{(1)}_{pq}c^{(r-1)}_{pp}-\frac{c^{(r-1)}_{pq}}{\overline {x}_p-\overline {x}_q}\Big]\nonumber\\
&&\;\;\;\;\;\;\;\;\;\; p,q=1,2,...,N,\;\; r=2,3,...,N-1,\;\;p\neq{q}\nonumber\\
&&\displaystyle C^{(r)}_{pq}=-\sum_{q=1,q\neq{p}}^{N}C^{(r)}_{pq}\nonumber\:\:\:\:\:\:\:\:\:\:\:\:\:\:\:\:\:\:\:\:\:\:\:\:\:\:\:\:\:\:\:\:\:\:\:\:\:\:\:\:\:\:\:\:\:\:\:\:\:\:\:\:\:\:\:\:\:\:\:\:\:\:\:\:\:\:\:\:\:\:\:\:\:\:\:\:\:\:\:\:\:\:\:\:\:\:\:\:\:\:\:\:\:\:\:\:\:\:\:\:\:\:\:\:\:\:\:\:\:\:\:\:\:\:\:\:\:\:\:\:\:\:\:\:\:\:\:\:\:\:\:\:\:\:\:\:\:\:\:\:\:\:\:\:\:\ (F3)
\end{eqnarray}

The arrangements of grid points used in the GDQ method is followed by the Chebyshev-Gauss-Lobatto distribution as follows
\begin{eqnarray}
&&  \overline {x}_p=\frac{1}{2}\Big[1-cos(\frac{\pi(p-1)}{N-1})\Big] \;\;\;\;
and\;\;\;\; p=1,2,...,N \nonumber\:\:\:\:\:\:\:\:\:\:\:\:\:\:\:\:\:\:\:\:\:\:\:\:\:\:\:\:\:\:\:\:\:\:\:\:\:\:\:\:\:\:\:\:\:\:\:\:\:\:\:\:\:\:\:\:\:\:\:\:\:\:\:\:\:\:\:\:\:\:\:\:\:\:\:\:\:\:\:\:\:\:\:\:\:\:\:\:\:\:\:\:\ (F4)
\end{eqnarray}
\indent The discretized form of Eqs. (8) and (9) are presented as
\begin{eqnarray}
&&\Big(-{A_2}^{\alpha}\omega^2-{B_2}^{\alpha}\Big){({{\overline{u}_t}})}_{p}^{\alpha}+\sum\limits_{q=1}^{N}\frac{{B_3}^{\alpha}}{{L_\alpha}^2}C_{pq}^{(2)}{({{\overline{u}_t}})}_{q}^{\alpha}+\Big(-{A_4}^{\alpha}\omega^2+{B_2}^{\alpha}\Big){({{\overline{u}_b}})}_{p}^{\alpha}+\sum\limits_{q=1}^{N}\frac{{B_6}^{\alpha}}{{L_{\alpha}}^2}C_{pq}^{(2)}{({{\overline{u}_b}})}_{q}^{\alpha}\nonumber\\
&&+\sum\limits_{q=1}^{N}\Big(\frac{{-A_5}^{\alpha}\omega^2+{B_4}^{\alpha}}{{L_\alpha}}C_{pq}^{(1)}+\frac{{B_7}^{\alpha}}{{L_\alpha}^3}C_{pq}^{(3)}\Big){({{\overline{w}}})}_{q}^{\alpha}=0  \nonumber\:\:\:\:\:\:\:\:\:\:\:\:\:\:\:\:\:\:\:\:\:\:\:\:\:\:\:\:\:\:\:\:\:\:\:\:\:\:\:\:\:\:\:\:\:\:\:\:\:\:\:\:\:\:\:\:\:\:\:\:\:\:\:\:\:\:\:\:\:\:\:\:\:\:\:\:\:\:\:\:\:\:\:\:\:\:\:\:\:\:\:\ (F5)
\end{eqnarray}

\begin{eqnarray}
&&\sum\limits_{q=1}^{N}\Big(\frac{A_3^{\alpha}\omega^2+B_4^{\alpha}}{L_\alpha}C_{pq}^{(1)}-\frac{B_5^{\alpha}}{{L_\alpha}^3}C_{pq}^{(2)}\Big){({{\overline{u}_t}})}_{q}^{\alpha}+\sum\limits_{q=1}^{N}\Big(\frac{A_5^{\alpha}\omega^2-B_4^{\alpha}}{L_\alpha}C_{pq}^{(1)}-\frac{B_7^{\alpha}}{{L_\alpha}^3}C_{pq}^{(3)}\Big){({{\overline{u}_b}})}_{q}^{\alpha}\nonumber\\
&&+\sum\limits_{q=1}^{N}\Big(\frac{-A_6^{\alpha}\omega^2+B_9^{\alpha}}{{L_\alpha}^2}C_{pq}^{(2)}-\frac{B_8^{\alpha}}{{L_\alpha}^4}C_{pq}^{(4)}\Big){({{\overline{w}}})}_{q}^{\alpha}+{M_T}^{\alpha}\omega^2{({{\overline{w}}})}_{p}^{\alpha}=0  \nonumber\:\:\:\:\:\:\:\:\:\:\:\:\:\:\:\:\:\:\:\:\:\:\:\:\:\:\:\:\:\:\:\:\:\:\:\:\:\:\:\:\:\:\:\:\:\:\:\:\:\:\:\:\:\:\:\:\:\:\:\:\:\:\:\:\:\ (F6)
\end{eqnarray}

Continuity conditions on $\overline{P}_t$, $\overline{P}_b$, $\overline{S}$, and $\overline{M}$, between the subsequent subunit cells $\alpha$ and $\alpha+1$ $(\alpha=1,..., \beta)$, in which $\beta$ is the total number of subunit cells, are as follows
\begin{eqnarray}
&&\frac{B_1^{\alpha}}{L_{\alpha}}\sum\limits_{q=1}^{N}C_{Nq}^{(1)}{({{\overline{u}_t}})}_{q}^{\alpha}+\frac{B_3^{\alpha}}{L_{\alpha}}\sum\limits_{q=1}^{N}C_{Nq}^{(1)}{({{\overline{u}_b}})}_{q}^{\alpha}+\frac{B_5^{\alpha}}{{L_{\alpha}}^2}\sum\limits_{q=1}^{N}C_{Nq}^{(2)}{({{\overline{w}}})}_{q}^{\alpha}=\nonumber\\
&&\frac{B_1^{\alpha+1}}{L_{\alpha+1}}\sum\limits_{q=1}^{N}C_{1q}^{(1)}{({{\overline{u}_t}})}_{q}^{\alpha+1}+\frac{B_3^{\alpha+1}}{L_{\alpha+1}}\sum\limits_{q=1}^{N}C_{1q}^{(1)}{({{\overline{u}_b}})}_{q}^{\alpha+1}+\frac{B_5^{\alpha+1}}{{L_{\alpha+1}}^2}\sum\limits_{q=1}^{N}C_{1q}^{(2)}{({{\overline{w}}})}_{q}^{\alpha+1}\nonumber\:\:\:\:\:\:\:\:\:\:\:\:\:\:\:\:\:\:\:\:\:\:\:\:\:\:\:\:\:\:\:\:\:\:\:\:\:\:\:\:\:\:\:\:\:\:\:\:\:\ (F7)
\end{eqnarray}

\begin{eqnarray}
&&\frac{B_6^{\alpha}}{L_{\alpha}}\sum\limits_{q=1}^{N}C_{Nq}^{(1)}{({{\overline{u}_b}})}_{q}^{\alpha}+\frac{B_3^{\alpha}}{L_{\alpha}}\sum\limits_{q=1}^{N}C_{Nq}^{(1)}{({{\overline{u}_t}})}_{q}^{\alpha}+\frac{B_7^{\alpha}}{{L_{\alpha}}^2}\sum\limits_{q=1}^{N}C_{Nq}^{(2)}{({{\overline{w}}})}_{q}^{\alpha}=\nonumber\\
&&\frac{B_6^{\alpha+1}}{L_{\alpha+1}}\sum\limits_{q=1}^{N}C_{1q}^{(1)}{({{\overline{u}_b}})}_{q}^{\alpha+1}+\frac{B_3^{\alpha+1}}{L_{\alpha+1}}\sum\limits_{q=1}^{N}C_{1q}^{(1)}{({{\overline{u}_t}})}_{q}^{\alpha+1}+\frac{B_7^{\alpha+1}}{{L_{\alpha+1}}^2}\sum\limits_{q=1}^{N}C_{1q}^{(2)}{({{\overline{w}}})}_{q}^{\alpha+1}\nonumber\:\:\:\:\:\:\:\:\:\:\:\:\:\:\:\:\:\:\:\:\:\:\:\:\:\:\:\:\:\:\:\:\:\:\:\:\:\:\:\:\:\:\:\:\:\:\:\:\ (F8)
\end{eqnarray}

\begin{eqnarray}
&&\left(-{A_3}^{\alpha}\omega^2-{B_4}^{\alpha}\right){({{\overline{u}_t}})}_{N}^{\alpha}+\frac{{B_5}^{\alpha}}{{L_{\alpha}}^2}\sum\limits_{q=1}^{N}C_{Nq}^{(2)}{({{\overline{u}_t}})}_{q}^{\alpha}+
\left(-{A_5}^{\alpha}\omega^2+{B_4}^{\alpha}\right){({{\overline{u}_b}})}_{N}^{\alpha}+\frac{{B_7}^{\alpha}}{{L_{\alpha}}^2}\sum\limits_{q=1}^{N}C_{Nq}^{(2)}{({{\overline{u}_b}})}_{q}^{\alpha}
\nonumber\\
&&+\frac{\left({A_6}^{\alpha}\omega^2-{B_9}^{{\alpha}}\right)}{L_{\alpha}}\sum\limits_{q=1}^{N}C_{Nq}^{(1)}{({{\overline{w}}})}_{q}^{\alpha} +\frac{{B_8}^{\alpha}}{{L_{\alpha}}^3}\sum\limits_{q=1}^{N}C_{Nq}^{(3)}{({{\overline{w}}})}_{q}^{\alpha}=\left(-A_3^{\alpha+1}\omega^2-{B_4}^{\alpha+1}\right){({{\overline{u}_t}})}_{1}^{\alpha+1}+\frac{{B_5}^{\alpha+1}}{{L_{\alpha+1}}^2}\sum\limits_{q=1}^{N}C_{1q}^{(2)}{({{\overline{u}_t}})}_{q}^{\alpha+1} \nonumber\\
&&+\left(-A_5^{\alpha+1}\omega^2+{B_4}^{\alpha+1}\right){({{\overline{u}_b}})}_{1}^{\alpha+1}+\frac{{B_7}^{\alpha+1}}{{L_{\alpha+1}}^2}\sum\limits_{q=1}^{N}C_{1q}^{(2)}{({{\overline{u}_b}})}_{q}^{\alpha+1}+\frac{\left({A_6}^{\alpha+1}\omega^2-{B_9}^{\alpha+1}\right)}{L_{\alpha+1}}\sum\limits_{q=1}^{N}C_{1q}^{(1)}{({{\overline{w}}})}_{q}^{\alpha+1}\nonumber\\ &&+\frac{{B_8}^{\alpha+1}}{{L_{\alpha+1}}^3}\sum\limits_{q=1}^{N}C_{1q}^{(3)}{({{\overline{w}}})}_{q}^{\alpha+1}\nonumber\:\:\:\:\:\:\:\:\:\:\:\:\:\:\:\:\:\:\:\:\:\:\:\:\:\:\:\:\:\:\:\:\:\:\:\:\:\:\:\:\:\:\:\:\:\:\:\:\:\:\:\:\:\:\:\:\:\:\:\:\:\:\:\:\:\:\:\:\:\:\:\:\:\:\:\:\:\:\:\:\:\:\:\:\:\:\:\:\:\:\:\:\:\:\:\:\:\:\:\:\:\:\:\:\:\:\:\:\:\:\:\:\:\:\:\:\:\:\:\:\:\:\:\:\:\:\:\:\:\:\:\:\:\:\:\:\:\:\:\:\:\:\:\:\:\:\:\:\:\:\ (F9)
\end{eqnarray}

\begin{eqnarray}
&&\frac{B_5^{\alpha}}{L_{\alpha}}\sum\limits_{q=1}^{N}C_{Nq}^{(1)}{({{\overline{u}_t}})}_{q}^{\alpha}+\frac{B_7^{\alpha}}{L_{\alpha}}\sum\limits_{q=1}^{N}C_{Nq}^{(1)}{({{\overline{u}_b}})}_{q}^{\alpha}+\frac{B_8^{\alpha}}{{L_{\alpha}}^2}\sum\limits_{q=1}^{N}C_{Nq}^{(2)}{({{\overline{w}}})}_{q}^{\alpha}=\nonumber\\
&&\frac{B_5^{\alpha+1}}{L_{\alpha+1}}\sum\limits_{q=1}^{N}C_{1q}^{(1)}{({{\overline{u}_t}})}_{q}^{\alpha+1}+\frac{B_7^{\alpha+1}}{L_{\alpha+1}}\sum\limits_{q=1}^{N}C_{1q}^{(1)}{({{\overline{u}_b}})}_{q}^{\alpha+1}+\frac{B_8^{\alpha+1}}{{L_{\alpha+1}}^2}\sum\limits_{q=1}^{N}C_{1q}^{(2)}{({{\overline{w}}})}_{q}^{\alpha+1}
\nonumber\:\:\:\:\:\:\:\:\:\:\:\:\:\:\:\:\:\:\:\:\:\:\:\:\:\:\:\:\:\:\:\:\:\:\:\:\:\:\:\:\:\:\:\:\:\:\:\:\ (F10)
\end{eqnarray}

Periodic boundary conditions on $\overline{P}_t$, $\overline{P}_b$, $\overline{S}$, and $\overline{M}$, between the left side of first subunit cell $I$ and right side of the last subunit cell $\beta$ are as 

\begin{eqnarray}
&&\lambda\big[{\frac{B_1^{I}}{L_{I}}\sum\limits_{q=1}^{N}C_{1q}^{(1)}{({{\overline{u}_t}})}_{q}^{I}+\frac{B_3^{I}}{L_{I}}\sum\limits_{q=1}^{N}C_{1q}^{(1)}{({{\overline{u}_b}})}_{q}^{I}+\frac{B_5^{1}}{{L_{I}}^2}\sum\limits_{q=1}^{N}C_{1q}^{(2)}{({{\overline{w}}})}_{q}^{I}}\big]=\nonumber\\
&&\frac{B_1^{\beta}}{L_{\beta}}\sum\limits_{q=1}^{N}C_{Nq}^{(1)}{({{\overline{u}_t}})}_{q}^{\beta}+\frac{B_3^{\beta}}{L_{\beta}}\sum\limits_{q=1}^{N}C_{Nq}^{(1)}{({{\overline{u}_b}})}_{q}^{\beta}+\frac{B_5^{\beta}}{{L_{\beta}}^2}\sum\limits_{q=1}^{N}C_{Nq}^{(2)}{({{\overline{w}}})}_{q}^{\beta}\nonumber\
\:\:\:\:\:\:\:\:\:\:\:\:\:\:\:\:\:\:\:\:\:\:\:\:\:\:\:\:\:\:\:\:\:\:\:\:\:\:\:\:\:\:\:\:\:\:\:\:\:\:\:\:\:\:\:\:\:\:\:\:\:\:\:\:\:\:\:\:\:\:\:\:\:\:\:\:\:\:\:\:\:\: (F11)
\end{eqnarray}

\begin{eqnarray}
&&\lambda\big[{\frac{B_6^{I}}{L_{I}}\sum\limits_{q=1}^{N}C_{1q}^{(1)}{({{\overline{u}_t}})}_{q}^{I}+\frac{B_3^{1}}{L_{I}}\sum\limits_{q=1}^{N}C_{1q}^{(1)}{({{\overline{u}_b}})}_{q}^{I}+\frac{B_7^{1}}{{L_{I}}^2}\sum\limits_{q=1}^{N}C_{1q}^{(2)}{({{\overline{w}}})}_{q}^{1}}\big]=\nonumber\\
&&\frac{B_6^{\beta}}{L_{\beta}}\sum\limits_{q=1}^{N}C_{Nq}^{(1)}{({{\overline{u}_t}})}_{q}^{\beta}+\frac{B_3^{\beta}}{L_{\beta}}\sum\limits_{q=1}^{N}C_{Nq}^{(1)}{({{\overline{u}_b}})}_{q}^{\beta}+\frac{B_7^{\beta}}{{L_{\beta}}^2}\sum\limits_{q=1}^{N}C_{Nq}^{(2)}{({{\overline{w}}})}_{q}^{\beta}\nonumber\:\:\:\:\:\:\:\:\:\:\:\:\:\:\:\:\:\:\:\:\:\:\:\:\:\:\:\:\:\:\:\:\:\:\:\:\:\:\:\:\:\:\:\:\:\:\:\:\:\:\:\:\:\:\:\:\:\:\:\:\:\:\:\:\:\:\:\:\:\:\:\:\:\:\:\:\:\:\:\:\:\: (F12)
\end{eqnarray}

\begin{eqnarray}
&&\lambda\big[\left(-{A_3^I}\omega^2-{B_4^I}\right){({{\overline{u}_t}})}_{1}^{I}+\frac{{B_5^I}}{{L_{I}}^2}\sum\limits_{q=1}^{N}C_{1q}^{(2)}{({{\overline{u}_t}})}_{q}^{I}+
\left(-{A_5^I}\omega^2+{B_4^I}\right){({{\overline{u}_b}})}_{1}^{I}+\frac{{B_7^I}}{{L_{I}}^2}\sum\limits_{q=1}^{N}C_{1q}^{(2)}{({{\overline{u}_b}})}_{q}^{I}
\nonumber\\
&&+\frac{\left({A_6}^{I}\omega^2-{B_9}^{{I}}\right)}{L_{I}}\sum\limits_{q=1}^{1}C_{1q}^{(1)}{({{\overline{w}}})}_{q}^{I} +\frac{{B_8}^{I}}{{L_{I}}^3}\sum\limits_{q=1}^{N}C_{1q}^{(3)}{({{\overline{w}}})}_{q}^{I}\big]=\left(-A_3^{\beta}\omega^2-{B_4}^{\beta}\right){({{\overline{u}_t}})}_{N}^{\beta}+\frac{{B_5}^{\beta}}{{L_{\beta}}^2}\sum\limits_{q=1}^{N}C_{Nq}^{(2)}{({{\overline{u}_t}})}_{q}^{\beta} \nonumber\\
&&+\left(-A_5^{\beta}\omega^2+{B_4}^{\beta}\right){({{\overline{u}_b}})}_{N}^{\beta}+\frac{{B_7}^{\beta}}{{L_{\beta}}^2}\sum\limits_{q=1}^{N}C_{Nq}^{(2)}{({{\overline{u}_b}})}_{q}^{\beta}+\frac{\left({A_6}^{\beta}\omega^2-{B_9}^{\beta}\right)}{L_{\beta}}\sum\limits_{q=1}^{N}C_{Nq}^{(1)}{({{\overline{w}}})}_{q}^{\beta}\nonumber\\ &&+\frac{{B_8}^{\beta}}{{L_{\beta}}^3}\sum\limits_{q=1}^{N}C_{Nq}^{(3)}{({{\overline{w}}})}_{q}^{\beta}\nonumber\:\:\:\:\:\:\:\:\:\:\:\:\:\:\:\:\:\:\:\:\:\:\:\:\:\:\:\:\:\:\:\:\:\:\:\:\:\:\:\:\:\:\:\:\:\:\:\:\:\:\:\:\:\:\:\:\:\:\:\:\:\:\:\:\:\:\:\:\:\:\:\:\:\:\:\:\:\:\:\:\:\:\:\:\:\:\:\:\:\:\:\:\:\:\:\:\:\:\:\:\:\:\:\:\:\:\:\:\:\:\:\:\:\:\:\:\:\:\:\:\:\:\:\:\:\:\:\:\:\:\:\:\:\:\:\:\:\:\:\:\:\:\:\:\:\:\:\:\:\:\:\:\:\:\ (F13)
\end{eqnarray}

\begin{eqnarray}
&&\lambda\big[{\frac{B_5^{I}}{L_{I}}\sum\limits_{q=1}^{N}C_{1q}^{(1)}{({{\overline{u}_t}})}_{q}^{I}+\frac{B_7^{1}}{L_{I}}\sum\limits_{q=1}^{N}C_{1q}^{(1)}{({{\overline{u}_b}})}_{q}^{I}+\frac{B_8^{1}}{{L_{I}}^2}\sum\limits_{q=1}^{N}C_{1q}^{(2)}{({{\overline{w}}})}_{q}^{I}}\big]=\nonumber\\
&&\frac{B_6^{\beta}}{L_{\beta}}\sum\limits_{q=1}^{N}C_{Nq}^{(1)}{({{\overline{u}_t}})}_{q}^{\beta}+\frac{B_3^{\beta}}{L_{\beta}}\sum\limits_{q=1}^{N}C_{Nq}^{(1)}{({{\overline{u}_b}})}_{q}^{\beta}+\frac{B_7^{\beta}}{{L_{\beta}}^2}\sum\limits_{q=1}^{N}C_{Nq}^{(2)}{({{\overline{w}}})}_{q}^{\beta}\nonumber\:\:\:\:\:\:\:\:\:\:\:\:\:\:\:\:\:\:\:\:\:\:\:\:\:\:\:\:\:\:\:\:\:\:\:\:\:\:\:\:\:\:\:\:\:\:\:\:\:\:\:\:\:\:\:\:\:\:\:\:\:\:\:\:\:\:\:\:\:\:\:\:\:\:\:\:\:\:\:\:\:\ (F14)
\end{eqnarray}

\noindent\textbf{Appendix G}\\

\indent Equivalent properties of the sandwich beam using Timoshenko beam theory can be obtained as \cite{chen2011dynamic}

\begin{eqnarray}
&& EI=E_fb(h_c^2hf/2+h_ch_f^2) \nonumber\\
&& GA=G_cb(h_c+2h_f)\nonumber\\
&& \rho A=2\rho_fbh_f+\rho_cbh_c\nonumber\\
&& \rho I=\rho_fb(h_c^2h_f/2+h_ch_f^2)+\rho_c bh_c^3/12
\nonumber\:\:\:\:\:\:\:\:\:\:\:\:\:\:\:\:\:\:\:\:\:\:\:\:\:\:\:\:\:\:\:\:\:\:\:\:\:\:\:\:\:\:\:\:\:\:\:\:\:\:\:\:\:\:\:\:\:\:\:\:\:\:\:\:\:\:\:\:\:\:\:\:\:\:\:\:\:\:\:\:\:\:\:\:\:\:\:\:\:\:\:\:\:\:\:\:\:\:\:\:\:\:\:\:\:\:\:\:\:\:\:\:\:\:\:\:\:\   (G1)
\end{eqnarray}

\noindent where $EI$, $GA$, $\rho A$, and $\rho I$ are the equivalent bending rigidity, transverse shear rigidity, mass per unit length, and rotary inertia of sandwich beam. $E_f$ is the young's modulus of the faces sheets, $G_c$ is the shear modulus of the core, $\rho_f$ and $\rho_c$ are densities of face sheets and core layers. $h_f$ and $h_c$ are the thickness of face sheets and core layers respectively.

\bibliographystyle{unsrt}  


\end{document}